\documentclass[aps,reprint,pra,amsmath,amssymb,nofootinbib,superscriptaddress]{revtex4-1}

\usepackage[english]{babel}
\usepackage[utf8x]{inputenc}
\usepackage[T1]{fontenc}

\usepackage{amsmath}
\newcommand{\angstrom}{\textup{\AA}}
\usepackage{physics}

\usepackage{mathtools}
\DeclarePairedDelimiter{\ceil}{\lceil}{\rceil}

\usepackage{float}
\restylefloat{table}

\usepackage{graphicx} 
\usepackage[colorlinks=true, allcolors=blue]{hyperref}
\usepackage{booktabs}

\usepackage[linesnumbered,ruled,vlined]{algorithm2e}

\SetCommentSty{mycommfont}
\SetKwInput{KwInput}{Input}    
\SetKwInput{KwOutput}{Output}  

\usepackage[dvipsnames]{xcolor}

\begin{document}

\title{Adaptive pruning-based optimization of parameterized quantum circuits}

\author{Sukin Sim}
\email{ssim@g.harvard.edu}
\affiliation{%
 Department of Chemistry and Chemical Biology, Harvard University, 12 Oxford Street, \protect\\ Cambridge, MA 02138, USA
}%
\affiliation{%
 Zapata Computing, Inc., 100 Federal Street, \protect\\ Boston, MA 02110, USA
}%

\author{Jonathan Romero}
\affiliation{%
 Zapata Computing, Inc., 100 Federal Street, \protect\\ Boston, MA 02110, USA
}%

\author{J\'er\^ome F. Gonthier}
\affiliation{%
 Zapata Computing, Inc., 100 Federal Street, \protect\\ Boston, MA 02110, USA
}%

\author{Alexander A. Kunitsa}
\affiliation{%
 Zapata Computing, Inc., 100 Federal Street, \protect\\ Boston, MA 02110, USA
}%

\begin{abstract}
Variational hybrid quantum-classical algorithms are powerful tools to maximize the use of Noisy Intermediate Scale Quantum devices.
While past studies have developed powerful and expressive ansatze, their near-term applications have been limited by the difficulty of optimizing in the vast parameter space.
In this work, we propose a heuristic optimization strategy for such ansatze used in variational quantum algorithms, which we call ``Parameter-Efficient Circuit Training'' (PECT).
Instead of optimizing all of the ansatz parameters at once, PECT launches a sequence of variational algorithms,
in which each iteration of the algorithm activates and optimizes a subset of the total parameter set.
To update the parameter subset between iterations, we adapt the \emph{dynamic sparse reparameterization} scheme by Mostafa \textit{et al.} (arXiv:1902.05967). 
We demonstrate PECT for the Variational Quantum Eigensolver, in which we benchmark unitary coupled-cluster ansatze
including UCCSD and $k$-UpCCGSD, as well as the low-depth circuit ansatz (LDCA), to estimate ground state energies of molecular systems.
We additionally use a layerwise variant of PECT to optimize a
hardware-efficient circuit for the 
Sycamore processor 
to estimate the ground state energy densities of the one-dimensional Fermi-Hubbard model.
From our numerical data, we find that PECT can enable optimizations of certain ansatze that were previously difficult to converge and more generally can improve the performance of variational algorithms by reducing the optimization runtime and/or 
the depth of
circuits that encode the solution candidate(s).
\end{abstract}

\maketitle

\section{Introduction}\label{sec:introduction}

Variational quantum algorithms (VQAs) were developed to maximize the capabilities of Noisy Intermediate-Scale Quantum (NISQ) computers \cite{Preskill2018}.
Two early examples of VQAs are the variational quantum eigensolver \cite{Peruzzo2014} and the quantum approximate optimization algorithm \cite{Farhi2014}. 
In addition to applications in quantum chemistry and combinatorial optimization, variational algorithms have more recently been applied to address a range of tasks in machine learning, including data classification \cite{Havlicek2018, Farhi2018, SchuldCircuitCentric}, data compression \cite{Romero2017}, and generative modeling \cite{Dallaire-DemersQGAN, Lloyd2018, Zeng2019, Situ2020, Zhu2019, Romero2019}.

Within the VQA framework, a trial wavefunction is prepared on a quantum device by executing a parameterized quantum circuit. 
This is followed by repeatedly measuring the state to estimate the expectation value of some Hermitian operator with respect to the current trial wavefunction. These expectation values are then used to evaluate an objective function which a classical optimizer maximizes or minimizes by varying the parameter values of the quantum circuit. 

While VQAs are promising candidates for 
demonstrating advantage 
of early quantum computers, one of the main challenges of realizing VQAs is effectively optimizing the parameters of the tunable ansatz.
In recent years, there have been significant efforts in various directions to better understand and improve this circuit optimization step.
Such directions include analyzing the trainability of parameterized quantum circuits (PQCs) \cite{McClean2018, Cerezo2020, Volkoff2020}, 
improving existing or proposing new optimizers \cite{Ostaszewski2019, Wilson2019, Grimsley2019, Tang2019, Stokes2019, Rattew2019, Kubler2020, Chivilikhin2020, Sung2020}, 
employing high-level optimization strategies on top of optimizers (e.g. adiabatically-assisted VQAs, layerwise training) \cite{Wecker2015, Garcia-Saez2018, Skolik2020}, 
devising more intelligent parameter initialization strategies \cite{Grant2019, Wilson2019, Verdon2019}, 
and running comprehensive optimizer benchmarks \cite{Lavrijsen2020}. 
Despite the rapid progress, optimization of PQCs remains a significant challenge especially in the NISQ era: we ideally want a \textit{low-depth} ansatz that is able to \textit{accurately} describe some quantum system-of-interest and is also relatively \textit{easy to optimize}.

To approach this problem, in this work we propose an optimization strategy that leverages past developments of powerful ansatze, which we call ``Parameter-Efficient Circuit Training'' (PECT). 
That is, numerous past studies have developed and analyzed ansatze that have the potential to express or well-describe quantum systems-of-interest, e.g. ground states of (strongly correlated) fermionic systems \cite{Shen2017, Babbush2017, Kivlichan2018, Barkoutsos2018, Dallaire-Demers2018, Lee2019, Choquette2020}.
Examples include unitary coupled-cluster with singles and doubles excitations (UCCSD), $k$-layered generalized variant of UCC wavefunction ($k$-UpCCGSD), and the low-depth circuit ansatz (LDCA) \cite{Yung2015, Lee2019, Dallaire-Demers2018}.
However, applications of these ansatze are often inhibited by the large numbers of variational parameters to tune.
We leverage the fact that these parameterized ansatze often contain superfluous parameters, i.e. parameters that are redundant and thus 
are not strictly necessary 
for preparing a state that 
achieves the objective(s). 
Our optimization strategy exploits this characteristic of PQCs, without prior knowledge of the target state, by 
limiting the optimization task to a subset of the total pool of ansatz parameters
and refining the parameter subset by updating both the corresponding gate composition and parameter values.

Using this strategy, we show that we can efficiently prepare ground states of various molecular systems using ansatze that have high parameter counts and are redundant in parameterization. 
For instance, using 30 out of 60 parameters in the $2$-UpCCGSD ansatz to estimate ground states of lithium hydride, we observed a 28\% reduction in circuit depth and 42\% reduction in two-qubit gate count on average when using PECT to optimize the parameterized ansatz.
We additionally estimate optimization runtimes using circuit depths and numbers of function evaluations.
We show that using PECT with ansatze such as $k$-UpCCGSD and LDCA, one can significantly reduce the optimization runtime estimates.
For instance, averaged over bond lengths considered in VQE calculations, 
using PECT to optimize LDCA circuits modeling ground states of lithium hydride led to a 73\% reduction in estimated optimization runtime, which in this case was achieved by reducing both the circuit depths and numbers of function calls.

The rest of the paper is organized as follows.
We first introduce our optimization strategy in detail in Section \ref{sec:pect_method}.
Our proposed strategy, 
PECT, is adapted from the ``dynamic sparse reparameterization'' (DSR) method that has been applied to train deep convolutional neural networks \cite{Mostafa2019}.
We then describe the PECT method applied in the context of the Variational Quantum Eigensolver (VQE) as an example of a VQA.
In Section \ref{sec:numerical_experiments}, we numerically demonstrate the utility of PECT by optimizing VQE circuits for lithium hydride and the symmetric stretching of O-H bonds in a water molecule using 
UCCSD, $k$-UpCCGSD, and LDCA.
While PECT can provide advantages in energy accuracy, circuit depths, and optimization runtimes, we show that certain ansatze are better suited for PECT than others.
In Section \ref{sec:discussion}, we further analyze our simulations and discuss our observations on the robustness of parameter initialization for LDCA, prunability of the three ansatze, reductions in circuit resources due to PECT, and estimation of optimization runtimes.
We additionally propose a method to combine PECT with a layerwise circuit optimization strategy.
Using the layerwise PECT approach, we show that we can efficiently optimize an ansatz that has recently been proposed to estimate the ground state energy densities of the one-dimensional Fermi-Hubbard model at various chain lengths \cite{Dallaire-Demers2020}.
Lastly, we comment on the observed parameter dynamics in the optimization of multi-layered PQCs, in which parameters of earlier layers appear to change less, on average, throughout the optimization compared to parameters of later layers.
We conclude with a summary and outlook of future directions for this work.

\section{Parameter-efficient circuit training (PECT)}\label{sec:pect_method}

\begin{figure*}[ht]
\centering
\includegraphics[width=0.45\textwidth]{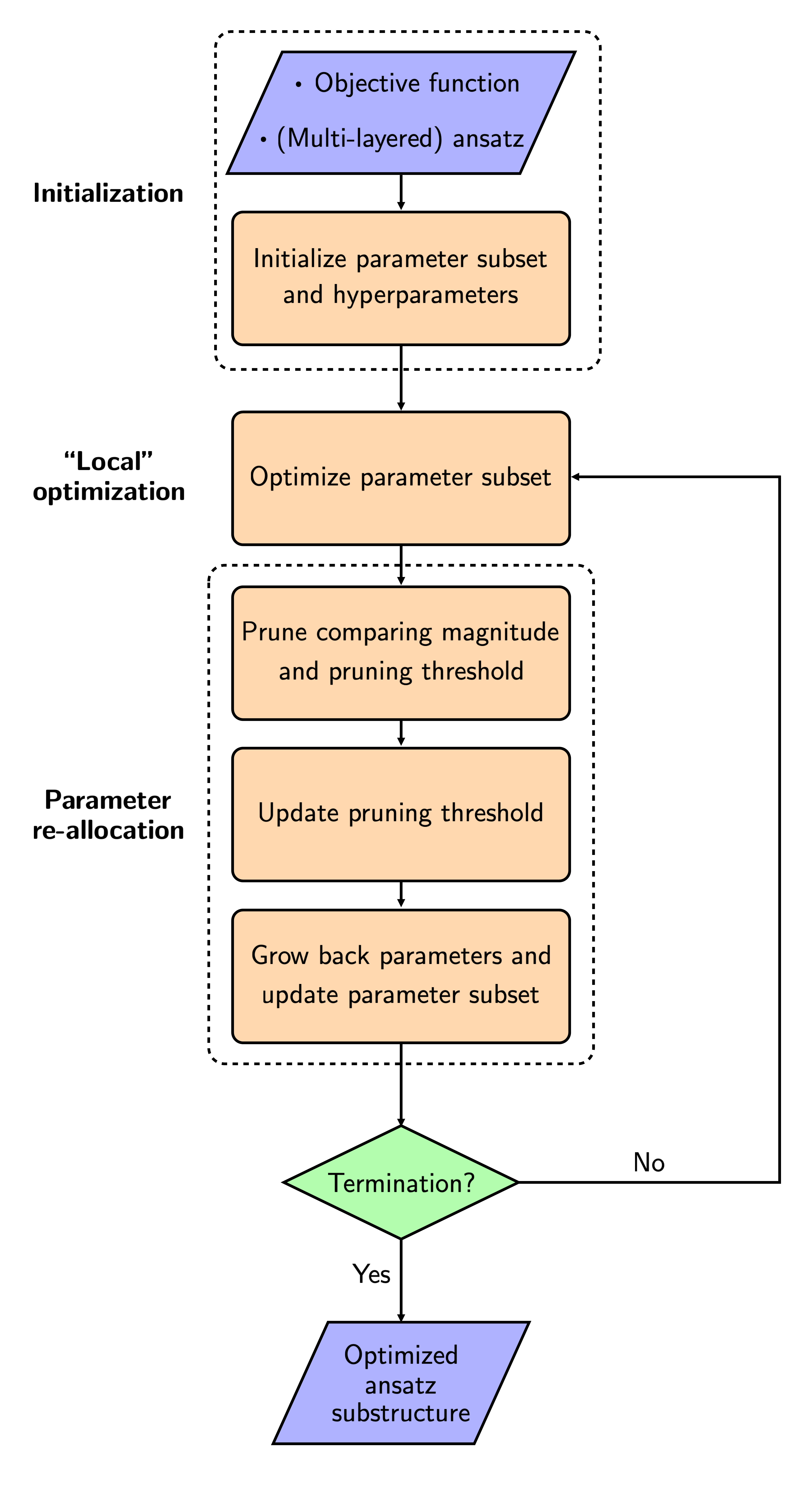}
\caption{
Flow diagram outlining main steps of the ``parameter-efficient circuit training'' (PECT) method.
In addition to inputs of a variational quantum algorithm, e.g. objective function and some (multi-layered) ansatz, several hyperparameters for PECT are provided by the user, e.g. global sparsity $s$ of the parameterized ansatz and the target number of parameters to prune at each iteration $N_p$.
After initializing the parameter subset, corresponding to an ansatz substructure, parameters within the subset are tuned using a classical optimizer, e.g. L-BFGS-B. We call this step a ``local'' optimization.
Once the parameter values of the subset have been tuned, the parameter subset is further updated in the parameter re-allocation step.
The re-allocation step consists of magnitude-based pruning of parameters/gates based on a threshold followed by updates of the pruning threshold for the next iteration and of the parameter subset by growing or adding the same number of parameters pruned. 
The local optimization and parameter re-allocation steps are repeated until some termination criteria are satisfied.
}
\label{fig:pect_scheme}
\end{figure*}

In the traditional implementation of a variational quantum algorithm, there is typically a single optimization, in which the classical optimization routine is tasked with tuning every angle of the PQC to find parameter values that correspond to a minimum (or maximum) of an objective function. While this strategy is viable for a limited number of parameters, as we scale the algorithm to larger systems, larger circuits with more variational parameters may be needed. As a direct result, the classical optimizers expend a substantially larger number of function evaluations from the quantum computer and be-come over-burdened when attempting to search over the vast parameter space.

The strategy employed by PECT turns this large optimization problem into a series of optimizations in which, at each iteration, an optimization is performed only on a subset of the whole parameter set. We show the overall flow diagram of PECT in Figure \ref{fig:pect_scheme}. In the following description of PECT, we adopt notations used in Ref. \cite{Mostafa2019} to provide an analogous description of their training method in the context and language of parameterized quantum circuits.

To initialize PECT, we first define the variational quantum algorithm and the hyperparameters for PECT. The variational algorithm can be defined by choosing an objective function and an $L$-layered parameterized quantum circuit. Many variational quantum algorithms make use of multi-layered parameterized quantum circuits, in which layers of a circuit template are repeated with the goal of increasing the expressiveness of circuits \cite{Sim2019}. We consider a case in which a PQC is described by the set of its circuit parameters, grouped by their corresponding layers: $\{ \boldsymbol\theta_l \}$, where $l \in [L]$ index the layers. We let $N$ be the total number of parameters in the PQC, with $N_l$ denoting the number of parameters in each circuit layer $l$. 
Before implementing PECT, there are two assumptions for the PQC-at-hand:
\begin{enumerate}
    \item The ansatz $\ket{\psi(\boldsymbol\theta)}$ is able to express the solution space/state and the ansatz is ``prunable,'' i.e. contains parameterized gate operations that are not necessary to generate the solution space.
    This may be due to redundant parameterization.
    We claim such redundancy is a valid assumption for many PQC designs with some commonly used patterns of gates, e.g. single-qubit rotations on every qubit.
    In this case, it was recently shown that not all single-qubit rotations are necessary for a PQC to be highly expressive \cite{Rasmussen2020}.
    \item Each gate operation of the PQC is constructed such that setting its corresponding parameter value to $0$ corresponds to implementing the trivial (identity) gate operation.
    This is a valid assumption for a wide class of parameterized gates of the form $G(\theta) = e^{-i a \theta P}$ for some Hermitian operator $P$ and constant $a$.
\end{enumerate}

While there are several hyperparameters in PECT, we introduce one particular hyperparameter, \textit{global sparsity}, here and reserve describing the others in the following subsection, where they are more relevant. The global sparsity, denoted by $s$, is defined as $s = 1 - \frac{M}{N}$, where $M$ is the size of the parameter subset to be optimized. Thus, for example, a low value for $s$ where $s \in (0, 1)$ implies a larger fraction of the total parameter set PECT optimizes in each iteration. After choosing the global sparsity, we define the parameter subset for the first iteration. Gates with parameters that are not included in the subset are replaced by the identity operation, for example by setting angles to 0 in the case of Pauli rotations, resulting in shorter circuits both in depth and number of operations after compilation.

To define the parameter subsets, we utilize the sparse parameter representation in Ref. \cite{Mostafa2019}, in which a sparse parameter vector for layer $l$, or $\boldsymbol\theta_l$, can be described using two vectors $\boldsymbol\psi_l$ and $\boldsymbol\phi_l$. Elements of $\boldsymbol\psi_l$ are the indices of $\boldsymbol\theta_l$ that correspond to parameters with nonzero values, and elements of $\boldsymbol\phi_l$ correspond to those nonzero values. 
We let $M_l$ be the size of $\boldsymbol\psi_l$ and $\boldsymbol\phi_l$, or the number of nonzero parameters in $\boldsymbol\theta_l$. The total number of nonzero parameters in the full parameter vector $\boldsymbol\theta$ is thus $M=\sum_l M_l$. This allows us to define the local sparsity $s_l = 1 - \frac{M_l}{N_l}$, the sparsity of parameters of each layer, in addition to the global sparsity $s$. While $s$ is kept constant throughout iterations in PECT, compositions of $\boldsymbol\phi_l$, $\boldsymbol\psi_l$, and $M_l$ (and thus $s_l$) may change between iterations.

To initialize the parameter subset, we uniformly sample $M_l^{(0)} = (1-s)N_l$ parameter positions in each circuit layer, which defines $\boldsymbol\psi_l$ at iteration $t=0$ or $\boldsymbol\psi_l^{(0)}$.
When generating $M_l^{(0)}$ for each layer, care must be taken in rounding $M_l$'s for all $l$ such that the global sparsity is kept constant.
The corresponding initial parameter values, $\boldsymbol\phi_l^{(0)}$ are either uniformly sampled, e.g. from the range $[0, 4\pi]$ for each rotation angle, or use informed guesses such as MP2 amplitudes for the unitary coupled-cluster ansatz \cite{Romero2017strategies}.
Once initialized, the parameter subset provides the description for an ansatz substructure, and the corresponding parameter values $\boldsymbol\phi$ ($\boldsymbol\phi_l$ for all $l$) are tuned to maximize or minimize the objective function using some optimizer, e.g. L-BFGS-B \cite{Byrd1995} or SLSQP \cite{Kraft1988}. 
We refer to such optimization of an ansatz substructure as a ``local optimization'' (as shown in Figure \ref{fig:pect_scheme}) and the classical optimizer used as a ``local optimizer.'' One of the design choices that is left to the user is the choice of local optimizer and the corresponding hyperparameters. Choice of the local optimizer may depend on factors such as the maximum number of function evaluations or measurements the user would like to allocate to each local optimization step or the availability of high-precision gradients.

Once the optimization is complete, both $\boldsymbol\psi_l$, which was defined before the local optimization step, and $\boldsymbol\phi_l$, which was just optimized, are fed as inputs to the \textit{parameter reallocation} algorithm to update the parameter subset. 
While we describe the parameter reallocation algorithm in greater detail in the following subsection, the goal of reallocating parameters is to update the parameter subset by removing and adding parameters the algorithm deems unimportant and important, respectively.
The sequence of local optimization of a parameter subset followed by the parameter reallocation subroutine is repeated until some termination criteria are met. Our implementation utilizes a combination of termination criteria including convergence in the value of the objective function,
limit on the overall maximum number of function calls (e.g. $5 \times 10^{5}$ for optimizing LDCA circuits for estimating ground states of LiH), and detection of oscillating behavior in the final objective function values of the four latest PECT iterations. By the end of the PECT procedure, we obtain the optimized ansatz substructure that is a solution candidate for the variational quantum algorithm.

Overall, one can think of PECT as a procedure that hops over and optimizes different sparse parameterizations of an ansatz, which correspond to different ansatz substructures. In the following subsection, we describe the parameter reallocation algorithm in greater detail.

\subsection{Parameter reallocation for parameterized quantum circuits}\label{sec:parameter_reallocation}

The parameter reallocation algorithm was proposed by Ref. \cite{Mostafa2019} to sparsely re-parameterize layers of deep convolutional neural networks every few hundred training epochs.
Similarly, PECT employs the algorithm to re-parameterize or update the subset of PQC parameters. 
In the previous section, the global sparsity $s$ was introduced as a hyperparameter for PECT. In addition to $s$, other hyperparameters of the parameter reallocation scheme include $N_p$, $H_{(0)}$, and $\delta$.
The hyperparameter $N_p$ is the target number of parameters to prune at each iteration. 
One can think of $N_p$ as a way of quantifying to what extent the user would like to ``shuffle'' the parameter subset per PECT iteration.
One would assign higher values to $N_p$ if the circuit at hand is highly redundant and thus the contribution of each parameter is not as significant.
While $N_p$ was fixed in the original formulation, one can in principle implement a schedule to vary $N_p$ as the iterations progress. 
Next, $H_{(0)}$ 
corresponds to the initial pruning threshold.
In our numerical experiments, values of $H_{(0)}$ were assigned based on the range over which random parameter values were sampled. That is, we chose a threshold value that was relatively small, e.g. $H_{(0)} = 0.01$ for range $[0, 4\pi]$ to prevent too many parameters being pruned in the first iteration.
For cases in which we employed informative parameter guesses, e.g. using MP2 amplitudes to initialize UCCSD operators \cite{Romero2017strategies}, we chose values of $H_{(0)}$ that were slightly less than the average of the parameter guess magnitudes.
Lastly, $\delta$, which defines the tolerance or neighborhood about $N_p$ (see Algorithm \ref{algo:threshold_update}), was fixed to be $0.1$ for all of our numerical simulations after observing through preliminary calculations that $\delta$ had no significant effect on the performance of PECT.

Inputs to the parameter reallocation algorithm are $\{ \boldsymbol\psi_l \}$, $\{ \boldsymbol\phi_l \}$, $M^{(t)}$, $H_{(t)}$, which correspond to (sparse) representations of the current parameter vector, the size of the parameter subset at the current iterate $t$, and the current pruning threshold respectively.
The re-definition or reallocation of the parameter subset comprises three main steps: (1) magnitude-based pruning, (2) adjustment of the pruning threshold, and (3) redistribution of free parameter slots.
In step (1), for each layer of the PQC at iterate $t$, parameters (and their corresponding gates) are pruned or removed from the parameter subset based on the current value of the pruning threshold, $H_{(t)}$.
That is, if the absolute value of each parameter in $\boldsymbol\phi_l$ is less than the value of $H_{(t)}$, the parameter is removed from the parameter subset (from both $\boldsymbol\psi_l$ and $\boldsymbol\phi_l$), and its corresponding gate operation is removed from the ansatz substructure.
Assuming convergence of the local optimization before the pruning step, the gradient should be zero or near-zero, justifying the removal of parameterized gates with near-zero parameter magnitudes.
At this step, the numbers of parameters that are pruned and survived in each circuit layer are stored and denoted as $K_l^{(t)}$ and $R_l^{(t)} = M_l^{(t)} - K_l^{(t)}$ respectively. 
In the following step of the parameter reallocation subroutine, we adjust the pruning threshold for the next iteration, $H_{(t+1)}$, by comparing $K = \sum_l K_l$, the total number of parameters pruned, to the target number $N_p$:
\begin{algorithm}[H]
\DontPrintSemicolon
  \KwInput{$K$, $H_{(t)}$, $N_p$, $\delta$}
  \KwOutput{$H_{(t+1)}$}
  \tcp{Case 1: Too many parameters pruned}
  \If{$K > (1+\delta) N_p$}
    {
        $H_{(t+1)} = \frac{H_{(t)}}{2}$
    }
    \tcp{Case 2: Too few parameters pruned}
    \ElseIf{$K < (1-\delta) N_p$}
    {
        $H_{(t+1)} = 2 H_{(t)}$
    }
    \Else
    {
    	$H_{(t+1)} = H_{(t)}$
    }
\caption{Pruning threshold update}
\label{algo:threshold_update}
\end{algorithm}
\noindent If too many parameters are pruned compared to the target number $N_p$, the pruning threshold for the next iteration is lowered, and if too few parameters are pruned, the pruning threshold is increased.
The user can customize the rate of increase or decrease for the pruning threshold. 
In our implementation, we multiply or divide by $2$ in respective cases.
In the final step of parameter reallocation subroutine, we grow or add $K$ parameter slots in total (and their corresponding gate operations) to the PQC and update $\boldsymbol\psi_l$, $\boldsymbol\phi_l$, and $M_l$ for the next iteration.
The growth rate for each circuit layer is determined by how many parameters in the original subset ``survived'' the pruning phase, which was quantified by $R_l$.
That is, the heuristic growth rate for each layer $l$ at iterate $t$ is defined as:
\begin{align}
\label{eq:growth_rate}
G_l^{(t)} = \frac{R_l^{(t)}}{\sum_l R_l^{(t)}} \sum_l K_l^{(t)}.
\end{align}
We note that care must be taken into rounding the growth rates for all $l$ such that the total number of grown parameters is equal to the total number of pruned parameters which was determined from the first step of the parameter reallocation scheme. This keeps the global sparsity $s$ constant.\footnote{For the case in which the updated number of parameters at layer $l$ for iteration $t+1$ exceeds the total number of parameters in that layer, i.e. $M_l^{(t+1)} > N_l$, we redistribute the excess parameters randomly to different circuit layers, as was done in \cite{Mostafa2019}.}
In practice, parameters are grown by updating $\boldsymbol\psi_l$, in which $G_l^{(t)}$ parameter positions or indices that have not been activated or selected in the current $\boldsymbol\psi_l$ are randomly sampled for circuit layer $l$. We then initialize the corresponding parameter values in $\boldsymbol\phi_l$ to $0$. Lastly, we update the parameter subset size for each layer: $M_l^{(t+1)} = M_l^{(t)} - K_l^{(t)} + G_l^{(t)}$. 
As previously discussed, according to Equation \ref{eq:growth_rate}, more parameter slots are to be allocated to PQC layers that have greater numbers of surviving parameters. 
The algorithm deems these layers as being more important to the optimization than circuit layers that had greater number of gates with parameters near $0$ and thus were pruned or removed. 
While further investigation is needed to determine 
if circuit layers with fewer pruned parameters are indeed more important, e.g. correspond to parameters with larger gradient magnitudes,
this heuristic approach affords us a way of adding free parameter slots to the parameter subset without making calls to the quantum computer, which, especially in the NISQ era, are the main bottlenecks in the performance of variational quantum algorithms.

\subsection{Prior work}\label{sec:prior_work}
PECT combines and leverages two ideas that have been explored in past studies: dynamic construction of the parameterized ansatz and ansatz reduction by pruning.
The idea of adaptively constructing an ansatz for variational quantum algorithms was proposed in ADAPT-VQE \cite{Grimsley2019}, in which the method was shown to be effective for preparing ground states of various molecular systems.
While ADAPT-VQE iteratively adds or grows a parameterized ansatz from an \textit{unordered} pool of quantum operations based on the corresponding gradients one operation at a time, 
the core objective in PECT is to find a compact representation of the solution state using a subset of gates from a pre-defined ansatz structure (i.e. an \textit{ordered} pool of quantum operations) and then pruning and growing multiple gates in a single iteration to refine the ansatz substructure. 
Second, the idea of pruning an ansatz was explored in a previous work that developed the ``Ansatz Architecture Search'' (AAS) to solve QAOA problem instances using fewer two-qubit gates \cite{Li2020}.
In each iteration, the AAS method reduces the current circuit architecture by generating all the unique architectures after removing one two-qubit operation. 
These pruned architectures are then ranked according to some defined scoring function, and circuit architectures with the best scores are chosen to be candidates for pruning for the following iteration. 
PECT also involves pruning or removal of gates, but in each iteration, multiple gates can be removed based on their optimized parameter values and the current pruning threshold.
In addition to pruning, PECT also involves an ansatz growth phase to reallocate a number of parameters/gates to update the ansatz substructure.

\section{Numerical experiments}\label{sec:numerical_experiments}

In this work, we use the Variational Quantum Eigensolver (VQE) as the VQA-of-choice to demonstrate optimization of parameterized ansatze using PECT.
While running a sequence of VQEs may seem like a costly task, each VQE instance likely has a shorter optimization runtime due to fewer circuit parameters to explore and update.
In addition, activating a fraction of the parameters likely corresponds to executing circuits with fewer gate operations and lower circuit depth on the quantum computer.
Although we did not apply this idea in our simulations, one might also consider establishing a maximum number of measurements to use in each local optimization, and applying techniques such as the one presented in Ref. \cite{Kubler2020} to optimize the use of measurements throughout the PECT execution. 
This way, one can guarantee the multiple VQE runs do not exceed a certain measurement budget.
For our numerical experiments, we investigate optimizations of three different VQE ansatze: the standard UCCSD ansatz \cite{Shen2017}, its low-depth and generalized variant $k$-UpCCGSD ansatz \cite{Lee2019}, and the LDCA \cite{Dallaire-Demers2018}.

\subsection{Ansatze}\label{sec:ansatze}

The UCC method, a unitary variant of the traditional coupled-cluster method, has been widely studied in the context of simulating quantum chemistry using quantum computers \cite{Yung2015, Shen2017, Romero2017strategies}.
However, the UCC wavefunction considering singles and doubles excitations (UCCSD) suffers from circuit depths that grow as $O(N^5)$ with system size. 
More recently, this depth was reduced by approximately a linear factor
using low-rank decomposition and the fermionic swap network while assuming linear connectivity \cite{Motta2018}. 
The number of parameters for UCCSD grows as $O(N_o^2 N_v^2)$, where $N_o$ and $N_v$ are numbers of occupied and virtual spin orbitals respectively, or in the worst case, as $O(N^4)$.
Moreover, UCCSD has been shown to be insufficient for describing particular molecular systems \cite{Cooper2010}.
Since then, a recent work developed a shallower and more flexible variant of the UCC method, called $k$-UpCCGSD \cite{Lee2019}.

The $k$-UpCCGSD ansatz comprises $k$ layers of a circuit that prepares a wavefunction considering generalized singles and paired doubles excitations.
Compared to UCCSD, the $k$-UpCCGSD ansatz has a significantly shorter circuit depth i.e. linear in the system size or $O(kN)$, making $k$-UpCCGSD better suited for NISQ applications. 
While the ansatz was able to well describe the ground and the first excited states of various strongly correlated molecular systems in Ref. \cite{Lee2019}, initialization and optimization of the circuit parameters appeared to be challenging tasks with increasing $k$.

Lastly, LDCA is another ansatz with depth scaling linearly with system size introduced by \citet{Dallaire-Demers2018}.
The LDCA uses nearest-neighbor parameterized two-qubit gates that are native to devices with tunable couplers.
These templates of nearest-neighbor two-qubit gates are repeated several times, producing a highly compact and nested circuit structure and thus accruing significant numbers of parameters.\footnote{We further explain and provide an example of the LDCA circuit structure in Appendix \ref{app:sec:ldca_details}.}
For instance, using the original formulation of LDCA, given a 12-qubit VQE problem using 5 superlayers or ``cycles'' of LDCA, this circuit already has 1662 parameters.
By contrast, for an instance of a 12-qubit VQE problem solving for ground state energies of LiH, UCCSD uses 56 parameters, and $2$-UpCCGSD uses 60 parameters.
While LDCA demonstrated potential to well describe strongly correlated fermionic systems, its application to larger systems has so far been limited by the difficulty in optimization.
In our implementation of LDCA, in order to reduce circuit depth, we apply the unitary $U_{\text{VarMG}}$ defined in Ref. \cite{Dallaire-Demers2018} to the Hartree-Fock state but do not apply $U^\dag_{\text{Bog}}$ afterwards.
To generate wavefunctions that preserve particle number and reduce parameter count, we employed parameter sharing (also known as correlating parameters \cite{Volkoff2020}), in which the $XX$ and $YY$ entanglers share the same parameter and the $XY$ and $YX$ entanglers share the same parameter with opposite signs, such that each two-qubit LDCA block has three free parameters, instead of five. 

To summarize, in terms of circuit depth, $k$-UpCCGSD and LDCA have comparable depths (up to some $k$) while UCCSD has a larger circuit depth for given system size.
In parameter count, UCCSD has the largest asymptotic scaling though for small systems such as ones considered in this study, it may correspond to the fewest number of parameters out of the three ansatze.
Even in cases where the number of parameters for UCCSD may be larger than those for $k$-UpCCGSD and LDCA, advantages of using UCCSD include well-motivated techniques for initializing the ansatz parameters and relative ease of optimization.
In terms of defining what constitutes a layer for each ansatz to implement PECT, 
we define UCCSD as having a single layer,
$k$-UpCCGSD having $k$ layers (each layer being a repetition of the UpCCGSD circuit),
and LDCA having $L*\frac{N}{2}*(N-1)$ layers, where $L$ is the number of LDCA superlayers and $N$ is the number of qubits.
For LDCA, this means that we define each layer for PECT as a two-qubit LDCA block which has three free parameters, as shown in Figure \ref{fig:ldca_circuit} in Appendix \ref{app:sec:ldca_details}. 

In the following subsections, we describe the computational details for VQE wavefunction simulations of LiH (STO-3G) and H$_2$O (STO-3G) and show our results comparing VQE optimizations with and without PECT.

\subsection{Computational details}

\begin{table*}[]
\centering
\def\arraystretch{1.1}\tabcolsep=4.5pt
\begin{tabular}{@{}cccccc@{}}
\toprule
\textbf{System} & \textbf{Ansatz} & \textbf{\begin{tabular}[c]{@{}c@{}}Global \\sparsity, $s$\end{tabular}} &
  \textbf{\begin{tabular}[c]{@{}c@{}}Parameter count \\ (PECT/non-PECT)\end{tabular}}  &\textbf{\begin{tabular}[c]{@{}c@{}}Initial pruning\\ threshold, $H_0$\end{tabular}} & \textbf{\begin{tabular}[c]{@{}c@{}}Target number of \\ parameters to prune, $N_p$\end{tabular}} \\ \midrule
LiH & UCCSD & 0.2 & 45/56 & 0.0001 & 8 \\
& 2-UpCCGSD & 0.5 & 30/60 & 0.001 & 6 \\
& LDCA (L-BFGS-B) & \begin{tabular}[c]{@{}c@{}}0.73 (rest)\\ 0.6 (0.9, 1.1 \angstrom)\end{tabular} & \begin{tabular}[c]{@{}c@{}} 291/1002 (rest)\\ 499/1200 (0.9, 1.1 \angstrom)\end{tabular} &  0.1 & \begin{tabular}[c]{@{}c@{}}15 (rest)\\ 30 (0.9, 1.1 \angstrom)\end{tabular} \\
 & LDCA (SLSQP) & 0.6 & 420/1002 & 0.1 & 30 \\
 \midrule
H$_2$O & UCCSD & \begin{tabular}[c]{@{}c@{}}0.15 (rest),\\ 0.1 (2.6 \angstrom)\end{tabular} & \begin{tabular}[c]{@{}c@{}}39/46 (rest),\\ 49/54 (2.6 \angstrom)\end{tabular} & 0.001 & 5 \\
 & 4-UpCCGSD & 0.3 & 84/120 & 0.01 & 10 \\
 & 5-UpCCGSD & 0.3 & 105/150 & 0.01 & 10 \\
 & LDCA & \begin{tabular}[c]{@{}c@{}}0.6 (rest)\\ 0.65 (2.2 \angstrom)\\ 0.5 (2.4 \angstrom)\end{tabular}& \begin{tabular}[c]{@{}c@{}}420/1002 (rest)\\ 440/1200 (2.2 \angstrom)\\ 618/1200 (2.4 \angstrom)\end{tabular} &  \begin{tabular}[c]{@{}c@{}}0.05 (rest)\\ 0.1 (2.2, 2.4 \angstrom)\end{tabular} & \begin{tabular}[c]{@{}c@{}}50 (rest)\\ 40 (2.2, 2.4 \angstrom)\end{tabular} \\ \bottomrule
\end{tabular}
\caption{Hyperparameters and parameter counts for PECT calculations reported in this study.}
\label{table:pect_hyperparameters}
\end{table*}

For testing PECT, we consider two molecular systems: (1) lithium hydride (4 electrons, 12 spin-orbitals/qubits) and (2) symmetric stretching of O-H bonds in a water molecule with a fixed H-O-H angle of 104.5 degrees (8 electrons, 12 spin-orbitals/qubits after freezing core orbitals), both in the STO-3G basis.
We run VQE calculations to estimate the ground state energies of both systems using two optimization methods: the traditional method of optimizing the entire ansatz, which we refer to as ``non-PECT'' calculations, and optimization using PECT. 
To fairly compare VQE optimizations with and without PECT, we use the same classical optimizers, L-BFGS-B \cite{Byrd1995} and SLSQP \cite{Kraft1988}, as well as initial parameters for a given ansatz.
In addition, we use the same optimizer arguments and options for PECT and non-PECT calculations, e.g. step size used for numerical gradients in L-BFGS-B or the maximum number of function evaluations.

For each ansatz, we employed different parameter initialization strategies.
For the UCCSD ansatz, we pre-computed MP2 amplitudes for parameter initialization \cite{Romero2017strategies}. 
With $k$-UpCCGSD ansatze for various $k$ values, we used MP2 amplitudes as initial parameter values for the first layer ($k=1$) then randomly initialized parameters of subsequent layers by uniformly sampling from the range [-0.1, 0.1].
We employed this parameter initialization technique 
due to advantages in using informed guesses such as MP2 amplitudes to initialize UCC-like wavefunctions \cite{Romero2017strategies}, 
and we assumed that parameter settings of the first circuit layer or $k=1$ would be more influential than those of subsequent layers. 
For LDCA, initial parameters were randomly generated by uniformly sampling from the range $[0, 4\pi]$.

We report the PECT hyperparameters for each system and ansatz type in Table \ref{table:pect_hyperparameters}. 
The fractional tolerance for $N_p$, or $\delta$, is fixed to 0.1 for all simulations.
We chose the global sparsities based on the number of parameters for each ansatz/problem and $H_0$ based on the initial guesses.
For example, for UCCSD, the pre-computed MP2 amplitudes for LiH were small in magnitude. 
To prevent too many parameters being pruned in the first few local optimizations, we set $H_0$ for those simulations to be slightly less than the average of the magnitudes of the MP2 amplitudes.
Lastly, we chose the values of $N_p$ based on the number of parameters and our assumptions on the relative parameter redundancy of each ansatz.

\subsection{LiH}\label{sec:lih}

\begin{figure*}[ht]
\centering
\includegraphics[scale=0.35]{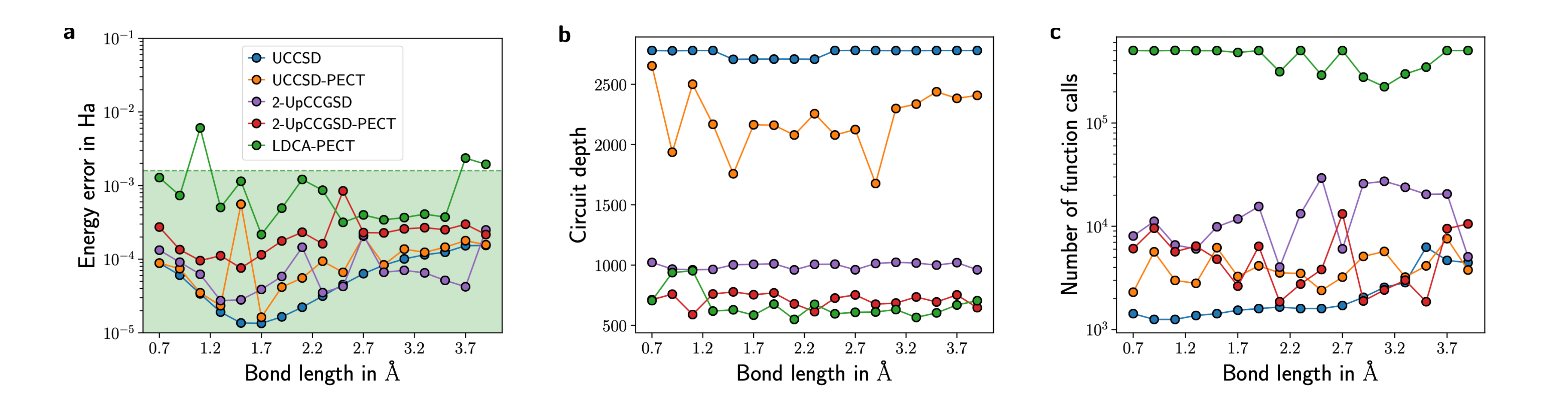}
\caption{VQE performance metrics computed for estimating ground state energies of LiH at various bond lengths. 
(a) Energy errors are shown for different ansatze that were optimized without or with the PECT strategy. The optimizer used is L-BFGS-B. 
The green shaded region indicates energies below chemical accuracy. 
(b) Depths of corresponding circuits. 
For circuits optimized using PECT, we report the depth averaged over PECT iterations.
(c) Number of total function calls for each ansatz. 
}
\label{fig:lih_results}
\end{figure*}

We show the results for estimating ground state energies of LiH over various bond lengths using the three types of ansatz in Figure \ref{fig:lih_results}. 
All of the calculations, both PECT and non-PECT optimizations, shown in this Figure employ the L-BFGS-B as the (local) optimizer.
We observe that the UCCSD ansatz optimized without PECT achieves the highest energy accuracy but corresponds to the deepest circuits across all bond lengths.
With PECT, it is possible to reduce the total depth by up to approximately 20\% in some cases, although at the cost of losing some accuracy in the final energy.
With $2$-UpCCGSD, we achieve energy values with errors that fall below the chemical accuracy threshold (i.e. $\leq$ 1 kcal/mol or $1.6 \times 10^{−3}$ Ha) while maintaining substantially lower circuit depths. 
With PECT, $2$-UpCCGSD further reduces circuit depths but also at the cost of higher ground state energy estimates. Still, for both UCCSD and $2$-UpCCGSD, errors using PECT are below the chemical accuracy threshold.
Lastly, LDCA optimized with PECT is shown to have comparable depths to $2$-UpCCGSD employing PECT.
Naively, this ansatz with 5-6 superlayers has 1002-1200 parameters respectively. 
A direct optimization of this number of parameters using the L-BFGS-B optimizer did not achieve convergence despite multiple attempts with randomly initialized parameters. 
In contrast, using PECT, we were able to reduce the number of parameters to 291 at most bond lengths (and 499 at two bond lengths), as shown in Table \ref{table:pect_hyperparameters}.
Although for a few bond lengths we were not able to achieve chemical accuracy with PECT, we note that without PECT, optimization of the LDCA ansatz was not possible at any bond length.
We also note that the number of function evaluations using this optimizer were often around the maximum number allocated, or $5 \times 10^5$.

\begin{figure*}[ht]
\centering
\includegraphics[scale=0.35]{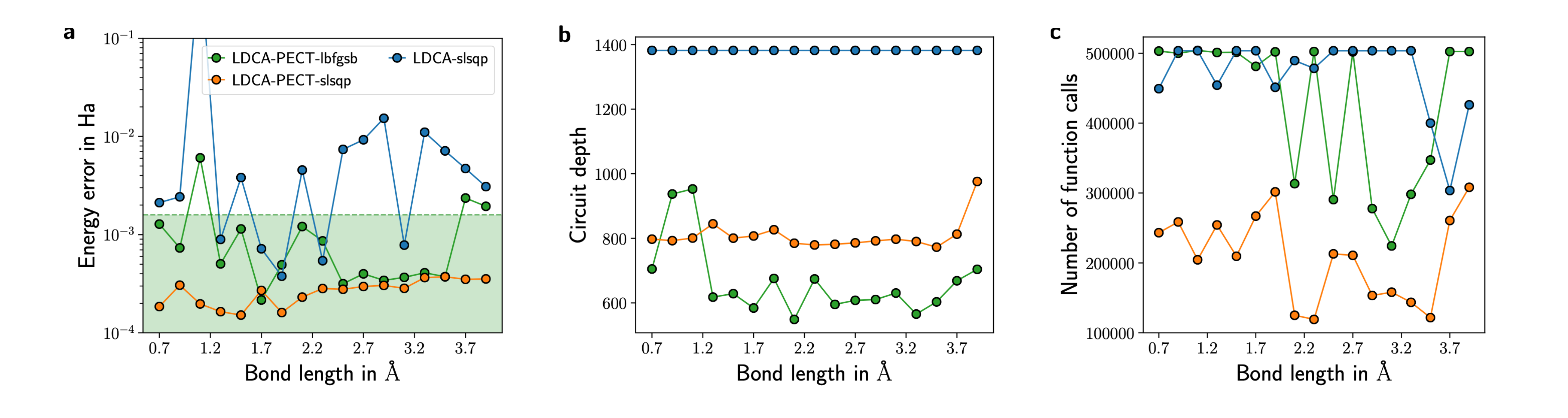}
\caption{Comparing VQE performance metrics computed for estimating ground state energies of LiH by optimizing LDCA with PECT but employing two different local optimizers: L-BFGS-B (green) and SLSQP (orange). 
We additionally plotted the VQE performance metrics for LDCA optimized without PECT (blue) using the SLSQP optimizer to highlight the advantages of using PECT for optimizing LDCA circuits.
(a) Energy errors for different PECT optimizers and global sparsities ($s=0.73$ and $0.6$ for L-BFGS-B and $s=0.6$ for SLSQP optimizers).
The green shaded region indicates energy errors below chemical accuracy. 
(b) Average depths of corresponding circuits. 
(c) Number of function calls for each PECT calculation.
}
\label{fig:lih_ldca_results}
\end{figure*}

To improve the performance of optimizing LDCA,
we explored an alternative local optimizer and global sparsities.
Using the SLSQP optimizer and lowering the global sparsity from 0.73 to 0.6 for all bond lengths (i.e. increasing parameter count from around 291 to 420 parameters), we were able to improve the energies at most bond lengths with all the energy errors along the potential energy surface falling below chemical accuracy, as shown in Figure \ref{fig:lih_ldca_results}.
In addition, using these PECT settings, we reduced the number of function calls on average by approximately half, compared to the LDCA calculations using L-BFGS-B with PECT. 
In Figure \ref{fig:lih_ldca_results}, we also tried optimizing LDCA with SLSQP but without PECT to demonstrate that optimization of LDCA at this system size and number of layers needs a strategy like PECT to achieve reasonable energies.

\subsection{Symmetric stretching of O-H bonds in H$_2$O}\label{sec:h2o}
\begin{figure*}[ht]
\centering
\includegraphics[scale=0.3]{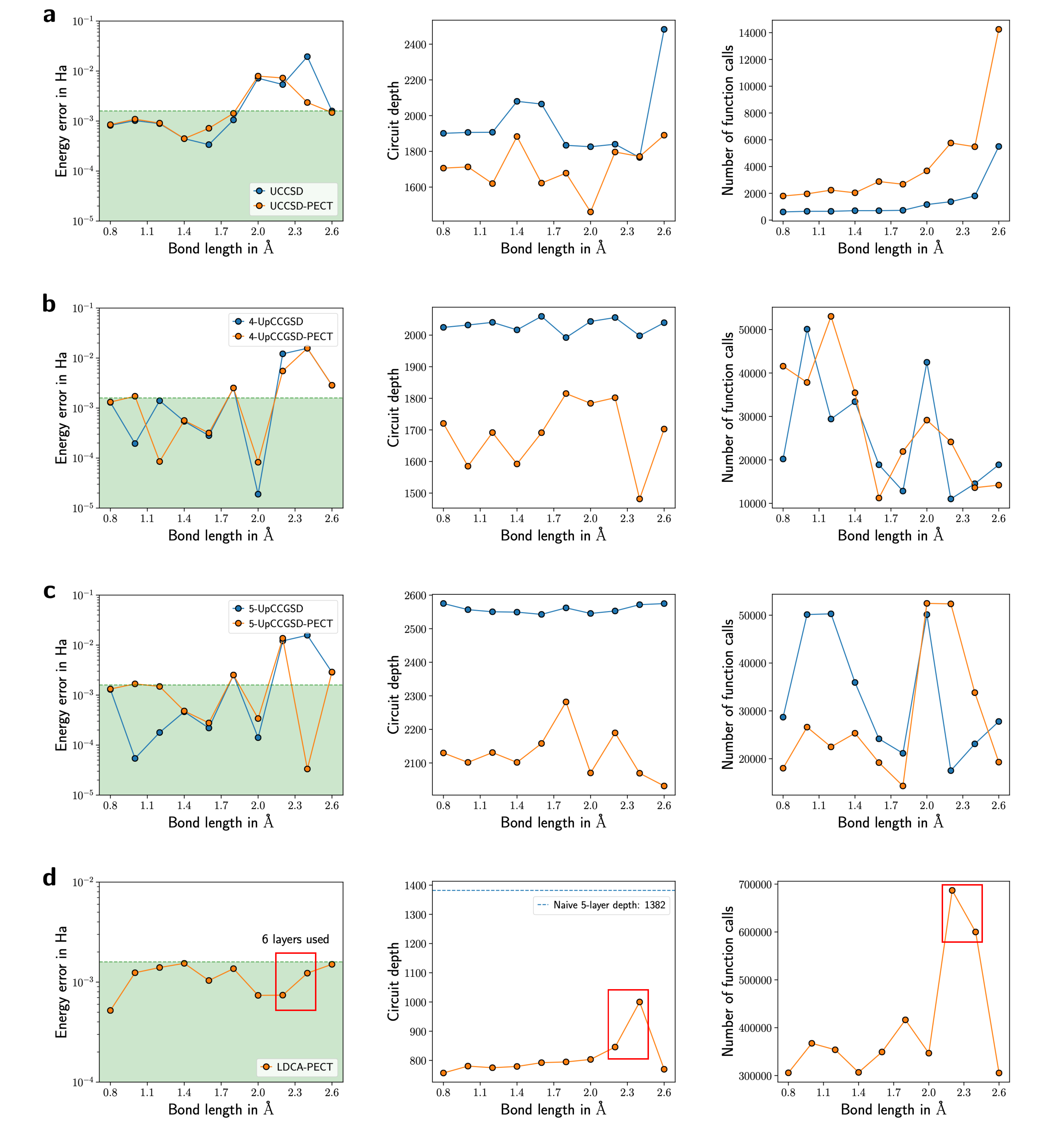}
\caption{
VQE performance metrics computed for estimating ground state energies of H$_2$O using various ansatze: (a) UCCSD, (b) 4-UpCCGSD, (c) 5-UpCCGSD, and (d) LDCA.
For two of the bond lengths as shown in the red boxes in (d), 6 superlayers of LDCA were used with PECT as using 5 superlayers led to poor final energies.
For each ansatz, the same initial parameters are used for PECT and non-PECT methods.
We show the resulting potential energy surface for each ansatz in Appendix \ref{app:sec:h2o_pes}.
}
\label{fig:h2o_results}
\end{figure*}

Symmetric dissociation of water is a common benchmark for the classical multi-reference methods~\cite{Ma2006,Li1998,Olsen1996,Brown1984} and has been recently studied in the context of $k$-UpCCGSD by Lee et al.~\cite{Lee2019} and by Sokolov et al.~\cite{Sokolov2020} as a test for the orbital optimized UCC methods.
Similarly structured as the previous subsection, we show the VQE optimization results for the symmetric stretching of O-H bonds in a water molecule in Figure \ref{fig:h2o_results}. 
For optimizing LDCA ansatze, SLSQP was employed as the local optimizer based on the relative performance of SLSQP compared to L-BFGS-B in VQE results of LiH (Figure \ref{fig:lih_ldca_results}).
On the other hand, for UCCSD and $k$-UpCCGSD where $k=4, 5$, we employed L-BFGS-B as the local optimizer after preliminary results showed lower final energies obtained using L-BFGS-B over SLSQP.
Plots in row (a) in Figure \ref{fig:h2o_results} show that in the case of UCCSD, the ground state energy estimates become worse as the O-H bonds are stretched, which is consistent with results from a past study investigating the performance of UCCSD in the same system
\cite{Cooper2010}.
In addition, some of the energy error may also come from an unfavorable ordering of cluster operators when the UCCSD unitary is Trotterized in our implementation \cite{Grimsley2020, Izmaylov2020}. 

Regarding the performance of PECT, it was interesting to note that, unlike in optimizations of LiH ansatze in which using PECT consistently produced higher ground state energy estimates, using PECT to optimize the UCCSD wavefunction yielded significantly lower energy than optimization without PECT when $R_{\text{O-H}} = 2.4 \angstrom$.
Additionally, at this bond length, PECT found a subset of gates with the same circuit depth as the un-pruned version of the ansatz.

Regarding the $k$-UpCCGSD ansatz, while $k=2$ was sufficient to reproduce the LiH ground state energy, we observed that up to $k=4$ and $k=5$ layers were needed in the case of water.
As noted in the original paper \cite{Lee2019}, optimizing $k$-UpCCGSD ansatz for higher $k$ proved to be challenging as the quality of the ansatz becomes highly sensitive to the initial values of the ansatz parameters.
In Figure \ref{fig:h2o_results}, plots in rows (b) and (c) show the energy errors, circuit depths, and number of function call for the $k=4$ and $k=5$ instances of the ansatz, respectively. 
These results show that PECT improves the accuracy for some bond lengths but not others. 
In all cases, however, it achieves reductions in the circuit depths.

For those geometries in which the optimizer was unable to produce final energies with errors below the chemical accuracy threshold, this may be due to unfavorable initial points, versus the limited expressiveness of the ansatz.
To support this claim, we ran multiple trials of a 4-UpCCGSD PECT-based optimization from ten different initial points, in which one of the trials achieved chemical accuracy, proving the sufficient expressiveness of the ansatz. 
We describe this numerical experiment in further detail in Section \ref{sec:prunability}.
Final energies using instances of the $k$-UpCCGSD ansatz were generally lower than energies using the UCCSD ansatz in plots (a), especially in regions of the potential energy surface where UCCSD had high errors. 
However, at $k=4$ and $k=5$, the resulting circuit depths of the $k$-UpCCGSD ansatz are comparable to or higher than the depths of the UCCSD ansatz.
Comparing optimizations of $k$-UpCCGSD with and without PECT, we observe that PECT converges to comparable or lower energies in the region where O-H bond is stretched (e.g. $R=2.4 \angstrom$ for $k=5$ in Figure \ref{fig:h2o_results}(c)) using a fraction of the total circuit depth and comparable number of function calls. 
To improve the performance of optimizing $k$-UpCCGSD using PECT, further investigation is needed to find better strategies for initializing parameters of the ansatz.
In this study, we observe that initializing the parameters of the first layer ($k=1$) with corresponding MP2 amplitudes performed better than when the entire parameter vector was randomly initialized. 
Nevertheless, even with this initialization scheme, we do not observe reliable convergence to low energies or energies within chemical accuracy.

Lastly, we considered the LDCA ansatz to estimate the ground state energies of water.
From the LDCA optimization results for LiH in Figure \ref{fig:lih_ldca_results} and initial numerical experiments, we observed high energy errors when optimizing LDCA without PECT.\footnote{The average VQE energy error across bond lengths of LiH using the LDCA wavefunction without PECT was approximately 0.04 Hartrees.}
This implies that to estimate the ground state potential energy curve of water, which is more difficult to describe than that of LiH, it is infeasible to optimize the corresponding LDCA circuit without PECT.
Thus, we limited our VQE optimizations of LDCA for water to ones employing PECT, as shown in Figure \ref{fig:h2o_results}(c). 
For all the bond lengths considered, we were able to generate VQE-optimized circuits that corresponded to energy errors falling below the chemical accuracy threshold.
With the exception of ground states at two O-H bond lengths, all PECT-based optimizations were executed on 5 superlayers of LDCA, which originally has depth 1382 and two-qubit gate count of 3300. 
Using PECT, the effective circuit depth and gate count are reduced, as discussed in more detail in Section \ref{sec:ansatz_reduction}.
For the two O-H bond lengths, indicated on Figure \ref{fig:lih_ldca_results}(d) in red boxes, 6 superlayers of LDCA were needed to prepare the ground state, which corresponded to simulating circuits with depths of around 1658 and two-qubit gate counts of around 3960. 
Compared to the other two types of ansatz, LDCA circuits had at least half the circuit depths though they required significantly more function evaluations to optimize. 
This is due to the greater number of parameters LDCA has compared to 
the other ansatze for the instances we studied. 
Computing gradients, analytical or numerical, requires two function evaluations per parameter.
Based on our observations and results, it appears that while LDCA circuits take longer to optimize, they are less sensitive to parameter initialization compared to $k$-UpCCGSD when optimized with PECT.
For each bond length, fewer than 5 random parameter initializations for LDCA were used to generate the plots in Figure \ref{fig:h2o_results}(d) which report the best results.\footnote{We set the PECT hyperparameters for optimizing LDCA for water based on corresponding values used for optimizing LDCA for LiH. Likewise, when optimizing e.g. two circuits of similar widths and depths using PECT, it may be possible to infer the hyperparameters for one run based on the other.}
We explore this further in Subsections \ref{sec:random_initialization} and \ref{sec:prunability}.

\section{Discussion}\label{sec:discussion}

From the previous section, we gained insight on each ansatz' ability to reproduce the potential energy surfaces of LiH and $\text{H}_2\text{O}$ using PECT.
Here, we provide further analyses on PECT, including its advantages in robustness of parameter initialization for LDCA as well as general reductions in circuit resources (i.e. depth and two-qubit gate count) and in optimization runtimes. 
Our PECT calculations also reveal the relative ``prunability'' of each of the three ansatze and provided numerical evidence of how parameters of early layers of multi-layered PQCs converge to their final values before parameters of later layers do. 
We then propose a way for extending PECT to circuits with more parameters by combining the method with the layerwise circuit training strategy \cite{Bengio2007, Dallaire-Demers2020, Skolik2020}. 
We test this combined strategy by optimizing VQE circuits to estimate the ground state densities of the one-dimensional Fermi-Hubbard model across various chain lengths.

\begin{figure*}[ht]
\centering
\includegraphics[scale=0.5]{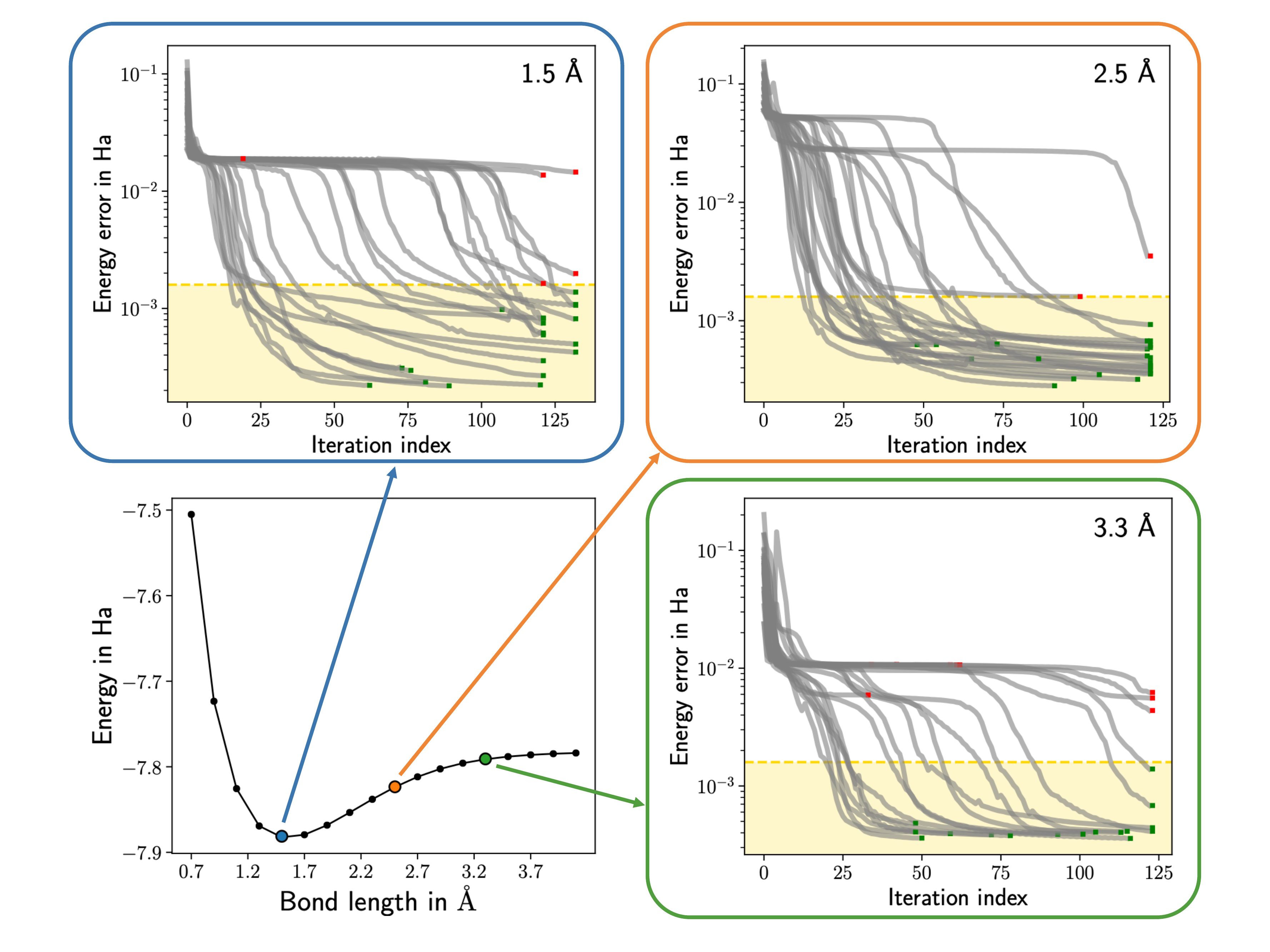}
\caption{
Energy convergence for finding the ground state of LiH using the LDCA circuit. Twenty-five random parameter initializations were executed for each of three bond lengths to ensure PECT can provide high-quality solutions with high probability.
The green square at the end of each optimization curve indicates a ``successful'' run, in which the final energy error fell below chemical accuracy.
Yellow shaded region indicates energy errors below chemical accuracy threshold. 
The x-axis refers to the iteration index of PECT.
For these simulations, we fixed the maximum number of function evaluations to $5 \times 10^5$.
}
\label{fig:lih_ldca_random_initialization}
\end{figure*}

\subsection{Robustness to random parameter initialization in LDCA}\label{sec:random_initialization}
In this section, we investigate the robustness to parameter initialization by applying PECT to optimize LDCA circuits.
While depth-efficient, LDCA accrues a large number of parameters.
Using PECT, however, we numerically show that optimization runs are robust to random initialization.
In Fig. \ref{fig:lih_ldca_random_initialization}, we consider three different bond lengths of LiH: $R=1.5, 2.5$, and $3.3 \angstrom$. 
For each bond length, we execute 25 randomly initialized optimizations of the 12-qubit, 5-superlayered LDCA which naively has 1002 parameters.
Using PECT with the L-BFGS-B optimizer, we fixed the global sparsity $s$ to 0.73 such that we were optimizing around 291 parameters during each local optimization. 
At bond length of 1.5 \AA, 20 out of 25 runs achieved final energies with errors that fell below the chemical accuracy threshold. 
For 2.5 \AA, 23 out of 25 runs succeeded, and for 3.3 \AA, 16 out of 25 runs.
The smaller success ratio for $R = 3.3 \angstrom$ is expected due to the increasing difficulty of describing the ground state as the Li-H bond is stretched. 
That is, as the interatomic distance increases, the ground state wave function acquires a multi-reference character.
This may have implications for the quality of the ansatz and likely changes the ``roughness'' of the objective function landscape, e.g. addition of local minima. 
For a parameter-heavy ansatz such as LDCA, we noted relatively high success probabilities at the three bond lengths.
This not only highlights the advantage of employing PECT for optimization but also sheds light on the amount of redundancy in the parameterization of LDCA, which we discuss in the following subsection.

\subsection{Prunability of ansatz}\label{sec:prunability}
From our optimizations of UCCSD, $k$-UpCCGSD, and LDCA using PECT, we were able to identify which ansatz is more ``prunable,'' that is, well-suited for optimization strategies such as PECT that utilizes pruning or removal of gates. 
Specifically, PECT seemed the most effective strategy for LDCA out of the three types of ansatze investigated.
For instance, in the case of estimating the ground state energy of the water molecule, for all the bond lengths considered, PECT was able to estimate energies with errors below the chemical accuracy threshold with fewer than 5 trials of random initial parameters, with several runs converging on the first trial.
For $k$-UpCCGSD with PECT, the optimization remained sensitive to initial parameter values as well as the sets of gates selected during the PECT procedure.
To test the sensitivity, in the case of $4$-UpCCGSD for $R_{\text{O-H}} = 2.2 \angstrom$, where in Figure \ref{fig:h2o_results}(b) we observed a high energy error, ten independent PECT optimizations were run with random parameter initializations.
These optimizations were originally designed to test whether the high energy error observed was due to the deficiency of the ansatz or the sensitivity to parameter initialization.
Out of the ten trials, only one run achieved a final energy with error below chemical accuracy, while the average energy error across the ten trials was approximately $0.01$ Hartrees with standard deviation of $0.005$ Hartrees.
With most trials producing energy errors that are close to the mean, this confirmed the sensitivity to initial parameters. 
One open question to investigate would be whether using PECT makes $k$-UpCCGSD \textit{less} sensitive to parameter initialization than optimizing the ansatz without PECT.

From our results, we determined the UCCSD ansatz is the least prunable ansatz of the three. 
For the sizes of problems investigated in this work, UCCSD, though high in depth, is the most compact in terms of its parameterization. 
That is, the numbers of parameters are already low enough such that optimizing subsets of parameters is unnecessary. 
In addition, the bigger challenge in using UCCSD is not in its optimization but in its expressibility or capability of describing ground states of (strongly correlated) molecular systems.
As noted earlier, effectiveness of UCCSD also highly depends on the ordering of Trotterized cluster operators \cite{Grimsley2020, Izmaylov2020}. 
Therefore, because PECT assumes a fixed ordering of operations (i.e. operates based on a pre-defined ansatz structure), a method such as ADAPT-VQE, which chooses operations from an unordered pool, may be better suited for optimizing an ansatz comprising cluster operators.

Comparing the performance of PECT for LDCA and $k$-UpCCGSD circuits in terms of energy error, the difference is likely due to the higher degree of redundancy in the LDCA circuit, compared to the $k$-UpCCGSD circuit.
That is, even if we remove some parameter $\theta$ in the LDCA circuit optimization, there are likely several other parameters that can replace $\theta$, i.e. that can be tuned with the same effect as tuning $\theta$.
In addition, this may help explain the robustness in parameter initialization for LDCA observed in Figure \ref{fig:lih_ldca_random_initialization}.
In conclusion, our analysis of the three ansatze indicates that PECT is \textit{not} an optimization strategy that will work well for any parameterized quantum circuit.
For optimal performance, the circuit should be parameter-heavy and contain some amount of redundancy in its parameterization. 
Fortunately, several PQC designs, especially ones that are efficient to particular hardware architectures and connectivities \cite{Kandala2017, Dallaire-Demers2020}, satisfy these conditions.

\subsection{Ansatz cost reduction}\label{sec:ansatz_reduction}

During the local optimization phase of PECT, a subset of parameters of the PQC is optimized while values of other parameters are set to 0 corresponding to identity operations.
This likely results in an ansatz substructure that has reduced circuit depth and lower (two-qubit) gate count compared to the original ansatz with full parameter count.
For instance, for optimization of the UCCSD wavefunction estimating LiH ground states, using 45 out of 56 parameters for PECT allowed for a 20\% reduction in circuit depth and 20\% reduction in two-qubit gate count, averaged over optimizations at different bond lengths. 
For optimization of $2$-UpCCGSD, using 30 out of 60 parameters produced a reduced ansatz with 28\% reduction in depth and 42\% fewer two-qubit gates.
Lastly, with LDCA, using 420 out of 1002 parameters led to a 42\% reduction in circuit depth and 58\% reduction in two-qubit gate count.

For optimizations of ground states of the water molecule, it is difficult to make equally meaningful statements on the reductions in circuit resources using UCCSD and $k$-UpCCGSD as we have done for LiH due to energy errors.
With the UCCSD ansatz, there were regions in the potential energy surface in which both SLSQP and SLSQP-with-PECT were not able to converge to chemically accurate energies.
With $k$-UpCCGSD where $k=4$ and $5$, PECT improved accuracies with respect to regular optimization only for certain geometries.
Nevertheless, we computed the percent reductions in circuit depths and two-qubit gate counts for UCCSD, $4$-UpCCGSD, and $5$-UpCCGSD averaged over all considered bond lengths.
Optimizing the UCCSD ansatz with PECT led to an average of 12\% reduction in depth and 12\% reduction in two-qubit gate count.
With 4-UpCCGSD, using PECT corresponded to an average of 17\% reduction in depth and 31\% reduction in two-qubit gate count.
Using PECT with the 5-UpCCGSD ansatz, we observed an average of 17\% reduction in depth and 32\% reduction in two-qubit gate count. 
To provide a fair analysis of the reductions in circuit resources, in Appendix \ref{app:sec:h2o_reductions} we report reductions in circuit depth and gate count for bond lengths at which both non-PECT and PECT based optimizations achieved final energies that are chemically accurate.
With LDCA, because we did not optimize the ansatz without PECT, we report the depths of the full 5- and 6-superlayered circuit and show that using PECT, the circuit depths were significantly reduced.
For bond lengths at which 5 superlayers of LDCA were used, there was an average of 42\% reduction in depth compared to the depth of the full 5-superlayered LDCA, and for 6 superlayers of LDCA, there was an average of 39\% reduction in depth.
The reader may wonder why for certain bond lengths, e.g. $R_{\text{O-H}} = 2.2$ and $2.4 \angstrom$, we used 6 superlayers instead of 5 superlayers with PECT if the resulting circuit depths at those bond lengths are still lower than those of a fully-dense 5-superlayered LDCA circuit.
We point out that the PECT optimization acting on 5 superlayers of LDCA did not achieve chemical accuracy for these geometries, while the PECT optimization on 6 superlayers did. 
The final circuit of the latter optimization contains gates from the sixth LDCA superlayer.

We showed in various instances that PECT is able to produce more depth-efficient circuits compared to the original ansatz.
In theory, such depth reductions make circuits generated using PECT more resilient to noise in the quantum device. We leave analysis of PECT under noise to future work.

\subsection{Optimization cost estimation}\label{sec:optimization_cost}
In this section, we discuss estimating the cost associated with optimizing a parameterized quantum circuit in terms of the number of objective function evaluations with the goal of \emph{comparing} optimization runtimes of different ansatze as opposed to providing accurate estimates of the actual runtimes.
The total time associated with the VQE optimization step can roughly be estimated as:

\begin{align}
\label{eq:single_feval_time}
t_{\text{opt, est}} = N * t_{\text{circuit}} * f_{\text{eval}},
\end{align}

\noindent where $N$ is the number of circuit repetitions or shots per function evaluation, $t_{\text{circuit}}$ is the time required to execute the circuit, 
and $f_{\text{eval}}$ is the number of function evaluations.
While small in magnitude, circuit execution time is a significant component in estimating the overall optimization runtime due to its relation to the circuit depth.
Deeper circuits, especially in the NISQ era, will execute with lower circuit fidelities, corresponding to lower-quality estimates of the objective function values. 
We expect optimizations using noisier estimates to be more difficult and expensive, e.g. requiring more function evaluations.
To reflect this, we can further approximate the total optimization time by replacing $t_{\text{circuit}}$ with the circuit depth, which indirectly accounts for the expected increase in cost associated with optimizing noisier circuits:

\begin{align}
\label{eq:single_vqe_time_proxy}
t_{\text{proxy}} \approx N * D * f_{\text{eval}},
\end{align}

\noindent where $D$ is the circuit depth.
The PECT scheme provides a way to potentially reduce this time by reducing the circuit depth required at each function evaluation though this may come at the cost of a greater number of function evaluations to search for an effective parameter subset.
For PECT, the time proxy can be defined as:
\begin{align}
\label{eq:pect_time_proxy}
t_{\text{proxy, PECT}} \approx \sum_i N_i * D_i * f_{\text{eval, i}},
\end{align}

\noindent where $i$ runs over the number of local optimizations.
Assuming the required precision for the cost function is the same for all local optimizations,
we simplify $N$ or $N_i$, the number of circuit repetitions or measurements, as being constant for each VQE sub-module and set it to $1$.
In practice, the number of circuit measurements for a VQE experiment is expected to be very high
(and thus time-intensive) for variational quantum algorithms and will likely be a large multiplicative factor.
For an in-depth runtime analysis especially in a cloud-based quantum computation setting, it may be necessary to use a more sophisticated runtime model that considers sampling, circuit batching, and network latency \cite{Sung2020}. 

\begin{figure*}[ht]
\centering
\includegraphics[scale=0.47]{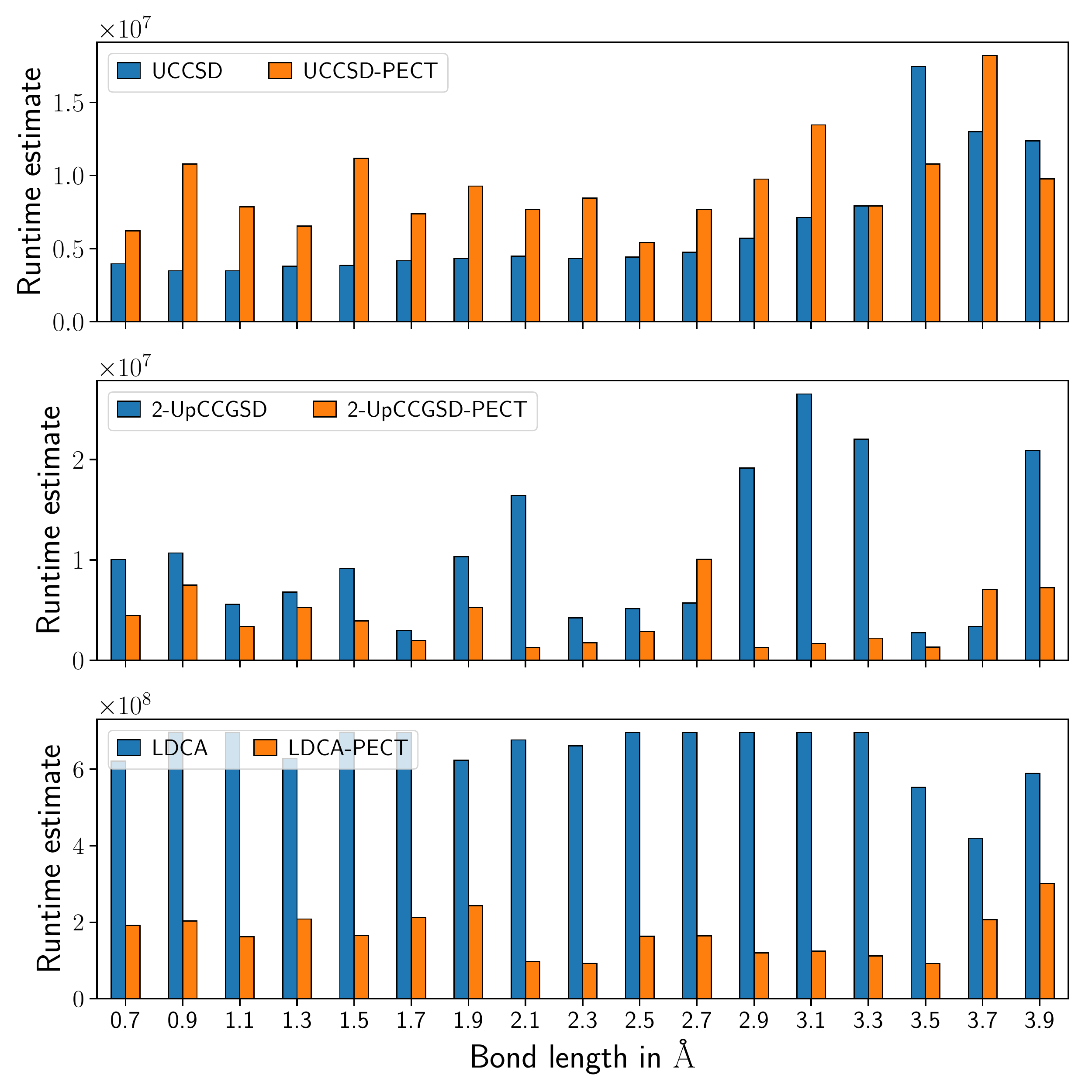}
\caption{
Runtime estimates (depth $\times$ number of function calls) 
of LiH VQE calculations for the three ansatze, with and without PECT, shown in orange and blue colored bars respectively.
Definitions for runtime estimates are shown in Equations \ref{eq:single_vqe_time_proxy} and \ref{eq:pect_time_proxy} for non-PECT and PECT optimizations respectively.
}
\label{fig:lih_runtime_estimates}
\end{figure*}

\begin{figure*}[ht]
\centering
\includegraphics[scale=0.47]{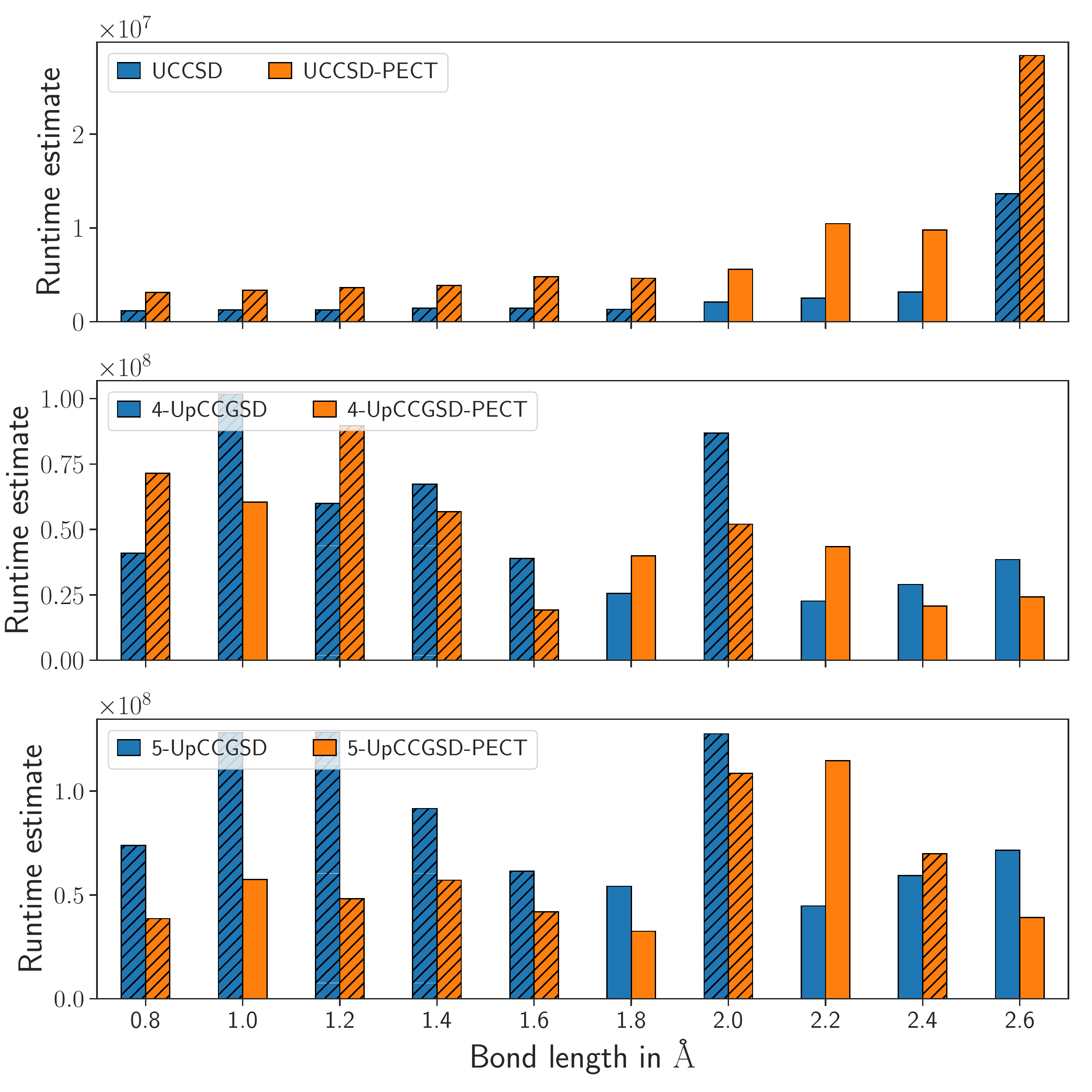}
\caption{
Runtime estimates (depth $\times$ number of function calls) of $\text{H}_2\text{O}$ VQE calculations for UCCSD, 4-UpCCGSD, and 5-UpCCGSD. For each type of ansatz, we report runtimes with and without PECT, shown using orange and blue colored bars respectively.
Definitions for runtime estimates are shown in Equations \ref{eq:single_vqe_time_proxy} and \ref{eq:pect_time_proxy} for non-PECT and PECT optimizations respectively.
Unlike LiH calculations, optimizations at certain bond lengths did not converge at chemically accurate energies.
We distinguish optimizations that have reached chemically accurate final energies with hash patterns on bars.}
\label{fig:h2o_runtime_estimates}
\end{figure*}

We compute the following runtime proxies (Equations \ref{eq:single_vqe_time_proxy} and \ref{eq:pect_time_proxy}) for non-PECT and PECT optimizations respectively for the three ansatze solving for the ground states of LiH.
These runtime estimates are shown in Figure \ref{fig:lih_runtime_estimates}.
This plot indicates that using PECT can help speed up optimizations of 2-UpCCGSD and LDCA for most or all bond lengths, but for an ansatz like UCCSD, it is better, in terms of runtime, to execute non-PECT optimization using the entire circuit.
However, as noted earlier, in the NISQ era, reducing circuit depth may be a priority over reducing the overall optimization runtime due to noise in the device, which increases with the number of gate operations and circuit depth. 
Because PECT reduces the effective circuit depth and gate counts during the optimization, it increases the fidelity of the ansatz circuit and therefore could improve the optimization results and their accuracy. 

For the water molecule, the UCCSD and $k$-UpCCGSD ansatze did not achieve chemically accurate energies for several bond lengths.
We report the estimated runtimes of UCCSD over different bond lengths in the top plot of Figure \ref{fig:h2o_runtime_estimates}.
For all the bond lengths at which both non-PECT and PECT calculations reached chemically accurate energies, indicated using bars with hash patterns, we observe that using PECT increased the optimization times.
This is expected due to the compact parameterization of UCCSD for the systems considered in this work.
Like UCCSD, both 4-UpCCGSD and 5-UpCCGSD were not able to produce chemically accurate energies for all considered bond lengths, with or without PECT.
However, while we believe the reason for the lack of convergence to accurate energies in UCCSD, especially as the O-H bond is stretched, is due to the inherent limitations in the expressiveness of the ansatz, 4-UpCCGSD and 5-UpCCGSD were unable to reproduce accurate energies likely due to the sensitivity in parameter initialization. 
With more trials testing different random parameter initializations, as we tried in Section \ref{sec:prunability}, $k$-UpCCGSD can converge to energies with errors below chemical accuracy, as was shown in Ref. \cite{Lee2019}.
The middle and bottom plots of Figure \ref{fig:h2o_runtime_estimates} show runtime estimates for 4-UpCCGSD and 5-UpCCGSD, respectively. 
These results suggest that the benefit of PECT in reducing runtime is more apparent for larger $k$ in the $k$-UpCCGSD ansatz.
Lastly, while we did not run optimizations for LDCA without PECT based on results for LiH, we expect significant reductions in the runtime, in addition to producing accurate energies, as was observed for LiH in the bottom plot of Figure \ref{fig:lih_runtime_estimates}.

\subsection{Scaling up: layerwise PECT}\label{sec:layerwise_pect}

\begin{figure*}[ht]
\centering
\includegraphics[scale=0.47]{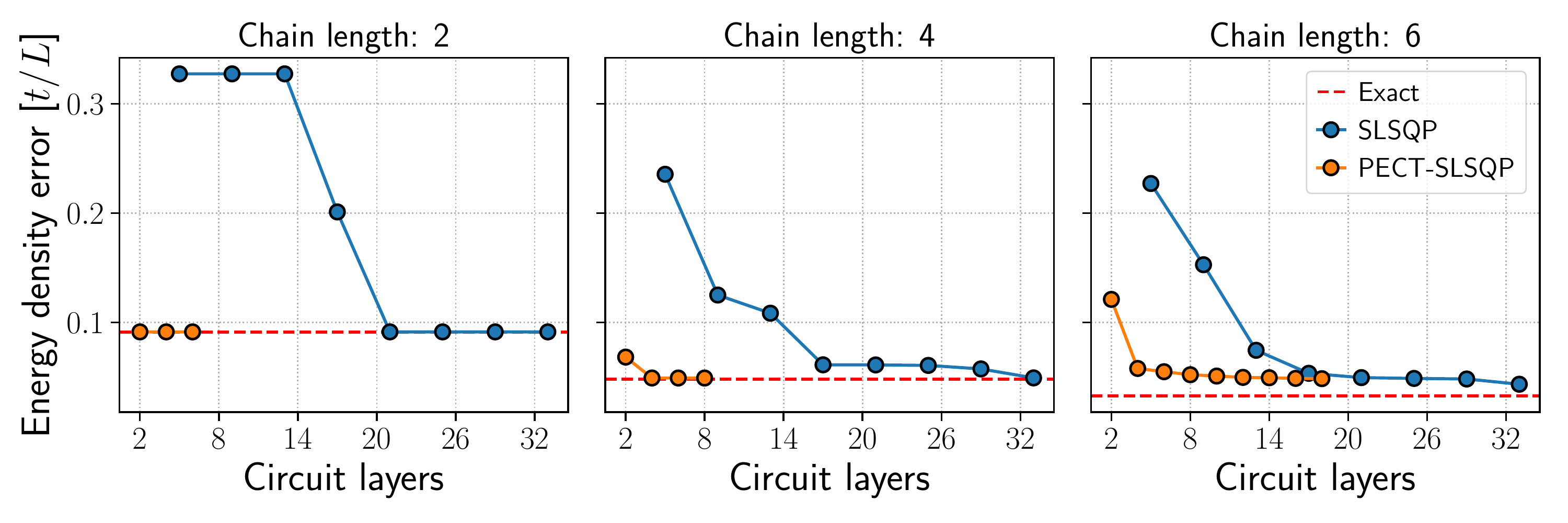}
\caption{
PECT results for optimizing VQE calculations for the one-dimensional Fermi-Hubbard model. 
Energy density errors (in units of kinetic energy, $t$ over $L$ sites) with respect to the infinite chain energy density are plotted against number of circuit layers of the ansatz proposed in Ref. \cite{Dallaire-Demers2020}.
Blue markers indicate addition of new circuit layers that are layerwise-optimized using the SLSQP optimizer (data points are obtained from Figure 9 of Ref. \cite{Dallaire-Demers2020}). 
Orange points indicate addition of new circuit layers that are layerwise-optimized using PECT employing SLSQP as the local optimizer. 
The red dashed line indicates the exact energy density at a given chain length.
}
\label{fig:fh_results}
\end{figure*}

As we approach VQE problems using ansatze with larger parameter counts, PECT will also suffer from inability to sufficiently explore and optimize in the parameter space.
In such cases, it may be useful to combine the ``layerwise'' training \cite{Bengio2007, Skolik2020} strategy with PECT.
To employ a simplified version of the layerwise training scheme, we first divide up the parameterized ansatz into $k$ layers and employ PECT to optimize each layer.
Once the PECT procedure is complete for the $j$-th layer ($j \in [k]$), the final parameter values of the layer are ``frozen,'' and the $j+1$-st layer is optimized with PECT. 
While fixing parameters of the previous layers may lead to local minima, we observe that relaxing these parameters in the optimization of subsequent layers can also lead to preparing a state corresponding to a higher energy.
It may then take many function calls to go back to the previous state, which can increase the overall cost of optimization.
We initialize the $j+1$-st layer with parameter values sampled uniformly from the range [0, 0.02] such that it constructs a near-identity operation to avoid potentially expending many function evaluations to return to the energy achieved with $j$ layers. 

We demonstrate the utility of layerwise PECT by running VQE optimizations for estimating the energy densities of the one-dimensional Fermi-Hubbard systems studied in Ref. \cite{Dallaire-Demers2020}. In this work, the one-dimensional Fermi-Hubbard model (FHM) at half-filling was proposed as a well-defined benchmark system to test the capabilities of NISQ devices \cite{Dallaire-Demers2020}. 
That is, VQEs on various chain lengths can be executed on a processor to determine the largest chain length for which the quantum computer can reliably approximate the ground state energy.
For the variational ansatz, the authors suggested a parameterized quantum circuit assuming a two-dimensional grid layout of qubits using qubit connectivity and gate operations that are native to the Sycamore processor, 
which has recently demonstrated the ability to outperform classical computers at certain tasks
\cite{Arute2019}. 
While we refer the readers to Ref. \cite{Dallaire-Demers2020} for further details on the system as well as the ansatz, in this work, we present VQE optimization results that show substantial improvement in the circuit depths while producing comparable energy density errors.

In Figure \ref{fig:fh_results}, we show the energy density error as a function of the number of circuit layers, where each circuit layer is defined in \cite{Dallaire-Demers2020}.
The original results from Ref. \cite{Dallaire-Demers2020}, shown in blue markers, employed layerwise training (without PECT) with the SLSQP optimizer. 
We additionally note that their layerwise training method did not freeze parameters of previous layers.
Each ``layer'' in layerwise training comprised four sub-layers of the proposed ansatz.
Using the layerwise PECT approach, where each ``layer'' comprises two sub-layers of the ansatz, the optimization can converge to comparable energy densities for different chain lengths using significantly fewer circuit layers, or equivalently, lower circuit depths.
Despite the reduction in circuit depth, as noted in previous results, PECT required greater numbers of function evaluations before convergence.
These numerical experiments exemplify the potential trade-off that exists between reducing the circuit depth and increasing the cost of evaluating cost function and gradient values (here represented by the number of function calls) when using PECT as the optimization strategy.
That is, if reducing circuit depth is an important objective in a VQE experiment and the large numbers of function calls is affordable, then PECT may be a promising optimization technique.
However, if reducing the number of function calls is more important, then PECT may not be as effective as an optimization strategy.
In the near term, due to the rapid accrual of noise in NISQ devices, we expect circuit depth reduction to be more critical in the application of variational algorithms. 
Furthermore, while potentially expensive in the number of function calls, the overall measurement cost of PECT could be mitigated using 
measurement-aware optimization strategies, such as the one proposed in Ref. \cite{Kubler2020}.

\subsection{Parameter dynamics of multi-layered PQCs}\label{sec:param_dynamics}
\begin{figure}[ht]
\centering
\includegraphics[scale=0.5]{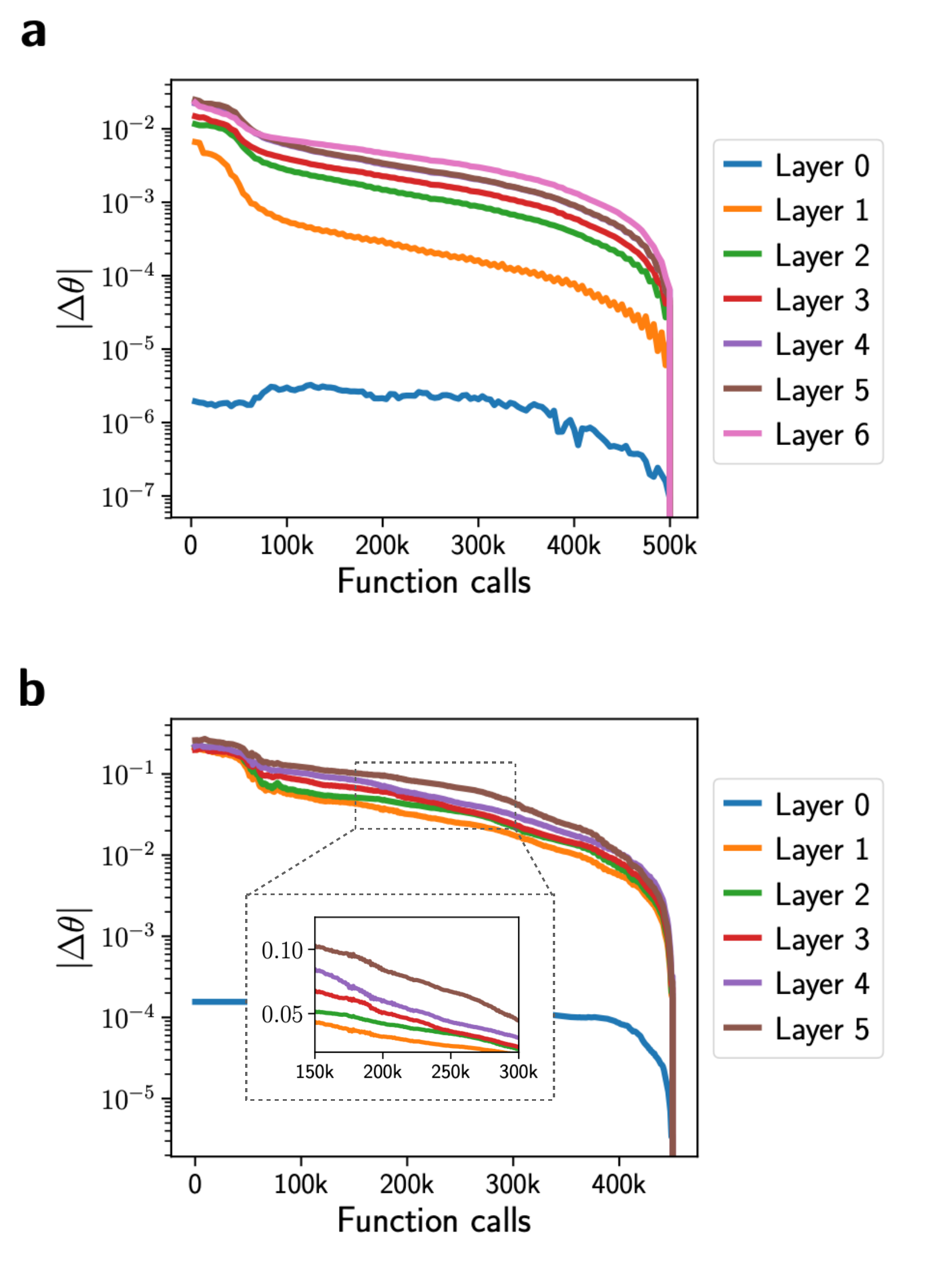}
\caption{
Plots tracking averaged parameter convergence to final values for each circuit layer.
Plot in (a) shows the averaged parameter convergence per circuit layer (superlayer) for a PECT-based optimization of the LDCA modeling the LiH ground state.
The averaged parameter convergence per layer is defined as in Equation \ref{eq:average absolute deviation}.
Plot in (b) shows the averaged parameter convergence per circuit layer for a non-PECT optimization of the LDCA modeling the LiH ground state.
We generally observe that parameters of earlier layers start closer and change less with respect to their final values than parameters of later layers do.
}
\label{fig:lih_ldca_pect_parameter_dynamics}
\end{figure}

While we expect earlier PQC layers to influence the optimization of later layers, we provide numerical evidence of this phenomenon by investigating how 
parameters of each layer of a PQC are optimized to their final values.
In Figure \ref{fig:lih_ldca_pect_parameter_dynamics}, we show the averaged convergence of parameters in each layer for two optimization instances, one using PECT and the other without.\footnote{These optimization results are from estimating the ground state energy of LiH at $R=2.5 \angstrom$ using the LDCA ansatz.}
We plot the absolute difference of parameter values at $t$-th optimization iteration from their corresponding final values, which are then averaged over parameters in circuit layer $l$.
That is, the averaged absolute difference of parameter values for the $l$-th circuit layer at the $t$-th iteration is defined as:
\begin{align}
\label{eq:average absolute deviation}
|\Delta \theta|_l^{(t)} = \frac{1}{N_l} \sum_{i=0}^{N_l} |\theta_i^{(t)} - \theta_{i}^{(t_{\text{final}})} |
\end{align}
\noindent where index $i$ enumerates parameters of layer $l$, and $N_l$ is the total number of parameters in layer $l$.
Thus, $\theta_i^{(t)}$ is the $i$-th parameter value in layer $l$ at iteration $t$, and $\theta_{i}^{(t_{\text{final}})}$ is the final/optimized $i$-th parameter value in layer $l$.
This quantity represents the average movement of parameter values per circuit layer. 
From Figure \ref{fig:lih_ldca_pect_parameter_dynamics},
we observe for both PECT and non-PECT optimizations that parameter values of earlier layers start closer and stray less from 
their final values, on average, compared to the movements of parameters of later layers.
This appears to suggest the importance of strategically initializing the first few layers such that parameters of later layers, which change values by greater magnitudes throughout the optimization, can more efficiently refine or improve upon circuit parameter settings of previous layers.
We clarify that the optimizations we consider in Figure \ref{fig:lih_ldca_pect_parameter_dynamics} do not employ layerwise training.
We performed this analysis for other PECT and non-PECT optimization runs, with several results shown in Appendix \ref{app:sec:param_dynamics_app}, and observed similar behavior.
The relative lack of change in earlier layers appears intuitive as we expect parameter settings of earlier layers to influence how parameters of later layers are tuned.
This observation also appears to justify layerwise training strategies for parameterized quantum circuits as well as suggest more carefully setting or initializing values of parameters of earlier layers.

\section{Future directions}\label{sec:future_directions}
PECT is a high-level optimization strategy for determining optimal parameter values for parameterized quantum circuits. 
Because PECT employs many subroutines, it remains suitable to leverage existing developments such as using the individual Coupled Adaptive Number of Shots (iCANs) optimizer \cite{Kubler2020} or the quantum natural gradient descent method \cite{Stokes2019} for local optimization.
Several advancements made to the optimization of PQCs are bottle-necked by the number of parameters in the circuit, e.g. quantum natural gradient's dependency on the computation and inversion of the Fubini-Study metric tensor which scales with the number of parameters.
Because PECT actively reduces the number of parameters in the PQC that are being optimized in each iteration, it also reduces the costs associated with these subroutines, thereby making the overall process more efficient.
Another improvement may be to employ the more sophisticated layerwise training schemes (e.g. in \cite{Skolik2020}) for layerwise PECT.

As with many heuristic methods, PECT's performance is sensitive to the choice of hyperparameters, especially global sparsity of the PQC $s$ and the target number of parameters to prune, $N_p$.
For instance, from our simulations, we observed that if the value of $s$ was assigned too high, i.e. the parameter subset at each iteration was too small, this often resulted in the PECT simulation terminating due to oscillating energies.
In such a case, by simply decreasing $s$, the performance often improved.
On the other hand, if $s$ was too low, PECT simulations would converge but expend more function calls due to the larger parameter space in each local optimization.
Thus, a closer investigation into properly tuning these hyperparameters may be necessary to more effectively use PECT for variational algorithms.

Our current implementation of PECT, using local optimizers such as L-BFGS-B and SLSQP, does not make effective use of circuit batching, which may result in a high overall experimental runtime due to network latency \cite{Sung2020}. 
One potential way to make PECT a better candidate for optimizing PQCs via the cloud is to employ local optimizers that support evaluations of circuit batches, e.g. particle swarm optimization \cite{Kennedy1995}. 

Lastly, significant effort has been dedicated to 
pruning or training sparse
networks in classical machine learning since the 1990's \cite{Le1990}.
Just as PECT adapted the dynamic sparse reparameterization method, we expect several neural network pruning methods to be applicable and beneficial for optimizing and generating more compact quantum circuits for variational quantum algorithms.

\section{Conclusion}\label{sec:conclusion}
Through random circuit sampling, quantum computers have recently been demonstrated to accomplish particular computational tasks exponentially faster than any classical computer can \cite{Arute2019}. 
A natural next step is designing and demonstrating practical applications of quantum computers; variational quantum algorithms are believed to be promising candidates for such experiments.
At this scale, optimization of parameterized quantum circuits will likely be a significant bottleneck among others, 
prompting a need for optimization strategies tailored to large circuits with many parameters.
Fortunately, there have been significant efforts in designing powerful parameterized ansatze for modeling ground states of strongly correlated systems. 
Here, we contribute to the set of ansatz optimization techniques by introducing the ``Parameter Efficient Circuit Training'' (PECT) method. 
By optimizing only a subset of gates of a parameterized ansatz and iteratively removing and adding gates to the subset to refine the circuit composition, PECT can construct an optimized circuit that additionally saves on resources such as circuit depth, two-qubit gate count and, in some cases, the optimization runtime.

Through noiseless simulations of VQE calculations on small molecular systems, we show that PECT is a promising approach for optimizing parameterized quantum circuits that have so far been limited by their difficulty in parameter exploration and optimization.
For instance, considering existing optimization strategies, PECT appears to be necessary for reliably converging calculations of molecular systems using LDCA and similar ansatze with high parameter counts and parameter redundancy.
As noted earlier, in addition to facilitating convergence to lower energies, using PECT may significantly reduce circuit depths, two-qubit gate counts, and optimization runtimes. 
Some of these features are particularly critical for realizing useful application of variational algorithms in NISQ devices, where noise imposes stringent limits to the size of the circuits we can execute.
PECT is compatible with other optimization strategies such as layerwise optimization, and could be used in conjunction with measurement-aware optimizers in order to minimize the number of circuit repetitions used during the local optimizations. 
We expect that, used jointly with several of these approaches, PECT would become an effective optimization strategy for refining and training parameterized quantum circuits across multiple quantum computing applications.

\section*{Acknowledgements}
S.S. is supported by the DOE Computational Science Graduate Fellowship under grant number DE-FG02-97ER25308.
All the authors would like to thank William Simon, Manuel Rudolph, Pierre-Luc Dallaire-Demers, and Peter Johnson for their valuable feedback. 
The optimization strategy and numerical simulations reported in this work were developed and executed through workflows managed by a pre-release version of the Orquestra platform \cite{orquestra}, using methods from OpenFermion \cite{openfermion} and Psi4 \cite{psi4} libraries to construct the problem Hamiltonians, Qiskit \cite{Qiskit} and Cirq libraries to estimate circuit resource requirements, and using the Intel Quantum Simulator \cite{Guerreschi2020} to simulate quantum circuits. 

\bibliographystyle{apsrev4-1}
\bibliography{references}

\begin{thebibliography}{66}%
\makeatletter
\providecommand \@ifxundefined [1]{%
 \@ifx{#1\undefined}
}%
\providecommand \@ifnum [1]{%
 \ifnum #1\expandafter \@firstoftwo
 \else \expandafter \@secondoftwo
 \fi
}%
\providecommand \@ifx [1]{%
 \ifx #1\expandafter \@firstoftwo
 \else \expandafter \@secondoftwo
 \fi
}%
\providecommand \natexlab [1]{#1}%
\providecommand \enquote  [1]{``#1''}%
\providecommand \bibnamefont  [1]{#1}%
\providecommand \bibfnamefont [1]{#1}%
\providecommand \citenamefont [1]{#1}%
\providecommand \href@noop [0]{\@secondoftwo}%
\providecommand \href [0]{\begingroup \@sanitize@url \@href}%
\providecommand \@href[1]{\@@startlink{#1}\@@href}%
\providecommand \@@href[1]{\endgroup#1\@@endlink}%
\providecommand \@sanitize@url [0]{\catcode `\\12\catcode `\$12\catcode
  `\&12\catcode `\#12\catcode `\^12\catcode `\_12\catcode `\%12\relax}%
\providecommand \@@startlink[1]{}%
\providecommand \@@endlink[0]{}%
\providecommand \url  [0]{\begingroup\@sanitize@url \@url }%
\providecommand \@url [1]{\endgroup\@href {#1}{\urlprefix }}%
\providecommand \urlprefix  [0]{URL }%
\providecommand \Eprint [0]{\href }%
\providecommand \doibase [0]{http://dx.doi.org/}%
\providecommand \selectlanguage [0]{\@gobble}%
\providecommand \bibinfo  [0]{\@secondoftwo}%
\providecommand \bibfield  [0]{\@secondoftwo}%
\providecommand \translation [1]{[#1]}%
\providecommand \BibitemOpen [0]{}%
\providecommand \bibitemStop [0]{}%
\providecommand \bibitemNoStop [0]{.\EOS\space}%
\providecommand \EOS [0]{\spacefactor3000\relax}%
\providecommand \BibitemShut  [1]{\csname bibitem#1\endcsname}%
\let\auto@bib@innerbib\@empty
\bibitem [{\citenamefont {Preskill}(2018)}]{Preskill2018}%
  \BibitemOpen
  \bibfield  {author} {\bibinfo {author} {\bibfnamefont {J.}~\bibnamefont
  {Preskill}},\ }\href {\doibase 10.22331/q-2018-08-06-79} {\bibfield
  {journal} {\bibinfo  {journal} {Quantum}\ }\textbf {\bibinfo {volume} {2}},\
  \bibinfo {pages} {79} (\bibinfo {year} {2018})}\BibitemShut {NoStop}%
\bibitem [{\citenamefont {Peruzzo}\ \emph {et~al.}(2014)\citenamefont
  {Peruzzo}, \citenamefont {McClean}, \citenamefont {Shadbolt}, \citenamefont
  {Yung}, \citenamefont {Zhou}, \citenamefont {Love}, \citenamefont
  {Aspuru-Guzik},\ and\ \citenamefont {O'Brien}}]{Peruzzo2014}%
  \BibitemOpen
  \bibfield  {author} {\bibinfo {author} {\bibfnamefont {A.}~\bibnamefont
  {Peruzzo}}, \bibinfo {author} {\bibfnamefont {J.}~\bibnamefont {McClean}},
  \bibinfo {author} {\bibfnamefont {P.}~\bibnamefont {Shadbolt}}, \bibinfo
  {author} {\bibfnamefont {M.-H.}\ \bibnamefont {Yung}}, \bibinfo {author}
  {\bibfnamefont {X.-Q.}\ \bibnamefont {Zhou}}, \bibinfo {author}
  {\bibfnamefont {P.~J.}\ \bibnamefont {Love}}, \bibinfo {author}
  {\bibfnamefont {A.}~\bibnamefont {Aspuru-Guzik}}, \ and\ \bibinfo {author}
  {\bibfnamefont {J.~L.}\ \bibnamefont {O'Brien}},\ }\href {\doibase
  10.1038/ncomms5213} {\bibfield  {journal} {\bibinfo  {journal} {Nat. Commun}\
  }\textbf {\bibinfo {volume} {5}},\ \bibinfo {pages} {4213} (\bibinfo {year}
  {2014})}\BibitemShut {NoStop}%
\bibitem [{\citenamefont {Farhi}\ \emph {et~al.}(2014)\citenamefont {Farhi},
  \citenamefont {Goldstone},\ and\ \citenamefont {Gutmann}}]{Farhi2014}%
  \BibitemOpen
  \bibfield  {author} {\bibinfo {author} {\bibfnamefont {E.}~\bibnamefont
  {Farhi}}, \bibinfo {author} {\bibfnamefont {J.}~\bibnamefont {Goldstone}}, \
  and\ \bibinfo {author} {\bibfnamefont {S.}~\bibnamefont {Gutmann}},\ }\href
  {http://arxiv.org/abs/1411.4028} {\enquote {\bibinfo {title} {{A Quantum
  Approximate Optimization Algorithm}},}\ } (\bibinfo {year} {2014}),\ \Eprint
  {http://arxiv.org/abs/1411.4028} {arXiv:1411.4028} \BibitemShut {NoStop}%
\bibitem [{\citenamefont {Havl{\'\i}{\v c}ek}\ \emph
  {et~al.}(2019)\citenamefont {Havl{\'\i}{\v c}ek}, \citenamefont
  {C{\'o}rcoles}, \citenamefont {Temme}, \citenamefont {Harrow}, \citenamefont
  {Kandala}, \citenamefont {Chow},\ and\ \citenamefont
  {Gambetta}}]{Havlicek2018}%
  \BibitemOpen
  \bibfield  {author} {\bibinfo {author} {\bibfnamefont {V.}~\bibnamefont
  {Havl{\'\i}{\v c}ek}}, \bibinfo {author} {\bibfnamefont {A.~D.}\ \bibnamefont
  {C{\'o}rcoles}}, \bibinfo {author} {\bibfnamefont {K.}~\bibnamefont {Temme}},
  \bibinfo {author} {\bibfnamefont {A.~W.}\ \bibnamefont {Harrow}}, \bibinfo
  {author} {\bibfnamefont {A.}~\bibnamefont {Kandala}}, \bibinfo {author}
  {\bibfnamefont {J.~M.}\ \bibnamefont {Chow}}, \ and\ \bibinfo {author}
  {\bibfnamefont {J.~M.}\ \bibnamefont {Gambetta}},\ }\href {\doibase
  10.1038/s41586-019-0980-2} {\bibfield  {journal} {\bibinfo  {journal}
  {Nature}\ }\textbf {\bibinfo {volume} {567}},\ \bibinfo {pages} {209}
  (\bibinfo {year} {2019})}\BibitemShut {NoStop}%
\bibitem [{\citenamefont {Farhi}\ and\ \citenamefont
  {Neven}(2018)}]{Farhi2018}%
  \BibitemOpen
  \bibfield  {author} {\bibinfo {author} {\bibfnamefont {E.}~\bibnamefont
  {Farhi}}\ and\ \bibinfo {author} {\bibfnamefont {H.}~\bibnamefont {Neven}},\
  }\href {http://arxiv.org/abs/1802.06002} {\enquote {\bibinfo {title}
  {{Classification with Quantum Neural Networks on Near Term Processors}},}\ }
  (\bibinfo {year} {2018}),\ \Eprint {http://arxiv.org/abs/1802.06002}
  {arXiv:1802.06002} \BibitemShut {NoStop}%
\bibitem [{\citenamefont {Schuld}\ \emph {et~al.}(2020)\citenamefont {Schuld},
  \citenamefont {Bocharov}, \citenamefont {Svore},\ and\ \citenamefont
  {Wiebe}}]{SchuldCircuitCentric}%
  \BibitemOpen
  \bibfield  {author} {\bibinfo {author} {\bibfnamefont {M.}~\bibnamefont
  {Schuld}}, \bibinfo {author} {\bibfnamefont {A.}~\bibnamefont {Bocharov}},
  \bibinfo {author} {\bibfnamefont {K.~M.}\ \bibnamefont {Svore}}, \ and\
  \bibinfo {author} {\bibfnamefont {N.}~\bibnamefont {Wiebe}},\ }\href
  {\doibase 10.1103/PhysRevA.101.032308} {\bibfield  {journal} {\bibinfo
  {journal} {Phys. Rev. A}\ }\textbf {\bibinfo {volume} {101}},\ \bibinfo
  {pages} {032308} (\bibinfo {year} {2020})}\BibitemShut {NoStop}%
\bibitem [{\citenamefont {Romero}\ \emph {et~al.}(2017)\citenamefont {Romero},
  \citenamefont {Olson},\ and\ \citenamefont {Aspuru-Guzik}}]{Romero2017}%
  \BibitemOpen
  \bibfield  {author} {\bibinfo {author} {\bibfnamefont {J.}~\bibnamefont
  {Romero}}, \bibinfo {author} {\bibfnamefont {J.~P.}\ \bibnamefont {Olson}}, \
  and\ \bibinfo {author} {\bibfnamefont {A.}~\bibnamefont {Aspuru-Guzik}},\
  }\href {\doibase 10.1088/2058-9565/aa8072} {\bibfield  {journal} {\bibinfo
  {journal} {Quantum Sci. Technol}\ }\textbf {\bibinfo {volume} {2}} (\bibinfo
  {year} {2017}),\ 10.1088/2058-9565/aa8072}\BibitemShut {NoStop}%
\bibitem [{\citenamefont {Dallaire-Demers}\ and\ \citenamefont
  {Killoran}(2018)}]{Dallaire-DemersQGAN}%
  \BibitemOpen
  \bibfield  {author} {\bibinfo {author} {\bibfnamefont {P.-L.}\ \bibnamefont
  {Dallaire-Demers}}\ and\ \bibinfo {author} {\bibfnamefont {N.}~\bibnamefont
  {Killoran}},\ }\href {\doibase 10.1103/PhysRevA.98.012324} {\bibfield
  {journal} {\bibinfo  {journal} {Phys. Rev. A}\ }\textbf {\bibinfo {volume}
  {98}},\ \bibinfo {pages} {012324} (\bibinfo {year} {2018})}\BibitemShut
  {NoStop}%
\bibitem [{\citenamefont {Lloyd}\ and\ \citenamefont
  {Weedbrook}(2018)}]{Lloyd2018}%
  \BibitemOpen
  \bibfield  {author} {\bibinfo {author} {\bibfnamefont {S.}~\bibnamefont
  {Lloyd}}\ and\ \bibinfo {author} {\bibfnamefont {C.}~\bibnamefont
  {Weedbrook}},\ }\href {\doibase 10.1103/PhysRevLett.121.040502} {\bibfield
  {journal} {\bibinfo  {journal} {Phys. Rev. Lett}\ }\textbf {\bibinfo {volume}
  {121}},\ \bibinfo {pages} {040502} (\bibinfo {year} {2018})}\BibitemShut
  {NoStop}%
\bibitem [{\citenamefont {Zeng}\ \emph {et~al.}(2019)\citenamefont {Zeng},
  \citenamefont {Wu}, \citenamefont {Liu}, \citenamefont {Wang},\ and\
  \citenamefont {Hu}}]{Zeng2019}%
  \BibitemOpen
  \bibfield  {author} {\bibinfo {author} {\bibfnamefont {J.}~\bibnamefont
  {Zeng}}, \bibinfo {author} {\bibfnamefont {Y.}~\bibnamefont {Wu}}, \bibinfo
  {author} {\bibfnamefont {J.~G.}\ \bibnamefont {Liu}}, \bibinfo {author}
  {\bibfnamefont {L.}~\bibnamefont {Wang}}, \ and\ \bibinfo {author}
  {\bibfnamefont {J.}~\bibnamefont {Hu}},\ }\href {\doibase
  10.1103/PhysRevA.99.052306} {\bibfield  {journal} {\bibinfo  {journal} {Phys.
  Rev. A}\ }\textbf {\bibinfo {volume} {99}} (\bibinfo {year} {2019}),\
  10.1103/PhysRevA.99.052306}\BibitemShut {NoStop}%
\bibitem [{\citenamefont {Situ}\ \emph {et~al.}(2020)\citenamefont {Situ},
  \citenamefont {He}, \citenamefont {Wang}, \citenamefont {Li},\ and\
  \citenamefont {Zheng}}]{Situ2020}%
  \BibitemOpen
  \bibfield  {author} {\bibinfo {author} {\bibfnamefont {H.}~\bibnamefont
  {Situ}}, \bibinfo {author} {\bibfnamefont {Z.}~\bibnamefont {He}}, \bibinfo
  {author} {\bibfnamefont {Y.}~\bibnamefont {Wang}}, \bibinfo {author}
  {\bibfnamefont {L.}~\bibnamefont {Li}}, \ and\ \bibinfo {author}
  {\bibfnamefont {S.}~\bibnamefont {Zheng}},\ }\href {\doibase
  10.1016/j.ins.2020.05.127} {\bibfield  {journal} {\bibinfo  {journal} {Inf.
  Sci.}\ }\textbf {\bibinfo {volume} {538}},\ \bibinfo {pages} {193} (\bibinfo
  {year} {2020})}\BibitemShut {NoStop}%
\bibitem [{\citenamefont {Zhu}\ \emph {et~al.}(2019)\citenamefont {Zhu},
  \citenamefont {Linke}, \citenamefont {Benedetti}, \citenamefont {Landsman},
  \citenamefont {Nguyen}, \citenamefont {Alderete}, \citenamefont
  {Perdomo-Ortiz}, \citenamefont {Korda}, \citenamefont {Garfoot},
  \citenamefont {Brecque}, \citenamefont {Egan}, \citenamefont {Perdomo},\ and\
  \citenamefont {Monroe}}]{Zhu2019}%
  \BibitemOpen
  \bibfield  {author} {\bibinfo {author} {\bibfnamefont {D.}~\bibnamefont
  {Zhu}}, \bibinfo {author} {\bibfnamefont {N.~M.}\ \bibnamefont {Linke}},
  \bibinfo {author} {\bibfnamefont {M.}~\bibnamefont {Benedetti}}, \bibinfo
  {author} {\bibfnamefont {K.~A.}\ \bibnamefont {Landsman}}, \bibinfo {author}
  {\bibfnamefont {N.~H.}\ \bibnamefont {Nguyen}}, \bibinfo {author}
  {\bibfnamefont {C.~H.}\ \bibnamefont {Alderete}}, \bibinfo {author}
  {\bibfnamefont {A.}~\bibnamefont {Perdomo-Ortiz}}, \bibinfo {author}
  {\bibfnamefont {N.}~\bibnamefont {Korda}}, \bibinfo {author} {\bibfnamefont
  {A.}~\bibnamefont {Garfoot}}, \bibinfo {author} {\bibfnamefont
  {C.}~\bibnamefont {Brecque}}, \bibinfo {author} {\bibfnamefont
  {L.}~\bibnamefont {Egan}}, \bibinfo {author} {\bibfnamefont {O.}~\bibnamefont
  {Perdomo}}, \ and\ \bibinfo {author} {\bibfnamefont {C.}~\bibnamefont
  {Monroe}},\ }\href {\doibase 10.1126/sciadv.aaw9918} {\bibfield  {journal}
  {\bibinfo  {journal} {Sci. Adv.}\ }\textbf {\bibinfo {volume} {5}},\ \bibinfo
  {pages} {eaaw9918} (\bibinfo {year} {2019})}\BibitemShut {NoStop}%
\bibitem [{\citenamefont {Romero}\ and\ \citenamefont
  {Aspuru-Guzik}(2019)}]{Romero2019}%
  \BibitemOpen
  \bibfield  {author} {\bibinfo {author} {\bibfnamefont {J.}~\bibnamefont
  {Romero}}\ and\ \bibinfo {author} {\bibfnamefont {A.}~\bibnamefont
  {Aspuru-Guzik}},\ }\href {http://arxiv.org/abs/1901.00848} {\enquote
  {\bibinfo {title} {{Variational quantum generators: Generative adversarial
  quantum machine learning for continuous distributions}},}\ } (\bibinfo {year}
  {2019}),\ \Eprint {http://arxiv.org/abs/1901.00848} {arXiv:1901.00848}
  \BibitemShut {NoStop}%
\bibitem [{\citenamefont {McClean}\ \emph {et~al.}(2018)\citenamefont
  {McClean}, \citenamefont {Boixo}, \citenamefont {Smelyanskiy}, \citenamefont
  {Babbush},\ and\ \citenamefont {Neven}}]{McClean2018}%
  \BibitemOpen
  \bibfield  {author} {\bibinfo {author} {\bibfnamefont {J.~R.}\ \bibnamefont
  {McClean}}, \bibinfo {author} {\bibfnamefont {S.}~\bibnamefont {Boixo}},
  \bibinfo {author} {\bibfnamefont {V.~N.}\ \bibnamefont {Smelyanskiy}},
  \bibinfo {author} {\bibfnamefont {R.}~\bibnamefont {Babbush}}, \ and\
  \bibinfo {author} {\bibfnamefont {H.}~\bibnamefont {Neven}},\ }\href
  {\doibase 10.1038/s41467-018-07090-4} {\bibfield  {journal} {\bibinfo
  {journal} {Nat. Commun}\ }\textbf {\bibinfo {volume} {9}},\ \bibinfo {pages}
  {4812} (\bibinfo {year} {2018})}\BibitemShut {NoStop}%
\bibitem [{\citenamefont {Cerezo}\ \emph {et~al.}(2020)\citenamefont {Cerezo},
  \citenamefont {Sone}, \citenamefont {Volkoff}, \citenamefont {Cincio},\ and\
  \citenamefont {Coles}}]{Cerezo2020}%
  \BibitemOpen
  \bibfield  {author} {\bibinfo {author} {\bibfnamefont {M.}~\bibnamefont
  {Cerezo}}, \bibinfo {author} {\bibfnamefont {A.}~\bibnamefont {Sone}},
  \bibinfo {author} {\bibfnamefont {T.}~\bibnamefont {Volkoff}}, \bibinfo
  {author} {\bibfnamefont {L.}~\bibnamefont {Cincio}}, \ and\ \bibinfo {author}
  {\bibfnamefont {P.~J.}\ \bibnamefont {Coles}},\ }\href
  {http://arxiv.org/abs/2001.00550} {\enquote {\bibinfo {title}
  {{Cost-Function-Dependent Barren Plateaus in Shallow Quantum Neural
  Networks}},}\ } (\bibinfo {year} {2020}),\ \Eprint
  {http://arxiv.org/abs/2001.00550} {arXiv:2001.00550} \BibitemShut {NoStop}%
\bibitem [{\citenamefont {Volkoff}\ and\ \citenamefont
  {Coles}(2020)}]{Volkoff2020}%
  \BibitemOpen
  \bibfield  {author} {\bibinfo {author} {\bibfnamefont {T.}~\bibnamefont
  {Volkoff}}\ and\ \bibinfo {author} {\bibfnamefont {P.~J.}\ \bibnamefont
  {Coles}},\ }\href {http://arxiv.org/abs/2005.12200} {\enquote {\bibinfo
  {title} {{Large gradients via correlation in random parameterized quantum
  circuits}},}\ } (\bibinfo {year} {2020}),\ \Eprint
  {http://arxiv.org/abs/2005.12200} {arXiv:2005.12200} \BibitemShut {NoStop}%
\bibitem [{\citenamefont {Ostaszewski}\ \emph {et~al.}(2019)\citenamefont
  {Ostaszewski}, \citenamefont {Grant},\ and\ \citenamefont
  {Benedetti}}]{Ostaszewski2019}%
  \BibitemOpen
  \bibfield  {author} {\bibinfo {author} {\bibfnamefont {M.}~\bibnamefont
  {Ostaszewski}}, \bibinfo {author} {\bibfnamefont {E.}~\bibnamefont {Grant}},
  \ and\ \bibinfo {author} {\bibfnamefont {M.}~\bibnamefont {Benedetti}},\
  }\href {http://arxiv.org/abs/1905.09692} {\enquote {\bibinfo {title}
  {{Quantum circuit structure learning}},}\ } (\bibinfo {year} {2019}),\
  \Eprint {http://arxiv.org/abs/1905.09692} {arXiv:1905.09692} \BibitemShut
  {NoStop}%
\bibitem [{\citenamefont {Wilson}\ \emph {et~al.}(2019)\citenamefont {Wilson},
  \citenamefont {Stromwold}, \citenamefont {Wudarski}, \citenamefont
  {Hadfield}, \citenamefont {Tubman},\ and\ \citenamefont
  {Rieffel}}]{Wilson2019}%
  \BibitemOpen
  \bibfield  {author} {\bibinfo {author} {\bibfnamefont {M.}~\bibnamefont
  {Wilson}}, \bibinfo {author} {\bibfnamefont {S.}~\bibnamefont {Stromwold}},
  \bibinfo {author} {\bibfnamefont {F.}~\bibnamefont {Wudarski}}, \bibinfo
  {author} {\bibfnamefont {S.}~\bibnamefont {Hadfield}}, \bibinfo {author}
  {\bibfnamefont {N.~M.}\ \bibnamefont {Tubman}}, \ and\ \bibinfo {author}
  {\bibfnamefont {E.~G.}\ \bibnamefont {Rieffel}},\ }\href
  {http://arxiv.org/abs/1908.03185} {\enquote {\bibinfo {title} {{Optimizing
  quantum heuristics with meta-learning}},}\ } (\bibinfo {year} {2019}),\
  \Eprint {http://arxiv.org/abs/1908.03185} {arXiv:1908.03185} \BibitemShut
  {NoStop}%
\bibitem [{\citenamefont {Grimsley}\ \emph {et~al.}(2019)\citenamefont
  {Grimsley}, \citenamefont {Economou}, \citenamefont {Barnes},\ and\
  \citenamefont {Mayhall}}]{Grimsley2019}%
  \BibitemOpen
  \bibfield  {author} {\bibinfo {author} {\bibfnamefont {H.~R.}\ \bibnamefont
  {Grimsley}}, \bibinfo {author} {\bibfnamefont {S.~E.}\ \bibnamefont
  {Economou}}, \bibinfo {author} {\bibfnamefont {E.}~\bibnamefont {Barnes}}, \
  and\ \bibinfo {author} {\bibfnamefont {N.~J.}\ \bibnamefont {Mayhall}},\
  }\href {\doibase 10.1038/s41467-019-10988-2} {\bibfield  {journal} {\bibinfo
  {journal} {Nat. Commun}\ }\textbf {\bibinfo {volume} {10}},\ \bibinfo {pages}
  {3007} (\bibinfo {year} {2019})}\BibitemShut {NoStop}%
\bibitem [{\citenamefont {Tang}\ \emph {et~al.}(2019)\citenamefont {Tang},
  \citenamefont {Barnes}, \citenamefont {Grimsley}, \citenamefont {Mayhall},\
  and\ \citenamefont {Economou}}]{Tang2019}%
  \BibitemOpen
  \bibfield  {author} {\bibinfo {author} {\bibfnamefont {H.~L.}\ \bibnamefont
  {Tang}}, \bibinfo {author} {\bibfnamefont {E.}~\bibnamefont {Barnes}},
  \bibinfo {author} {\bibfnamefont {H.~R.}\ \bibnamefont {Grimsley}}, \bibinfo
  {author} {\bibfnamefont {N.~J.}\ \bibnamefont {Mayhall}}, \ and\ \bibinfo
  {author} {\bibfnamefont {S.~E.}\ \bibnamefont {Economou}},\ }\href
  {http://arxiv.org/abs/1911.10205} {\enquote {\bibinfo {title}
  {{qubit-ADAPT-VQE: An adaptive algorithm for constructing hardware-efficient
  ansatze on a quantum processor}},}\ } (\bibinfo {year} {2019}),\ \Eprint
  {http://arxiv.org/abs/1911.10205} {arXiv:1911.10205} \BibitemShut {NoStop}%
\bibitem [{\citenamefont {Stokes}\ \emph {et~al.}(2020)\citenamefont {Stokes},
  \citenamefont {Izaac}, \citenamefont {Killoran},\ and\ \citenamefont
  {Carleo}}]{Stokes2019}%
  \BibitemOpen
  \bibfield  {author} {\bibinfo {author} {\bibfnamefont {J.}~\bibnamefont
  {Stokes}}, \bibinfo {author} {\bibfnamefont {J.}~\bibnamefont {Izaac}},
  \bibinfo {author} {\bibfnamefont {N.}~\bibnamefont {Killoran}}, \ and\
  \bibinfo {author} {\bibfnamefont {G.}~\bibnamefont {Carleo}},\ }\href
  {\doibase 10.22331/q-2020-05-25-269} {\bibfield  {journal} {\bibinfo
  {journal} {Quantum}\ }\textbf {\bibinfo {volume} {4}},\ \bibinfo {pages}
  {269} (\bibinfo {year} {2020})}\BibitemShut {NoStop}%
\bibitem [{\citenamefont {Rattew}\ \emph {et~al.}(2019)\citenamefont {Rattew},
  \citenamefont {Hu}, \citenamefont {Pistoia}, \citenamefont {Chen},\ and\
  \citenamefont {Wood}}]{Rattew2019}%
  \BibitemOpen
  \bibfield  {author} {\bibinfo {author} {\bibfnamefont {A.~G.}\ \bibnamefont
  {Rattew}}, \bibinfo {author} {\bibfnamefont {S.}~\bibnamefont {Hu}}, \bibinfo
  {author} {\bibfnamefont {M.}~\bibnamefont {Pistoia}}, \bibinfo {author}
  {\bibfnamefont {R.}~\bibnamefont {Chen}}, \ and\ \bibinfo {author}
  {\bibfnamefont {S.}~\bibnamefont {Wood}},\ }\href
  {http://arxiv.org/abs/1910.09694} {\enquote {\bibinfo {title} {{A
  Domain-agnostic, Noise-resistant, Hardware-efficient Evolutionary Variational
  Quantum Eigensolver}},}\ } (\bibinfo {year} {2019}),\ \Eprint
  {http://arxiv.org/abs/1910.09694} {arXiv:1910.09694} \BibitemShut {NoStop}%
\bibitem [{\citenamefont {K{\"{u}}bler}\ \emph {et~al.}(2020)\citenamefont
  {K{\"{u}}bler}, \citenamefont {Arrasmith}, \citenamefont {Cincio},\ and\
  \citenamefont {Coles}}]{Kubler2020}%
  \BibitemOpen
  \bibfield  {author} {\bibinfo {author} {\bibfnamefont {J.~M.}\ \bibnamefont
  {K{\"{u}}bler}}, \bibinfo {author} {\bibfnamefont {A.}~\bibnamefont
  {Arrasmith}}, \bibinfo {author} {\bibfnamefont {L.}~\bibnamefont {Cincio}}, \
  and\ \bibinfo {author} {\bibfnamefont {P.~J.}\ \bibnamefont {Coles}},\ }\href
  {\doibase 10.22331/q-2020-05-11-263} {\bibfield  {journal} {\bibinfo
  {journal} {Quantum}\ }\textbf {\bibinfo {volume} {4}},\ \bibinfo {pages}
  {263} (\bibinfo {year} {2020})},\ \Eprint {http://arxiv.org/abs/1909.09083}
  {1909.09083} \BibitemShut {NoStop}%
\bibitem [{\citenamefont {Chivilikhin}\ \emph {et~al.}(2020)\citenamefont
  {Chivilikhin}, \citenamefont {Samarin}, \citenamefont {Ulyantsev},
  \citenamefont {Iorsh}, \citenamefont {Oganov},\ and\ \citenamefont
  {Kyriienko}}]{Chivilikhin2020}%
  \BibitemOpen
  \bibfield  {author} {\bibinfo {author} {\bibfnamefont {D.}~\bibnamefont
  {Chivilikhin}}, \bibinfo {author} {\bibfnamefont {A.}~\bibnamefont
  {Samarin}}, \bibinfo {author} {\bibfnamefont {V.}~\bibnamefont {Ulyantsev}},
  \bibinfo {author} {\bibfnamefont {I.}~\bibnamefont {Iorsh}}, \bibinfo
  {author} {\bibfnamefont {A.~R.}\ \bibnamefont {Oganov}}, \ and\ \bibinfo
  {author} {\bibfnamefont {O.}~\bibnamefont {Kyriienko}},\ }\href
  {http://arxiv.org/abs/2007.04424} {\enquote {\bibinfo {title} {{MoG-VQE:
  Multiobjective genetic variational quantum eigensolver}},}\ } (\bibinfo
  {year} {2020}),\ \Eprint {http://arxiv.org/abs/2007.04424} {arXiv:2007.04424}
  \BibitemShut {NoStop}%
\bibitem [{\citenamefont {Sung}\ \emph {et~al.}(2020)\citenamefont {Sung},
  \citenamefont {Harrigan}, \citenamefont {Rubin}, \citenamefont {Jiang},
  \citenamefont {Babbush},\ and\ \citenamefont {McClean}}]{Sung2020}%
  \BibitemOpen
  \bibfield  {author} {\bibinfo {author} {\bibfnamefont {K.~J.}\ \bibnamefont
  {Sung}}, \bibinfo {author} {\bibfnamefont {M.~P.}\ \bibnamefont {Harrigan}},
  \bibinfo {author} {\bibfnamefont {N.~C.}\ \bibnamefont {Rubin}}, \bibinfo
  {author} {\bibfnamefont {Z.}~\bibnamefont {Jiang}}, \bibinfo {author}
  {\bibfnamefont {R.}~\bibnamefont {Babbush}}, \ and\ \bibinfo {author}
  {\bibfnamefont {J.~R.}\ \bibnamefont {McClean}},\ }\href
  {http://arxiv.org/abs/2005.11011} {\enquote {\bibinfo {title} {{An
  Exploration of Practical Optimizers for Variational Quantum Algorithms on
  Superconducting Qubit Processors}},}\ } (\bibinfo {year} {2020}),\ \Eprint
  {http://arxiv.org/abs/2005.11011} {arXiv:2005.11011} \BibitemShut {NoStop}%
\bibitem [{\citenamefont {Wecker}\ \emph {et~al.}(2015)\citenamefont {Wecker},
  \citenamefont {Hastings},\ and\ \citenamefont {Troyer}}]{Wecker2015}%
  \BibitemOpen
  \bibfield  {author} {\bibinfo {author} {\bibfnamefont {D.}~\bibnamefont
  {Wecker}}, \bibinfo {author} {\bibfnamefont {M.~B.}\ \bibnamefont
  {Hastings}}, \ and\ \bibinfo {author} {\bibfnamefont {M.}~\bibnamefont
  {Troyer}},\ }\href {\doibase 10.1103/PhysRevA.92.042303} {\bibfield
  {journal} {\bibinfo  {journal} {Phys. Rev. A}\ }\textbf {\bibinfo {volume}
  {92}},\ \bibinfo {pages} {042303} (\bibinfo {year} {2015})}\BibitemShut
  {NoStop}%
\bibitem [{\citenamefont {Garcia-Saez}\ and\ \citenamefont
  {Latorre}(2018)}]{Garcia-Saez2018}%
  \BibitemOpen
  \bibfield  {author} {\bibinfo {author} {\bibfnamefont {A.}~\bibnamefont
  {Garcia-Saez}}\ and\ \bibinfo {author} {\bibfnamefont {J.~I.}\ \bibnamefont
  {Latorre}},\ }\href {http://arxiv.org/abs/1806.02287} {\enquote {\bibinfo
  {title} {{Addressing hard classical problems with Adiabatically Assisted
  Variational Quantum Eigensolvers}},}\ } (\bibinfo {year} {2018}),\ \Eprint
  {http://arxiv.org/abs/1806.02287} {arXiv:1806.02287} \BibitemShut {NoStop}%
\bibitem [{\citenamefont {Skolik}\ \emph {et~al.}(2020)\citenamefont {Skolik},
  \citenamefont {McClean}, \citenamefont {Mohseni}, \citenamefont {van~der
  Smagt},\ and\ \citenamefont {Leib}}]{Skolik2020}%
  \BibitemOpen
  \bibfield  {author} {\bibinfo {author} {\bibfnamefont {A.}~\bibnamefont
  {Skolik}}, \bibinfo {author} {\bibfnamefont {J.~R.}\ \bibnamefont {McClean}},
  \bibinfo {author} {\bibfnamefont {M.}~\bibnamefont {Mohseni}}, \bibinfo
  {author} {\bibfnamefont {P.}~\bibnamefont {van~der Smagt}}, \ and\ \bibinfo
  {author} {\bibfnamefont {M.}~\bibnamefont {Leib}},\ }\href
  {http://arxiv.org/abs/2006.14904} {\enquote {\bibinfo {title} {{Layerwise
  learning for quantum neural networks}},}\ } (\bibinfo {year} {2020}),\
  \Eprint {http://arxiv.org/abs/2006.14904} {arXiv:2006.14904} \BibitemShut
  {NoStop}%
\bibitem [{\citenamefont {Grant}\ \emph {et~al.}(2019)\citenamefont {Grant},
  \citenamefont {Wossnig}, \citenamefont {Ostaszewski},\ and\ \citenamefont
  {Benedetti}}]{Grant2019}%
  \BibitemOpen
  \bibfield  {author} {\bibinfo {author} {\bibfnamefont {E.}~\bibnamefont
  {Grant}}, \bibinfo {author} {\bibfnamefont {L.}~\bibnamefont {Wossnig}},
  \bibinfo {author} {\bibfnamefont {M.}~\bibnamefont {Ostaszewski}}, \ and\
  \bibinfo {author} {\bibfnamefont {M.}~\bibnamefont {Benedetti}},\ }\href
  {\doibase 10.22331/q-2019-12-09-214} {\bibfield  {journal} {\bibinfo
  {journal} {Quantum}\ }\textbf {\bibinfo {volume} {3}},\ \bibinfo {pages}
  {214} (\bibinfo {year} {2019})}\BibitemShut {NoStop}%
\bibitem [{\citenamefont {Verdon}\ \emph {et~al.}(2019)\citenamefont {Verdon},
  \citenamefont {Broughton}, \citenamefont {McClean}, \citenamefont {Sung},
  \citenamefont {Babbush}, \citenamefont {Jiang}, \citenamefont {Neven},\ and\
  \citenamefont {Mohseni}}]{Verdon2019}%
  \BibitemOpen
  \bibfield  {author} {\bibinfo {author} {\bibfnamefont {G.}~\bibnamefont
  {Verdon}}, \bibinfo {author} {\bibfnamefont {M.}~\bibnamefont {Broughton}},
  \bibinfo {author} {\bibfnamefont {J.~R.}\ \bibnamefont {McClean}}, \bibinfo
  {author} {\bibfnamefont {K.~J.}\ \bibnamefont {Sung}}, \bibinfo {author}
  {\bibfnamefont {R.}~\bibnamefont {Babbush}}, \bibinfo {author} {\bibfnamefont
  {Z.}~\bibnamefont {Jiang}}, \bibinfo {author} {\bibfnamefont
  {H.}~\bibnamefont {Neven}}, \ and\ \bibinfo {author} {\bibfnamefont
  {M.}~\bibnamefont {Mohseni}},\ }\href {http://arxiv.org/abs/1907.05415}
  {\enquote {\bibinfo {title} {{Learning to learn with quantum neural networks
  via classical neural networks}},}\ } (\bibinfo {year} {2019}),\ \Eprint
  {http://arxiv.org/abs/1907.05415} {arXiv:1907.05415} \BibitemShut {NoStop}%
\bibitem [{\citenamefont {Lavrijsen}\ \emph {et~al.}(2020)\citenamefont
  {Lavrijsen}, \citenamefont {Tudor}, \citenamefont {M{\"{u}}ller},
  \citenamefont {Iancu},\ and\ \citenamefont {de~Jong}}]{Lavrijsen2020}%
  \BibitemOpen
  \bibfield  {author} {\bibinfo {author} {\bibfnamefont {W.}~\bibnamefont
  {Lavrijsen}}, \bibinfo {author} {\bibfnamefont {A.}~\bibnamefont {Tudor}},
  \bibinfo {author} {\bibfnamefont {J.}~\bibnamefont {M{\"{u}}ller}}, \bibinfo
  {author} {\bibfnamefont {C.}~\bibnamefont {Iancu}}, \ and\ \bibinfo {author}
  {\bibfnamefont {W.}~\bibnamefont {de~Jong}},\ }\href
  {http://arxiv.org/abs/2004.03004} {\enquote {\bibinfo {title} {{Classical
  Optimizers for Noisy Intermediate-Scale Quantum Devices}},}\ } (\bibinfo
  {year} {2020}),\ \Eprint {http://arxiv.org/abs/2004.03004} {arXiv:2004.03004}
  \BibitemShut {NoStop}%
\bibitem [{\citenamefont {Shen}\ \emph {et~al.}(2017)\citenamefont {Shen},
  \citenamefont {Zhang}, \citenamefont {Zhang}, \citenamefont {Zhang},
  \citenamefont {Yung},\ and\ \citenamefont {Kim}}]{Shen2017}%
  \BibitemOpen
  \bibfield  {author} {\bibinfo {author} {\bibfnamefont {Y.}~\bibnamefont
  {Shen}}, \bibinfo {author} {\bibfnamefont {X.}~\bibnamefont {Zhang}},
  \bibinfo {author} {\bibfnamefont {S.}~\bibnamefont {Zhang}}, \bibinfo
  {author} {\bibfnamefont {J.~N.}\ \bibnamefont {Zhang}}, \bibinfo {author}
  {\bibfnamefont {M.~H.}\ \bibnamefont {Yung}}, \ and\ \bibinfo {author}
  {\bibfnamefont {K.}~\bibnamefont {Kim}},\ }\href {\doibase
  10.1103/PhysRevA.95.020501} {\bibfield  {journal} {\bibinfo  {journal} {Phys.
  Rev. A}\ }\textbf {\bibinfo {volume} {95}} (\bibinfo {year} {2017}),\
  10.1103/PhysRevA.95.020501}\BibitemShut {NoStop}%
\bibitem [{\citenamefont {Babbush}\ \emph {et~al.}(2018)\citenamefont
  {Babbush}, \citenamefont {Wiebe}, \citenamefont {McClean}, \citenamefont
  {McClain}, \citenamefont {Neven},\ and\ \citenamefont {Chan}}]{Babbush2017}%
  \BibitemOpen
  \bibfield  {author} {\bibinfo {author} {\bibfnamefont {R.}~\bibnamefont
  {Babbush}}, \bibinfo {author} {\bibfnamefont {N.}~\bibnamefont {Wiebe}},
  \bibinfo {author} {\bibfnamefont {J.}~\bibnamefont {McClean}}, \bibinfo
  {author} {\bibfnamefont {J.}~\bibnamefont {McClain}}, \bibinfo {author}
  {\bibfnamefont {H.}~\bibnamefont {Neven}}, \ and\ \bibinfo {author}
  {\bibfnamefont {G.~K.-L.}\ \bibnamefont {Chan}},\ }\href {\doibase
  10.1103/PhysRevX.8.011044} {\bibfield  {journal} {\bibinfo  {journal} {Phys.
  Rev. X}\ }\textbf {\bibinfo {volume} {8}},\ \bibinfo {pages} {011044}
  (\bibinfo {year} {2018})}\BibitemShut {NoStop}%
\bibitem [{\citenamefont {Kivlichan}\ \emph {et~al.}(2018)\citenamefont
  {Kivlichan}, \citenamefont {McClean}, \citenamefont {Wiebe}, \citenamefont
  {Gidney}, \citenamefont {Aspuru-Guzik}, \citenamefont {Chan},\ and\
  \citenamefont {Babbush}}]{Kivlichan2018}%
  \BibitemOpen
  \bibfield  {author} {\bibinfo {author} {\bibfnamefont {I.~D.}\ \bibnamefont
  {Kivlichan}}, \bibinfo {author} {\bibfnamefont {J.}~\bibnamefont {McClean}},
  \bibinfo {author} {\bibfnamefont {N.}~\bibnamefont {Wiebe}}, \bibinfo
  {author} {\bibfnamefont {C.}~\bibnamefont {Gidney}}, \bibinfo {author}
  {\bibfnamefont {A.}~\bibnamefont {Aspuru-Guzik}}, \bibinfo {author}
  {\bibfnamefont {G.~K.~L.}\ \bibnamefont {Chan}}, \ and\ \bibinfo {author}
  {\bibfnamefont {R.}~\bibnamefont {Babbush}},\ }\href {\doibase
  10.1103/PhysRevLett.120.110501} {\bibfield  {journal} {\bibinfo  {journal}
  {Phys. Rev. Lett}\ }\textbf {\bibinfo {volume} {120}} (\bibinfo {year}
  {2018}),\ 10.1103/PhysRevLett.120.110501}\BibitemShut {NoStop}%
\bibitem [{\citenamefont {Barkoutsos}\ \emph {et~al.}(2018)\citenamefont
  {Barkoutsos}, \citenamefont {Gonthier}, \citenamefont {Sokolov},
  \citenamefont {Moll}, \citenamefont {Salis}, \citenamefont {Fuhrer},
  \citenamefont {Ganzhorn}, \citenamefont {Egger}, \citenamefont {Troyer},
  \citenamefont {Mezzacapo}, \citenamefont {Filipp},\ and\ \citenamefont
  {Tavernelli}}]{Barkoutsos2018}%
  \BibitemOpen
  \bibfield  {author} {\bibinfo {author} {\bibfnamefont {P.~K.}\ \bibnamefont
  {Barkoutsos}}, \bibinfo {author} {\bibfnamefont {J.~F.}\ \bibnamefont
  {Gonthier}}, \bibinfo {author} {\bibfnamefont {I.}~\bibnamefont {Sokolov}},
  \bibinfo {author} {\bibfnamefont {N.}~\bibnamefont {Moll}}, \bibinfo {author}
  {\bibfnamefont {G.}~\bibnamefont {Salis}}, \bibinfo {author} {\bibfnamefont
  {A.}~\bibnamefont {Fuhrer}}, \bibinfo {author} {\bibfnamefont
  {M.}~\bibnamefont {Ganzhorn}}, \bibinfo {author} {\bibfnamefont {D.~J.}\
  \bibnamefont {Egger}}, \bibinfo {author} {\bibfnamefont {M.}~\bibnamefont
  {Troyer}}, \bibinfo {author} {\bibfnamefont {A.}~\bibnamefont {Mezzacapo}},
  \bibinfo {author} {\bibfnamefont {S.}~\bibnamefont {Filipp}}, \ and\ \bibinfo
  {author} {\bibfnamefont {I.}~\bibnamefont {Tavernelli}},\ }\href {\doibase
  10.1103/PhysRevA.98.022322} {\bibfield  {journal} {\bibinfo  {journal} {Phys.
  Rev. A}\ }\textbf {\bibinfo {volume} {98}},\ \bibinfo {pages} {022322}
  (\bibinfo {year} {2018})}\BibitemShut {NoStop}%
\bibitem [{\citenamefont {Dallaire-Demers}\ \emph {et~al.}(2019)\citenamefont
  {Dallaire-Demers}, \citenamefont {Romero}, \citenamefont {Veis},
  \citenamefont {Sim},\ and\ \citenamefont
  {Aspuru-Guzik}}]{Dallaire-Demers2018}%
  \BibitemOpen
  \bibfield  {author} {\bibinfo {author} {\bibfnamefont {P.-L.}\ \bibnamefont
  {Dallaire-Demers}}, \bibinfo {author} {\bibfnamefont {J.}~\bibnamefont
  {Romero}}, \bibinfo {author} {\bibfnamefont {L.}~\bibnamefont {Veis}},
  \bibinfo {author} {\bibfnamefont {S.}~\bibnamefont {Sim}}, \ and\ \bibinfo
  {author} {\bibfnamefont {A.}~\bibnamefont {Aspuru-Guzik}},\ }\href {\doibase
  10.1088/2058-9565/ab3951} {\bibfield  {journal} {\bibinfo  {journal} {Quantum
  Sci. Technol.}\ }\textbf {\bibinfo {volume} {4}},\ \bibinfo {pages} {045005}
  (\bibinfo {year} {2019})}\BibitemShut {NoStop}%
\bibitem [{\citenamefont {Lee}\ \emph {et~al.}(2019)\citenamefont {Lee},
  \citenamefont {Huggins}, \citenamefont {Head-Gordon},\ and\ \citenamefont
  {Whaley}}]{Lee2019}%
  \BibitemOpen
  \bibfield  {author} {\bibinfo {author} {\bibfnamefont {J.}~\bibnamefont
  {Lee}}, \bibinfo {author} {\bibfnamefont {W.~J.}\ \bibnamefont {Huggins}},
  \bibinfo {author} {\bibfnamefont {M.}~\bibnamefont {Head-Gordon}}, \ and\
  \bibinfo {author} {\bibfnamefont {K.~B.}\ \bibnamefont {Whaley}},\ }\href
  {\doibase 10.1021/acs.jctc.8b01004} {\bibfield  {journal} {\bibinfo
  {journal} {J. Chem. Theory Comput}\ }\textbf {\bibinfo {volume} {15}},\
  \bibinfo {pages} {311} (\bibinfo {year} {2019})}\BibitemShut {NoStop}%
\bibitem [{\citenamefont {Choquette}\ \emph {et~al.}(2020)\citenamefont
  {Choquette}, \citenamefont {{Di Paolo}}, \citenamefont {Barkoutsos},
  \citenamefont {S{\'{e}}n{\'{e}}chal}, \citenamefont {Tavernelli},\ and\
  \citenamefont {Blais}}]{Choquette2020}%
  \BibitemOpen
  \bibfield  {author} {\bibinfo {author} {\bibfnamefont {A.}~\bibnamefont
  {Choquette}}, \bibinfo {author} {\bibfnamefont {A.}~\bibnamefont {{Di
  Paolo}}}, \bibinfo {author} {\bibfnamefont {P.~K.}\ \bibnamefont
  {Barkoutsos}}, \bibinfo {author} {\bibfnamefont {D.}~\bibnamefont
  {S{\'{e}}n{\'{e}}chal}}, \bibinfo {author} {\bibfnamefont {I.}~\bibnamefont
  {Tavernelli}}, \ and\ \bibinfo {author} {\bibfnamefont {A.}~\bibnamefont
  {Blais}},\ }\href {http://arxiv.org/abs/2008.01098} {\enquote {\bibinfo
  {title} {{Quantum-optimal-control-inspired ansatz for variational quantum
  algorithms}},}\ } (\bibinfo {year} {2020}),\ \Eprint
  {http://arxiv.org/abs/2008.01098} {arXiv:2008.01098} \BibitemShut {NoStop}%
\bibitem [{\citenamefont {Yung}\ \emph {et~al.}(2015)\citenamefont {Yung},
  \citenamefont {Casanova}, \citenamefont {Mezzacapo}, \citenamefont {McClean},
  \citenamefont {Lamata}, \citenamefont {Aspuru-Guzik},\ and\ \citenamefont
  {Solano}}]{Yung2015}%
  \BibitemOpen
  \bibfield  {author} {\bibinfo {author} {\bibfnamefont {M.-H.}\ \bibnamefont
  {Yung}}, \bibinfo {author} {\bibfnamefont {J.}~\bibnamefont {Casanova}},
  \bibinfo {author} {\bibfnamefont {A.}~\bibnamefont {Mezzacapo}}, \bibinfo
  {author} {\bibfnamefont {J.}~\bibnamefont {McClean}}, \bibinfo {author}
  {\bibfnamefont {L.}~\bibnamefont {Lamata}}, \bibinfo {author} {\bibfnamefont
  {A.}~\bibnamefont {Aspuru-Guzik}}, \ and\ \bibinfo {author} {\bibfnamefont
  {E.}~\bibnamefont {Solano}},\ }\href {\doibase 10.1038/srep03589} {\bibfield
  {journal} {\bibinfo  {journal} {Sci. Rep}\ }\textbf {\bibinfo {volume} {4}},\
  \bibinfo {pages} {3589} (\bibinfo {year} {2015})}\BibitemShut {NoStop}%
\bibitem [{\citenamefont {Mostafa}\ and\ \citenamefont
  {Wang}(2019)}]{Mostafa2019}%
  \BibitemOpen
  \bibfield  {author} {\bibinfo {author} {\bibfnamefont {H.}~\bibnamefont
  {Mostafa}}\ and\ \bibinfo {author} {\bibfnamefont {X.}~\bibnamefont {Wang}},\
  }\href {http://arxiv.org/abs/1902.05967} {\enquote {\bibinfo {title}
  {{Parameter Efficient Training of Deep Convolutional Neural Networks by
  Dynamic Sparse Reparameterization}},}\ } (\bibinfo {year} {2019}),\ \Eprint
  {http://arxiv.org/abs/1902.05967} {arXiv:1902.05967} \BibitemShut {NoStop}%
\bibitem [{\citenamefont {Dallaire-Demers}\ \emph {et~al.}(2020)\citenamefont
  {Dallaire-Demers}, \citenamefont {St{\c{e}}ch{\l}y}, \citenamefont
  {Gonthier}, \citenamefont {Bashige}, \citenamefont {Romero},\ and\
  \citenamefont {Cao}}]{Dallaire-Demers2020}%
  \BibitemOpen
  \bibfield  {author} {\bibinfo {author} {\bibfnamefont {P.-L.}\ \bibnamefont
  {Dallaire-Demers}}, \bibinfo {author} {\bibfnamefont {M.}~\bibnamefont
  {St{\c{e}}ch{\l}y}}, \bibinfo {author} {\bibfnamefont {J.~F.}\ \bibnamefont
  {Gonthier}}, \bibinfo {author} {\bibfnamefont {N.~T.}\ \bibnamefont
  {Bashige}}, \bibinfo {author} {\bibfnamefont {J.}~\bibnamefont {Romero}}, \
  and\ \bibinfo {author} {\bibfnamefont {Y.}~\bibnamefont {Cao}},\ }\href
  {http://arxiv.org/abs/2003.01862} {\enquote {\bibinfo {title} {{An
  application benchmark for fermionic quantum simulations}},}\ } (\bibinfo
  {year} {2020}),\ \Eprint {http://arxiv.org/abs/2003.01862} {arXiv:2003.01862}
  \BibitemShut {NoStop}%
\bibitem [{\citenamefont {Sim}\ \emph {et~al.}(2019)\citenamefont {Sim},
  \citenamefont {Johnson},\ and\ \citenamefont {Aspuru-Guzik}}]{Sim2019}%
  \BibitemOpen
  \bibfield  {author} {\bibinfo {author} {\bibfnamefont {S.}~\bibnamefont
  {Sim}}, \bibinfo {author} {\bibfnamefont {P.~D.}\ \bibnamefont {Johnson}}, \
  and\ \bibinfo {author} {\bibfnamefont {A.}~\bibnamefont {Aspuru-Guzik}},\
  }\href {\doibase 10.1002/qute.201900070} {\bibfield  {journal} {\bibinfo
  {journal} {Adv. Quantum Technol.}\ }\textbf {\bibinfo {volume} {2}},\
  \bibinfo {pages} {1900070} (\bibinfo {year} {2019})}\BibitemShut {NoStop}%
\bibitem [{\citenamefont {Rasmussen}\ \emph {et~al.}(2020)\citenamefont
  {Rasmussen}, \citenamefont {Loft}, \citenamefont {B{\ae}kkegaard},
  \citenamefont {Kues},\ and\ \citenamefont {Zinner}}]{Rasmussen2020}%
  \BibitemOpen
  \bibfield  {author} {\bibinfo {author} {\bibfnamefont {S.~E.}\ \bibnamefont
  {Rasmussen}}, \bibinfo {author} {\bibfnamefont {N.~J.~S.}\ \bibnamefont
  {Loft}}, \bibinfo {author} {\bibfnamefont {T.}~\bibnamefont
  {B{\ae}kkegaard}}, \bibinfo {author} {\bibfnamefont {M.}~\bibnamefont
  {Kues}}, \ and\ \bibinfo {author} {\bibfnamefont {N.~T.}\ \bibnamefont
  {Zinner}},\ }\href {http://arxiv.org/abs/2005.13548} {\enquote {\bibinfo
  {title} {{Single-qubit rotations in parameterized quantum circuits}},}\ }
  (\bibinfo {year} {2020}),\ \Eprint {http://arxiv.org/abs/2005.13548}
  {arXiv:2005.13548} \BibitemShut {NoStop}%
\bibitem [{\citenamefont {Romero}\ \emph {et~al.}(2018)\citenamefont {Romero},
  \citenamefont {Babbush}, \citenamefont {McClean}, \citenamefont {Hempel},
  \citenamefont {Love},\ and\ \citenamefont
  {Aspuru-Guzik}}]{Romero2017strategies}%
  \BibitemOpen
  \bibfield  {author} {\bibinfo {author} {\bibfnamefont {J.}~\bibnamefont
  {Romero}}, \bibinfo {author} {\bibfnamefont {R.}~\bibnamefont {Babbush}},
  \bibinfo {author} {\bibfnamefont {J.~R.}\ \bibnamefont {McClean}}, \bibinfo
  {author} {\bibfnamefont {C.}~\bibnamefont {Hempel}}, \bibinfo {author}
  {\bibfnamefont {P.~J.}\ \bibnamefont {Love}}, \ and\ \bibinfo {author}
  {\bibfnamefont {A.}~\bibnamefont {Aspuru-Guzik}},\ }\href {\doibase
  10.1088/2058-9565/aad3e4} {\bibfield  {journal} {\bibinfo  {journal} {Quantum
  Sci. Technol.}\ }\textbf {\bibinfo {volume} {4}},\ \bibinfo {pages} {014008}
  (\bibinfo {year} {2018})}\BibitemShut {NoStop}%
\bibitem [{\citenamefont {Byrd}\ \emph {et~al.}(1995)\citenamefont {Byrd},
  \citenamefont {Lu}, \citenamefont {Nocedal},\ and\ \citenamefont
  {Zhu}}]{Byrd1995}%
  \BibitemOpen
  \bibfield  {author} {\bibinfo {author} {\bibfnamefont {R.~H.}\ \bibnamefont
  {Byrd}}, \bibinfo {author} {\bibfnamefont {P.}~\bibnamefont {Lu}}, \bibinfo
  {author} {\bibfnamefont {J.}~\bibnamefont {Nocedal}}, \ and\ \bibinfo
  {author} {\bibfnamefont {C.}~\bibnamefont {Zhu}},\ }\href {\doibase
  10.1137/0916069} {\bibfield  {journal} {\bibinfo  {journal} {SIAM J. Sci.
  Comput}\ }\textbf {\bibinfo {volume} {16}},\ \bibinfo {pages} {1190}
  (\bibinfo {year} {1995})}\BibitemShut {NoStop}%
\bibitem [{\citenamefont {Kraft}(1988)}]{Kraft1988}%
  \BibitemOpen
  \bibfield  {author} {\bibinfo {author} {\bibfnamefont {D.}~\bibnamefont
  {Kraft}},\ }\href@noop {} {\bibfield  {journal} {\bibinfo  {journal}
  {DFVLR-FB}\ }\textbf {\bibinfo {volume} {88}},\ \bibinfo {pages} {33}
  (\bibinfo {year} {1988})}\BibitemShut {NoStop}%
\bibitem [{\citenamefont {Li}\ \emph {et~al.}(2020)\citenamefont {Li},
  \citenamefont {Fan}, \citenamefont {Coram}, \citenamefont {Riley},\ and\
  \citenamefont {Leichenauer}}]{Li2020}%
  \BibitemOpen
  \bibfield  {author} {\bibinfo {author} {\bibfnamefont {L.}~\bibnamefont
  {Li}}, \bibinfo {author} {\bibfnamefont {M.}~\bibnamefont {Fan}}, \bibinfo
  {author} {\bibfnamefont {M.}~\bibnamefont {Coram}}, \bibinfo {author}
  {\bibfnamefont {P.}~\bibnamefont {Riley}}, \ and\ \bibinfo {author}
  {\bibfnamefont {S.}~\bibnamefont {Leichenauer}},\ }\href {\doibase
  10.1103/PhysRevResearch.2.023074} {\bibfield  {journal} {\bibinfo  {journal}
  {Phys. Rev. Res.}\ }\textbf {\bibinfo {volume} {2}},\ \bibinfo {pages}
  {023074} (\bibinfo {year} {2020})}\BibitemShut {NoStop}%
\bibitem [{\citenamefont {Motta}\ \emph {et~al.}(2018)\citenamefont {Motta},
  \citenamefont {Ye}, \citenamefont {McClean}, \citenamefont {Li},
  \citenamefont {Minnich}, \citenamefont {Babbush},\ and\ \citenamefont
  {Chan}}]{Motta2018}%
  \BibitemOpen
  \bibfield  {author} {\bibinfo {author} {\bibfnamefont {M.}~\bibnamefont
  {Motta}}, \bibinfo {author} {\bibfnamefont {E.}~\bibnamefont {Ye}}, \bibinfo
  {author} {\bibfnamefont {J.~R.}\ \bibnamefont {McClean}}, \bibinfo {author}
  {\bibfnamefont {Z.}~\bibnamefont {Li}}, \bibinfo {author} {\bibfnamefont
  {A.~J.}\ \bibnamefont {Minnich}}, \bibinfo {author} {\bibfnamefont
  {R.}~\bibnamefont {Babbush}}, \ and\ \bibinfo {author} {\bibfnamefont
  {G.~K.-L.}\ \bibnamefont {Chan}},\ }\href {http://arxiv.org/abs/1808.02625}
  {\enquote {\bibinfo {title} {{Low rank representations for quantum simulation
  of electronic structure}},}\ } (\bibinfo {year} {2018}),\ \Eprint
  {http://arxiv.org/abs/1808.02625} {arXiv:1808.02625} \BibitemShut {NoStop}%
\bibitem [{\citenamefont {Cooper}\ and\ \citenamefont
  {Knowles}(2010)}]{Cooper2010}%
  \BibitemOpen
  \bibfield  {author} {\bibinfo {author} {\bibfnamefont {B.}~\bibnamefont
  {Cooper}}\ and\ \bibinfo {author} {\bibfnamefont {P.~J.}\ \bibnamefont
  {Knowles}},\ }\href {\doibase 10.1063/1.3520564} {\bibfield  {journal}
  {\bibinfo  {journal} {J. Chem. Phys}\ }\textbf {\bibinfo {volume} {133}}
  (\bibinfo {year} {2010}),\ 10.1063/1.3520564}\BibitemShut {NoStop}%
\bibitem [{\citenamefont {Ma}\ \emph {et~al.}(2006)\citenamefont {Ma},
  \citenamefont {Li},\ and\ \citenamefont {Li}}]{Ma2006}%
  \BibitemOpen
  \bibfield  {author} {\bibinfo {author} {\bibfnamefont {J.}~\bibnamefont
  {Ma}}, \bibinfo {author} {\bibfnamefont {S.}~\bibnamefont {Li}}, \ and\
  \bibinfo {author} {\bibfnamefont {W.}~\bibnamefont {Li}},\ }\href {\doibase
  10.1002/jcc.20319} {\bibfield  {journal} {\bibinfo  {journal} {J. Comput.
  Chem}\ }\textbf {\bibinfo {volume} {27}},\ \bibinfo {pages} {39} (\bibinfo
  {year} {2006})}\BibitemShut {NoStop}%
\bibitem [{\citenamefont {Li}\ and\ \citenamefont {Paldus}(1998)}]{Li1998}%
  \BibitemOpen
  \bibfield  {author} {\bibinfo {author} {\bibfnamefont {X.}~\bibnamefont
  {Li}}\ and\ \bibinfo {author} {\bibfnamefont {J.}~\bibnamefont {Paldus}},\
  }\href {\doibase 10.1063/1.475425} {\bibfield  {journal} {\bibinfo  {journal}
  {J. Chem. Phys}\ }\textbf {\bibinfo {volume} {108}},\ \bibinfo {pages} {637}
  (\bibinfo {year} {1998})}\BibitemShut {NoStop}%
\bibitem [{\citenamefont {Olsen}\ \emph {et~al.}(1996)\citenamefont {Olsen},
  \citenamefont {J{\o}rgensen}, \citenamefont {Koch}, \citenamefont {Balkova},\
  and\ \citenamefont {Bartlett}}]{Olsen1996}%
  \BibitemOpen
  \bibfield  {author} {\bibinfo {author} {\bibfnamefont {J.}~\bibnamefont
  {Olsen}}, \bibinfo {author} {\bibfnamefont {P.}~\bibnamefont {J{\o}rgensen}},
  \bibinfo {author} {\bibfnamefont {H.}~\bibnamefont {Koch}}, \bibinfo {author}
  {\bibfnamefont {A.}~\bibnamefont {Balkova}}, \ and\ \bibinfo {author}
  {\bibfnamefont {R.~J.}\ \bibnamefont {Bartlett}},\ }\href {\doibase
  10.1063/1.471518} {\bibfield  {journal} {\bibinfo  {journal} {J. Chem. Phys}\
  }\textbf {\bibinfo {volume} {104}},\ \bibinfo {pages} {8007} (\bibinfo {year}
  {1996})}\BibitemShut {NoStop}%
\bibitem [{\citenamefont {Brown}\ \emph {et~al.}(1984)\citenamefont {Brown},
  \citenamefont {Shavitt},\ and\ \citenamefont {Shepard}}]{Brown1984}%
  \BibitemOpen
  \bibfield  {author} {\bibinfo {author} {\bibfnamefont {F.~B.}\ \bibnamefont
  {Brown}}, \bibinfo {author} {\bibfnamefont {I.}~\bibnamefont {Shavitt}}, \
  and\ \bibinfo {author} {\bibfnamefont {R.}~\bibnamefont {Shepard}},\ }\href
  {\doibase 10.1016/0009-2614(84)80042-1} {\bibfield  {journal} {\bibinfo
  {journal} {Chem. Phys. Lett}\ }\textbf {\bibinfo {volume} {105}},\ \bibinfo
  {pages} {363} (\bibinfo {year} {1984})}\BibitemShut {NoStop}%
\bibitem [{\citenamefont {Sokolov}\ \emph {et~al.}(2020)\citenamefont
  {Sokolov}, \citenamefont {Barkoutsos}, \citenamefont {Ollitrault},
  \citenamefont {Greenberg}, \citenamefont {Rice}, \citenamefont {Pistoia},\
  and\ \citenamefont {Tavernelli}}]{Sokolov2020}%
  \BibitemOpen
  \bibfield  {author} {\bibinfo {author} {\bibfnamefont {I.~O.}\ \bibnamefont
  {Sokolov}}, \bibinfo {author} {\bibfnamefont {P.~K.}\ \bibnamefont
  {Barkoutsos}}, \bibinfo {author} {\bibfnamefont {P.~J.}\ \bibnamefont
  {Ollitrault}}, \bibinfo {author} {\bibfnamefont {D.}~\bibnamefont
  {Greenberg}}, \bibinfo {author} {\bibfnamefont {J.}~\bibnamefont {Rice}},
  \bibinfo {author} {\bibfnamefont {M.}~\bibnamefont {Pistoia}}, \ and\
  \bibinfo {author} {\bibfnamefont {I.}~\bibnamefont {Tavernelli}},\ }\href
  {\doibase 10.1063/1.5141835} {\bibfield  {journal} {\bibinfo  {journal} {J.
  Chem. Phys}\ }\textbf {\bibinfo {volume} {152}} (\bibinfo {year} {2020}),\
  10.1063/1.5141835}\BibitemShut {NoStop}%
\bibitem [{\citenamefont {Grimsley}\ \emph {et~al.}(2020)\citenamefont
  {Grimsley}, \citenamefont {Claudino}, \citenamefont {Economou}, \citenamefont
  {Barnes},\ and\ \citenamefont {Mayhall}}]{Grimsley2020}%
  \BibitemOpen
  \bibfield  {author} {\bibinfo {author} {\bibfnamefont {H.~R.}\ \bibnamefont
  {Grimsley}}, \bibinfo {author} {\bibfnamefont {D.}~\bibnamefont {Claudino}},
  \bibinfo {author} {\bibfnamefont {S.~E.}\ \bibnamefont {Economou}}, \bibinfo
  {author} {\bibfnamefont {E.}~\bibnamefont {Barnes}}, \ and\ \bibinfo {author}
  {\bibfnamefont {N.~J.}\ \bibnamefont {Mayhall}},\ }\href {\doibase
  10.1021/acs.jctc.9b01083} {\bibfield  {journal} {\bibinfo  {journal} {J.
  Chem. Theory Comput}\ }\textbf {\bibinfo {volume} {16}},\ \bibinfo {pages}
  {1} (\bibinfo {year} {2020})}\BibitemShut {NoStop}%
\bibitem [{\citenamefont {Izmaylov}\ \emph {et~al.}(2020)\citenamefont
  {Izmaylov}, \citenamefont {D{\'{i}}az-Tinoco},\ and\ \citenamefont
  {Lang}}]{Izmaylov2020}%
  \BibitemOpen
  \bibfield  {author} {\bibinfo {author} {\bibfnamefont {A.~F.}\ \bibnamefont
  {Izmaylov}}, \bibinfo {author} {\bibfnamefont {M.}~\bibnamefont
  {D{\'{i}}az-Tinoco}}, \ and\ \bibinfo {author} {\bibfnamefont {R.~A.}\
  \bibnamefont {Lang}},\ }\href {\doibase 10.1039/D0CP01707H} {\bibfield
  {journal} {\bibinfo  {journal} {Phys. Chem. Chem. Phys}\ }\textbf {\bibinfo
  {volume} {22}},\ \bibinfo {pages} {12980} (\bibinfo {year} {2020})},\ \Eprint
  {http://arxiv.org/abs/2003.07351} {2003.07351} \BibitemShut {NoStop}%
\bibitem [{\citenamefont {Bengio}\ \emph {et~al.}(2007)\citenamefont {Bengio},
  \citenamefont {Lamblin}, \citenamefont {Popovici},\ and\ \citenamefont
  {Larochelle}}]{Bengio2007}%
  \BibitemOpen
  \bibfield  {author} {\bibinfo {author} {\bibfnamefont {Y.}~\bibnamefont
  {Bengio}}, \bibinfo {author} {\bibfnamefont {P.}~\bibnamefont {Lamblin}},
  \bibinfo {author} {\bibfnamefont {D.}~\bibnamefont {Popovici}}, \ and\
  \bibinfo {author} {\bibfnamefont {H.}~\bibnamefont {Larochelle}},\
  }\href@noop {} {\bibfield  {journal} {\bibinfo  {journal} {Adv Neural Inf
  Process Syst}\ ,\ \bibinfo {pages} {153}} (\bibinfo {year}
  {2007})}\BibitemShut {NoStop}%
\bibitem [{\citenamefont {Kandala}\ \emph {et~al.}(2017)\citenamefont
  {Kandala}, \citenamefont {Mezzacapo}, \citenamefont {Temme}, \citenamefont
  {Takita}, \citenamefont {Brink}, \citenamefont {Chow},\ and\ \citenamefont
  {Gambetta}}]{Kandala2017}%
  \BibitemOpen
  \bibfield  {author} {\bibinfo {author} {\bibfnamefont {A.}~\bibnamefont
  {Kandala}}, \bibinfo {author} {\bibfnamefont {A.}~\bibnamefont {Mezzacapo}},
  \bibinfo {author} {\bibfnamefont {K.}~\bibnamefont {Temme}}, \bibinfo
  {author} {\bibfnamefont {M.}~\bibnamefont {Takita}}, \bibinfo {author}
  {\bibfnamefont {M.}~\bibnamefont {Brink}}, \bibinfo {author} {\bibfnamefont
  {J.~M.}\ \bibnamefont {Chow}}, \ and\ \bibinfo {author} {\bibfnamefont
  {J.~M.}\ \bibnamefont {Gambetta}},\ }\href {\doibase 10.1038/nature23879}
  {\bibfield  {journal} {\bibinfo  {journal} {Nature}\ }\textbf {\bibinfo
  {volume} {549}},\ \bibinfo {pages} {242} (\bibinfo {year}
  {2017})}\BibitemShut {NoStop}%
\bibitem [{\citenamefont {Arute}\ \emph {et~al.}(2019)\citenamefont {Arute},
  \citenamefont {Arya}, \citenamefont {Babbush}, \citenamefont {Bacon},
  \citenamefont {Bardin}, \citenamefont {Barends}, \citenamefont {Biswas},
  \citenamefont {Boixo}, \citenamefont {Brandao}, \citenamefont {Buell},
  \citenamefont {Burkett}, \citenamefont {Chen}, \citenamefont {Chen},
  \citenamefont {Chiaro}, \citenamefont {Collins}, \citenamefont {Courtney},
  \citenamefont {Dunsworth}, \citenamefont {Farhi}, \citenamefont {Foxen},
  \citenamefont {Fowler}, \citenamefont {Gidney}, \citenamefont {Giustina},
  \citenamefont {Graff}, \citenamefont {Guerin}, \citenamefont {Habegger},
  \citenamefont {Harrigan}, \citenamefont {Hartmann}, \citenamefont {Ho},
  \citenamefont {Hoffmann}, \citenamefont {Huang}, \citenamefont {Humble},
  \citenamefont {Isakov}, \citenamefont {Jeffrey}, \citenamefont {Jiang},
  \citenamefont {Kafri}, \citenamefont {Kechedzhi}, \citenamefont {Kelly},
  \citenamefont {Klimov}, \citenamefont {Knysh}, \citenamefont {Korotkov},
  \citenamefont {Kostritsa}, \citenamefont {Landhuis}, \citenamefont
  {Lindmark}, \citenamefont {Lucero}, \citenamefont {Lyakh}, \citenamefont
  {Mandr{\`{a}}}, \citenamefont {McClean}, \citenamefont {McEwen},
  \citenamefont {Megrant}, \citenamefont {Mi}, \citenamefont {Michielsen},
  \citenamefont {Mohseni}, \citenamefont {Mutus}, \citenamefont {Naaman},
  \citenamefont {Neeley}, \citenamefont {Neill}, \citenamefont {Niu},
  \citenamefont {Ostby}, \citenamefont {Petukhov}, \citenamefont {Platt},
  \citenamefont {Quintana}, \citenamefont {Rieffel}, \citenamefont {Roushan},
  \citenamefont {Rubin}, \citenamefont {Sank}, \citenamefont {Satzinger},
  \citenamefont {Smelyanskiy}, \citenamefont {Sung}, \citenamefont
  {Trevithick}, \citenamefont {Vainsencher}, \citenamefont {Villalonga},
  \citenamefont {White}, \citenamefont {Yao}, \citenamefont {Yeh},
  \citenamefont {Zalcman}, \citenamefont {Neven},\ and\ \citenamefont
  {Martinis}}]{Arute2019}%
  \BibitemOpen
  \bibfield  {author} {\bibinfo {author} {\bibfnamefont {F.}~\bibnamefont
  {Arute}}, \bibinfo {author} {\bibfnamefont {K.}~\bibnamefont {Arya}},
  \bibinfo {author} {\bibfnamefont {R.}~\bibnamefont {Babbush}}, \bibinfo
  {author} {\bibfnamefont {D.}~\bibnamefont {Bacon}}, \bibinfo {author}
  {\bibfnamefont {J.~C.}\ \bibnamefont {Bardin}}, \bibinfo {author}
  {\bibfnamefont {R.}~\bibnamefont {Barends}}, \bibinfo {author} {\bibfnamefont
  {R.}~\bibnamefont {Biswas}}, \bibinfo {author} {\bibfnamefont
  {S.}~\bibnamefont {Boixo}}, \bibinfo {author} {\bibfnamefont {F.~G. S.~L.}\
  \bibnamefont {Brandao}}, \bibinfo {author} {\bibfnamefont {D.~A.}\
  \bibnamefont {Buell}}, \bibinfo {author} {\bibfnamefont {B.}~\bibnamefont
  {Burkett}}, \bibinfo {author} {\bibfnamefont {Y.}~\bibnamefont {Chen}},
  \bibinfo {author} {\bibfnamefont {Z.}~\bibnamefont {Chen}}, \bibinfo {author}
  {\bibfnamefont {B.}~\bibnamefont {Chiaro}}, \bibinfo {author} {\bibfnamefont
  {R.}~\bibnamefont {Collins}}, \bibinfo {author} {\bibfnamefont
  {W.}~\bibnamefont {Courtney}}, \bibinfo {author} {\bibfnamefont
  {A.}~\bibnamefont {Dunsworth}}, \bibinfo {author} {\bibfnamefont
  {E.}~\bibnamefont {Farhi}}, \bibinfo {author} {\bibfnamefont
  {B.}~\bibnamefont {Foxen}}, \bibinfo {author} {\bibfnamefont
  {A.}~\bibnamefont {Fowler}}, \bibinfo {author} {\bibfnamefont
  {C.}~\bibnamefont {Gidney}}, \bibinfo {author} {\bibfnamefont
  {M.}~\bibnamefont {Giustina}}, \bibinfo {author} {\bibfnamefont
  {R.}~\bibnamefont {Graff}}, \bibinfo {author} {\bibfnamefont
  {K.}~\bibnamefont {Guerin}}, \bibinfo {author} {\bibfnamefont
  {S.}~\bibnamefont {Habegger}}, \bibinfo {author} {\bibfnamefont {M.~P.}\
  \bibnamefont {Harrigan}}, \bibinfo {author} {\bibfnamefont {M.~J.}\
  \bibnamefont {Hartmann}}, \bibinfo {author} {\bibfnamefont {A.}~\bibnamefont
  {Ho}}, \bibinfo {author} {\bibfnamefont {M.}~\bibnamefont {Hoffmann}},
  \bibinfo {author} {\bibfnamefont {T.}~\bibnamefont {Huang}}, \bibinfo
  {author} {\bibfnamefont {T.~S.}\ \bibnamefont {Humble}}, \bibinfo {author}
  {\bibfnamefont {S.~V.}\ \bibnamefont {Isakov}}, \bibinfo {author}
  {\bibfnamefont {E.}~\bibnamefont {Jeffrey}}, \bibinfo {author} {\bibfnamefont
  {Z.}~\bibnamefont {Jiang}}, \bibinfo {author} {\bibfnamefont
  {D.}~\bibnamefont {Kafri}}, \bibinfo {author} {\bibfnamefont
  {K.}~\bibnamefont {Kechedzhi}}, \bibinfo {author} {\bibfnamefont
  {J.}~\bibnamefont {Kelly}}, \bibinfo {author} {\bibfnamefont {P.~V.}\
  \bibnamefont {Klimov}}, \bibinfo {author} {\bibfnamefont {S.}~\bibnamefont
  {Knysh}}, \bibinfo {author} {\bibfnamefont {A.}~\bibnamefont {Korotkov}},
  \bibinfo {author} {\bibfnamefont {F.}~\bibnamefont {Kostritsa}}, \bibinfo
  {author} {\bibfnamefont {D.}~\bibnamefont {Landhuis}}, \bibinfo {author}
  {\bibfnamefont {M.}~\bibnamefont {Lindmark}}, \bibinfo {author}
  {\bibfnamefont {E.}~\bibnamefont {Lucero}}, \bibinfo {author} {\bibfnamefont
  {D.}~\bibnamefont {Lyakh}}, \bibinfo {author} {\bibfnamefont
  {S.}~\bibnamefont {Mandr{\`{a}}}}, \bibinfo {author} {\bibfnamefont {J.~R.}\
  \bibnamefont {McClean}}, \bibinfo {author} {\bibfnamefont {M.}~\bibnamefont
  {McEwen}}, \bibinfo {author} {\bibfnamefont {A.}~\bibnamefont {Megrant}},
  \bibinfo {author} {\bibfnamefont {X.}~\bibnamefont {Mi}}, \bibinfo {author}
  {\bibfnamefont {K.}~\bibnamefont {Michielsen}}, \bibinfo {author}
  {\bibfnamefont {M.}~\bibnamefont {Mohseni}}, \bibinfo {author} {\bibfnamefont
  {J.}~\bibnamefont {Mutus}}, \bibinfo {author} {\bibfnamefont
  {O.}~\bibnamefont {Naaman}}, \bibinfo {author} {\bibfnamefont
  {M.}~\bibnamefont {Neeley}}, \bibinfo {author} {\bibfnamefont
  {C.}~\bibnamefont {Neill}}, \bibinfo {author} {\bibfnamefont {M.~Y.}\
  \bibnamefont {Niu}}, \bibinfo {author} {\bibfnamefont {E.}~\bibnamefont
  {Ostby}}, \bibinfo {author} {\bibfnamefont {A.}~\bibnamefont {Petukhov}},
  \bibinfo {author} {\bibfnamefont {J.~C.}\ \bibnamefont {Platt}}, \bibinfo
  {author} {\bibfnamefont {C.}~\bibnamefont {Quintana}}, \bibinfo {author}
  {\bibfnamefont {E.~G.}\ \bibnamefont {Rieffel}}, \bibinfo {author}
  {\bibfnamefont {P.}~\bibnamefont {Roushan}}, \bibinfo {author} {\bibfnamefont
  {N.~C.}\ \bibnamefont {Rubin}}, \bibinfo {author} {\bibfnamefont
  {D.}~\bibnamefont {Sank}}, \bibinfo {author} {\bibfnamefont {K.~J.}\
  \bibnamefont {Satzinger}}, \bibinfo {author} {\bibfnamefont {V.}~\bibnamefont
  {Smelyanskiy}}, \bibinfo {author} {\bibfnamefont {K.~J.}\ \bibnamefont
  {Sung}}, \bibinfo {author} {\bibfnamefont {M.~D.}\ \bibnamefont
  {Trevithick}}, \bibinfo {author} {\bibfnamefont {A.}~\bibnamefont
  {Vainsencher}}, \bibinfo {author} {\bibfnamefont {B.}~\bibnamefont
  {Villalonga}}, \bibinfo {author} {\bibfnamefont {T.}~\bibnamefont {White}},
  \bibinfo {author} {\bibfnamefont {Z.~J.}\ \bibnamefont {Yao}}, \bibinfo
  {author} {\bibfnamefont {P.}~\bibnamefont {Yeh}}, \bibinfo {author}
  {\bibfnamefont {A.}~\bibnamefont {Zalcman}}, \bibinfo {author} {\bibfnamefont
  {H.}~\bibnamefont {Neven}}, \ and\ \bibinfo {author} {\bibfnamefont {J.~M.}\
  \bibnamefont {Martinis}},\ }\href {\doibase 10.1038/s41586-019-1666-5}
  {\bibfield  {journal} {\bibinfo  {journal} {Nature}\ }\textbf {\bibinfo
  {volume} {574}},\ \bibinfo {pages} {505} (\bibinfo {year}
  {2019})}\BibitemShut {NoStop}%
\bibitem [{\citenamefont {Kennedy}\ and\ \citenamefont
  {Eberhart}(1995)}]{Kennedy1995}%
  \BibitemOpen
  \bibfield  {author} {\bibinfo {author} {\bibfnamefont {J.}~\bibnamefont
  {Kennedy}}\ and\ \bibinfo {author} {\bibfnamefont {R.}~\bibnamefont
  {Eberhart}},\ }in\ \href {\doibase 10.1109/ICNN.1995.488968} {\emph {\bibinfo
  {booktitle} {Proceedings of ICNN'95 - International Conference on Neural
  Networks}}},\ Vol.~\bibinfo {volume} {4}\ (\bibinfo  {publisher} {IEEE},\
  \bibinfo {year} {1995})\ pp.\ \bibinfo {pages} {1942--1948}\BibitemShut
  {NoStop}%
\bibitem [{\citenamefont {Le}(1990)}]{Le1990}%
  \BibitemOpen
  \bibfield  {author} {\bibinfo {author} {\bibfnamefont {C.}~\bibnamefont
  {Le}},\ }\href {http://yann.lecun.com/exdb/publis/pdf/lecun-90b.pdf}
  {\bibfield  {journal} {\bibinfo  {journal} {Adv Neural Inf Process Syst}\
  }\textbf {\bibinfo {volume} {2}},\ \bibinfo {pages} {598} (\bibinfo {year}
  {1990})}\BibitemShut {NoStop}%
\bibitem [{orq(2018)}]{orquestra}%
  \BibitemOpen
  \href@noop {} {\enquote {\bibinfo {title} {{Orquestra}},}\ }\bibinfo
  {howpublished} {\url{https://www.orquestra.io/}} (\bibinfo {year}
  {2018})\BibitemShut {NoStop}%
\bibitem [{\citenamefont {McClean}\ \emph {et~al.}(2020)\citenamefont
  {McClean}, \citenamefont {Rubin}, \citenamefont {Sung}, \citenamefont
  {Kivlichan}, \citenamefont {Bonet-Monroig}, \citenamefont {Cao},
  \citenamefont {Dai}, \citenamefont {Fried}, \citenamefont {Gidney},
  \citenamefont {Gimby}, \citenamefont {Gokhale}, \citenamefont {H{\"{a}}ner},
  \citenamefont {Hardikar}, \citenamefont {Havl{\'{i}}{\v{c}}ek}, \citenamefont
  {Higgott}, \citenamefont {Huang}, \citenamefont {Izaac}, \citenamefont
  {Jiang}, \citenamefont {Liu}, \citenamefont {McArdle}, \citenamefont
  {Neeley}, \citenamefont {O'Brien}, \citenamefont {O'Gorman}, \citenamefont
  {Ozfidan}, \citenamefont {Radin}, \citenamefont {Romero}, \citenamefont
  {Sawaya}, \citenamefont {Senjean}, \citenamefont {Setia}, \citenamefont
  {Sim}, \citenamefont {Steiger}, \citenamefont {Steudtner}, \citenamefont
  {Sun}, \citenamefont {Sun}, \citenamefont {Wang}, \citenamefont {Zhang},\
  and\ \citenamefont {Babbush}}]{openfermion}%
  \BibitemOpen
  \bibfield  {author} {\bibinfo {author} {\bibfnamefont {J.~R.}\ \bibnamefont
  {McClean}}, \bibinfo {author} {\bibfnamefont {N.~C.}\ \bibnamefont {Rubin}},
  \bibinfo {author} {\bibfnamefont {K.~J.}\ \bibnamefont {Sung}}, \bibinfo
  {author} {\bibfnamefont {I.~D.}\ \bibnamefont {Kivlichan}}, \bibinfo {author}
  {\bibfnamefont {X.}~\bibnamefont {Bonet-Monroig}}, \bibinfo {author}
  {\bibfnamefont {Y.}~\bibnamefont {Cao}}, \bibinfo {author} {\bibfnamefont
  {C.}~\bibnamefont {Dai}}, \bibinfo {author} {\bibfnamefont {E.~S.}\
  \bibnamefont {Fried}}, \bibinfo {author} {\bibfnamefont {C.}~\bibnamefont
  {Gidney}}, \bibinfo {author} {\bibfnamefont {B.}~\bibnamefont {Gimby}},
  \bibinfo {author} {\bibfnamefont {P.}~\bibnamefont {Gokhale}}, \bibinfo
  {author} {\bibfnamefont {T.}~\bibnamefont {H{\"{a}}ner}}, \bibinfo {author}
  {\bibfnamefont {T.}~\bibnamefont {Hardikar}}, \bibinfo {author}
  {\bibfnamefont {V.}~\bibnamefont {Havl{\'{i}}{\v{c}}ek}}, \bibinfo {author}
  {\bibfnamefont {O.}~\bibnamefont {Higgott}}, \bibinfo {author} {\bibfnamefont
  {C.}~\bibnamefont {Huang}}, \bibinfo {author} {\bibfnamefont
  {J.}~\bibnamefont {Izaac}}, \bibinfo {author} {\bibfnamefont
  {Z.}~\bibnamefont {Jiang}}, \bibinfo {author} {\bibfnamefont
  {X.}~\bibnamefont {Liu}}, \bibinfo {author} {\bibfnamefont {S.}~\bibnamefont
  {McArdle}}, \bibinfo {author} {\bibfnamefont {M.}~\bibnamefont {Neeley}},
  \bibinfo {author} {\bibfnamefont {T.}~\bibnamefont {O'Brien}}, \bibinfo
  {author} {\bibfnamefont {B.}~\bibnamefont {O'Gorman}}, \bibinfo {author}
  {\bibfnamefont {I.}~\bibnamefont {Ozfidan}}, \bibinfo {author} {\bibfnamefont
  {M.~D.}\ \bibnamefont {Radin}}, \bibinfo {author} {\bibfnamefont
  {J.}~\bibnamefont {Romero}}, \bibinfo {author} {\bibfnamefont {N.~P.~D.}\
  \bibnamefont {Sawaya}}, \bibinfo {author} {\bibfnamefont {B.}~\bibnamefont
  {Senjean}}, \bibinfo {author} {\bibfnamefont {K.}~\bibnamefont {Setia}},
  \bibinfo {author} {\bibfnamefont {S.}~\bibnamefont {Sim}}, \bibinfo {author}
  {\bibfnamefont {D.~S.}\ \bibnamefont {Steiger}}, \bibinfo {author}
  {\bibfnamefont {M.}~\bibnamefont {Steudtner}}, \bibinfo {author}
  {\bibfnamefont {Q.}~\bibnamefont {Sun}}, \bibinfo {author} {\bibfnamefont
  {W.}~\bibnamefont {Sun}}, \bibinfo {author} {\bibfnamefont {D.}~\bibnamefont
  {Wang}}, \bibinfo {author} {\bibfnamefont {F.}~\bibnamefont {Zhang}}, \ and\
  \bibinfo {author} {\bibfnamefont {R.}~\bibnamefont {Babbush}},\ }\href
  {\doibase 10.1088/2058-9565/ab8ebc} {\bibfield  {journal} {\bibinfo
  {journal} {Quantum Sci. Technol.}\ }\textbf {\bibinfo {volume} {5}},\
  \bibinfo {pages} {034014} (\bibinfo {year} {2020})}\BibitemShut {NoStop}%
\bibitem [{\citenamefont {Parrish}\ \emph {et~al.}(2017)\citenamefont
  {Parrish}, \citenamefont {Burns}, \citenamefont {Smith}, \citenamefont
  {Simmonett}, \citenamefont {DePrince}, \citenamefont {Hohenstein},
  \citenamefont {Bozkaya}, \citenamefont {Sokolov}, \citenamefont {{Di
  Remigio}}, \citenamefont {Richard}, \citenamefont {Gonthier}, \citenamefont
  {James}, \citenamefont {McAlexander}, \citenamefont {Kumar}, \citenamefont
  {Saitow}, \citenamefont {Wang}, \citenamefont {Pritchard}, \citenamefont
  {Verma}, \citenamefont {Schaefer}, \citenamefont {Patkowski}, \citenamefont
  {King}, \citenamefont {Valeev}, \citenamefont {Evangelista}, \citenamefont
  {Turney}, \citenamefont {Crawford},\ and\ \citenamefont {Sherrill}}]{psi4}%
  \BibitemOpen
  \bibfield  {author} {\bibinfo {author} {\bibfnamefont {R.~M.}\ \bibnamefont
  {Parrish}}, \bibinfo {author} {\bibfnamefont {L.~A.}\ \bibnamefont {Burns}},
  \bibinfo {author} {\bibfnamefont {D.~G.~A.}\ \bibnamefont {Smith}}, \bibinfo
  {author} {\bibfnamefont {A.~C.}\ \bibnamefont {Simmonett}}, \bibinfo {author}
  {\bibfnamefont {A.~E.}\ \bibnamefont {DePrince}}, \bibinfo {author}
  {\bibfnamefont {E.~G.}\ \bibnamefont {Hohenstein}}, \bibinfo {author}
  {\bibfnamefont {U.}~\bibnamefont {Bozkaya}}, \bibinfo {author} {\bibfnamefont
  {A.~Y.}\ \bibnamefont {Sokolov}}, \bibinfo {author} {\bibfnamefont
  {R.}~\bibnamefont {{Di Remigio}}}, \bibinfo {author} {\bibfnamefont {R.~M.}\
  \bibnamefont {Richard}}, \bibinfo {author} {\bibfnamefont {J.~F.}\
  \bibnamefont {Gonthier}}, \bibinfo {author} {\bibfnamefont {A.~M.}\
  \bibnamefont {James}}, \bibinfo {author} {\bibfnamefont {H.~R.}\ \bibnamefont
  {McAlexander}}, \bibinfo {author} {\bibfnamefont {A.}~\bibnamefont {Kumar}},
  \bibinfo {author} {\bibfnamefont {M.}~\bibnamefont {Saitow}}, \bibinfo
  {author} {\bibfnamefont {X.}~\bibnamefont {Wang}}, \bibinfo {author}
  {\bibfnamefont {B.~P.}\ \bibnamefont {Pritchard}}, \bibinfo {author}
  {\bibfnamefont {P.}~\bibnamefont {Verma}}, \bibinfo {author} {\bibfnamefont
  {H.~F.}\ \bibnamefont {Schaefer}}, \bibinfo {author} {\bibfnamefont
  {K.}~\bibnamefont {Patkowski}}, \bibinfo {author} {\bibfnamefont {R.~A.}\
  \bibnamefont {King}}, \bibinfo {author} {\bibfnamefont {E.~F.}\ \bibnamefont
  {Valeev}}, \bibinfo {author} {\bibfnamefont {F.~A.}\ \bibnamefont
  {Evangelista}}, \bibinfo {author} {\bibfnamefont {J.~M.}\ \bibnamefont
  {Turney}}, \bibinfo {author} {\bibfnamefont {T.~D.}\ \bibnamefont
  {Crawford}}, \ and\ \bibinfo {author} {\bibfnamefont {C.~D.}\ \bibnamefont
  {Sherrill}},\ }\href {\doibase 10.1021/acs.jctc.7b00174} {\bibfield
  {journal} {\bibinfo  {journal} {J. Chem. Theory Comput}\ }\textbf {\bibinfo
  {volume} {13}},\ \bibinfo {pages} {3185} (\bibinfo {year}
  {2017})}\BibitemShut {NoStop}%
\bibitem [{\citenamefont {Abraham}\ \emph {et~al.}(2019)\citenamefont
  {Abraham}, \citenamefont {Akhalwaya}, \citenamefont {Aleksandrowicz},
  \citenamefont {Alexander}, \citenamefont {Alexandrowics}, \citenamefont
  {Arbel}, \citenamefont {Asfaw}, \citenamefont {Azaustre}, \citenamefont
  {AzizNgoueya}, \citenamefont {Barkoutsos}, \citenamefont {Barron},
  \citenamefont {Bello}, \citenamefont {Ben-Haim}, \citenamefont {Bevenius},
  \citenamefont {Bishop}, \citenamefont {Bosch}, \citenamefont {Bravyi},
  \citenamefont {Bucher}, \citenamefont {Cabrera}, \citenamefont {Calpin},
  \citenamefont {Capelluto}, \citenamefont {Carballo}, \citenamefont
  {Carrascal}, \citenamefont {Chen}, \citenamefont {Chen}, \citenamefont
  {Chen}, \citenamefont {Chow}, \citenamefont {Claus}, \citenamefont {Clauss},
  \citenamefont {Cross}, \citenamefont {Cross}, \citenamefont {Cross},
  \citenamefont {Cruz-Benito}, \citenamefont {Culver}, \citenamefont
  {C{\'o}rcoles-Gonzales}, \citenamefont {Dague}, \citenamefont {Dandachi},
  \citenamefont {Dartiailh}, \citenamefont {DavideFrr}, \citenamefont {Davila},
  \citenamefont {Ding}, \citenamefont {Doi}, \citenamefont {Drechsler},
  \citenamefont {Drew}, \citenamefont {Dumitrescu}, \citenamefont {Dumon},
  \citenamefont {Duran}, \citenamefont {EL-Safty}, \citenamefont {Eastman},
  \citenamefont {Eendebak}, \citenamefont {Egger}, \citenamefont {Everitt},
  \citenamefont {Fern{\'a}ndez}, \citenamefont {Ferrera}, \citenamefont
  {Frisch}, \citenamefont {Fuhrer}, \citenamefont {GEORGE}, \citenamefont
  {Gacon}, \citenamefont {Gadi}, \citenamefont {Gago}, \citenamefont
  {Gambetta}, \citenamefont {Gammanpila}, \citenamefont {Garcia}, \citenamefont
  {Garion}, \citenamefont {Gomez-Mosquera}, \citenamefont {de~la
  Puente~Gonz{\'a}lez}, \citenamefont {Gould}, \citenamefont {Greenberg},
  \citenamefont {Grinko}, \citenamefont {Guan}, \citenamefont {Gunnels},
  \citenamefont {Haide}, \citenamefont {Hamamura}, \citenamefont {Havlicek},
  \citenamefont {Hellmers}, \citenamefont {Herok}, \citenamefont {Hillmich},
  \citenamefont {Horii}, \citenamefont {Howington}, \citenamefont {Hu},
  \citenamefont {Hu}, \citenamefont {Imai}, \citenamefont {Imamichi},
  \citenamefont {Ishizaki}, \citenamefont {Iten}, \citenamefont {Itoko},
  \citenamefont {Javadi-Abhari}, \citenamefont {Jessica}, \citenamefont
  {Johns}, \citenamefont {Kachmann}, \citenamefont {Kanazawa}, \citenamefont
  {Kang-Bae}, \citenamefont {Karazeev}, \citenamefont {Kassebaum},
  \citenamefont {King}, \citenamefont {Knabberjoe}, \citenamefont {Kovyrshin},
  \citenamefont {Krishnan}, \citenamefont {Krsulich}, \citenamefont {Kus},
  \citenamefont {LaRose}, \citenamefont {Lambert}, \citenamefont {Latone},
  \citenamefont {Lawrence}, \citenamefont {Liu}, \citenamefont {Liu},
  \citenamefont {Maeng}, \citenamefont {Malyshev}, \citenamefont {Marecek},
  \citenamefont {Marques}, \citenamefont {Mathews}, \citenamefont {Matsuo},
  \citenamefont {McClure}, \citenamefont {McGarry}, \citenamefont {McKay},
  \citenamefont {Meesala}, \citenamefont {Mevissen}, \citenamefont {Mezzacapo},
  \citenamefont {Midha}, \citenamefont {Minev}, \citenamefont {Moll},
  \citenamefont {Mooring}, \citenamefont {Morales}, \citenamefont {Moran},
  \citenamefont {Murali}, \citenamefont {M{\"u}ggenburg}, \citenamefont
  {Nadlinger}, \citenamefont {Nannicini}, \citenamefont {Nation}, \citenamefont
  {Naveh}, \citenamefont {Neuweiler}, \citenamefont {Niroula}, \citenamefont
  {Norlen}, \citenamefont {O'Riordan}, \citenamefont {Ogunbayo}, \citenamefont
  {Ollitrault}, \citenamefont {Oud}, \citenamefont {Padilha}, \citenamefont
  {Paik}, \citenamefont {Perriello}, \citenamefont {Phan}, \citenamefont
  {Pistoia}, \citenamefont {Pozas-iKerstjens}, \citenamefont {Prutyanov},
  \citenamefont {Puzzuoli}, \citenamefont {P{\'e}rez}, \citenamefont
  {Quintiii}, \citenamefont {Raymond}, \citenamefont {Redondo}, \citenamefont
  {Reuter}, \citenamefont {Rice}, \citenamefont {Rodr{\'\i}guez}, \citenamefont
  {Rossmannek}, \citenamefont {Ryu}, \citenamefont {SAPV}, \citenamefont
  {SamFerracin}, \citenamefont {Sandberg}, \citenamefont {Sathaye},
  \citenamefont {Schmitt}, \citenamefont {Schnabel}, \citenamefont
  {Schoenfeld}, \citenamefont {Scholten}, \citenamefont {Schoute},
  \citenamefont {Sertage}, \citenamefont {Setia}, \citenamefont {Shammah},
  \citenamefont {Shi}, \citenamefont {Silva}, \citenamefont {Simonetto},
  \citenamefont {Singstock}, \citenamefont {Siraichi}, \citenamefont
  {Sitdikov}, \citenamefont {Sivarajah}, \citenamefont {Sletfjerding},
  \citenamefont {Smolin}, \citenamefont {Soeken}, \citenamefont {Sokolov},
  \citenamefont {Steenken}, \citenamefont {Stypulkoski}, \citenamefont
  {Takahashi}, \citenamefont {Tavernelli}, \citenamefont {Taylor},
  \citenamefont {Taylour}, \citenamefont {Thomas}, \citenamefont {Tillet},
  \citenamefont {Tod}, \citenamefont {de~la Torre}, \citenamefont {Trabing},
  \citenamefont {Treinish}, \citenamefont {TrishaPe}, \citenamefont {Turner},
  \citenamefont {Vaknin}, \citenamefont {Valcarce}, \citenamefont {Varchon},
  \citenamefont {Vazquez}, \citenamefont {Vogt-Lee}, \citenamefont {Vuillot},
  \citenamefont {Weaver}, \citenamefont {Wieczorek}, \citenamefont {Wildstrom},
  \citenamefont {Wille}, \citenamefont {Winston}, \citenamefont {Woehr},
  \citenamefont {Woerner}, \citenamefont {Woo}, \citenamefont {Wood},
  \citenamefont {Wood}, \citenamefont {Wood}, \citenamefont {Wootton},
  \citenamefont {Yeralin}, \citenamefont {Young}, \citenamefont {Yu},
  \citenamefont {Zachow}, \citenamefont {Zdanski}, \citenamefont {Zoufal},
  \citenamefont {Zoufalc}, \citenamefont {azulehner}, \citenamefont
  {bcamorrison}, \citenamefont {brandhsn}, \citenamefont {chlorophyll zz},
  \citenamefont {dime10}, \citenamefont {drholmie}, \citenamefont
  {elfrocampeador}, \citenamefont {faisaldebouni}, \citenamefont
  {fanizzamarco}, \citenamefont {gruu}, \citenamefont {kanejess}, \citenamefont
  {klinvill}, \citenamefont {kurarrr}, \citenamefont {lerongil}, \citenamefont
  {ma5x}, \citenamefont {merav aharoni}, \citenamefont {ordmoj}, \citenamefont
  {sethmerkel}, \citenamefont {strickroman}, \citenamefont {sumitpuri},
  \citenamefont {tigerjack}, \citenamefont {toural}, \citenamefont {vvilpas},
  \citenamefont {willhbang}, \citenamefont {yang.luh},\ and\ \citenamefont
  {yotamvakninibm}}]{Qiskit}%
  \BibitemOpen
  \bibfield  {author} {\bibinfo {author} {\bibfnamefont {H.}~\bibnamefont
  {Abraham}}, \bibinfo {author} {\bibfnamefont {I.~Y.}\ \bibnamefont
  {Akhalwaya}}, \bibinfo {author} {\bibfnamefont {G.}~\bibnamefont
  {Aleksandrowicz}}, \bibinfo {author} {\bibfnamefont {T.}~\bibnamefont
  {Alexander}}, \bibinfo {author} {\bibfnamefont {G.}~\bibnamefont
  {Alexandrowics}}, \bibinfo {author} {\bibfnamefont {E.}~\bibnamefont
  {Arbel}}, \bibinfo {author} {\bibfnamefont {A.}~\bibnamefont {Asfaw}},
  \bibinfo {author} {\bibfnamefont {C.}~\bibnamefont {Azaustre}}, \bibinfo
  {author} {\bibnamefont {AzizNgoueya}}, \bibinfo {author} {\bibfnamefont
  {P.}~\bibnamefont {Barkoutsos}}, \bibinfo {author} {\bibfnamefont
  {G.}~\bibnamefont {Barron}}, \bibinfo {author} {\bibfnamefont
  {L.}~\bibnamefont {Bello}}, \bibinfo {author} {\bibfnamefont
  {Y.}~\bibnamefont {Ben-Haim}}, \bibinfo {author} {\bibfnamefont
  {D.}~\bibnamefont {Bevenius}}, \bibinfo {author} {\bibfnamefont {L.~S.}\
  \bibnamefont {Bishop}}, \bibinfo {author} {\bibfnamefont {S.}~\bibnamefont
  {Bosch}}, \bibinfo {author} {\bibfnamefont {S.}~\bibnamefont {Bravyi}},
  \bibinfo {author} {\bibfnamefont {D.}~\bibnamefont {Bucher}}, \bibinfo
  {author} {\bibfnamefont {F.}~\bibnamefont {Cabrera}}, \bibinfo {author}
  {\bibfnamefont {P.}~\bibnamefont {Calpin}}, \bibinfo {author} {\bibfnamefont
  {L.}~\bibnamefont {Capelluto}}, \bibinfo {author} {\bibfnamefont
  {J.}~\bibnamefont {Carballo}}, \bibinfo {author} {\bibfnamefont
  {G.}~\bibnamefont {Carrascal}}, \bibinfo {author} {\bibfnamefont
  {A.}~\bibnamefont {Chen}}, \bibinfo {author} {\bibfnamefont {C.-F.}\
  \bibnamefont {Chen}}, \bibinfo {author} {\bibfnamefont {R.}~\bibnamefont
  {Chen}}, \bibinfo {author} {\bibfnamefont {J.~M.}\ \bibnamefont {Chow}},
  \bibinfo {author} {\bibfnamefont {C.}~\bibnamefont {Claus}}, \bibinfo
  {author} {\bibfnamefont {C.}~\bibnamefont {Clauss}}, \bibinfo {author}
  {\bibfnamefont {A.~J.}\ \bibnamefont {Cross}}, \bibinfo {author}
  {\bibfnamefont {A.~W.}\ \bibnamefont {Cross}}, \bibinfo {author}
  {\bibfnamefont {S.}~\bibnamefont {Cross}}, \bibinfo {author} {\bibfnamefont
  {J.}~\bibnamefont {Cruz-Benito}}, \bibinfo {author} {\bibfnamefont
  {C.}~\bibnamefont {Culver}}, \bibinfo {author} {\bibfnamefont {A.~D.}\
  \bibnamefont {C{\'o}rcoles-Gonzales}}, \bibinfo {author} {\bibfnamefont
  {S.}~\bibnamefont {Dague}}, \bibinfo {author} {\bibfnamefont {T.~E.}\
  \bibnamefont {Dandachi}}, \bibinfo {author} {\bibfnamefont {M.}~\bibnamefont
  {Dartiailh}}, \bibinfo {author} {\bibnamefont {DavideFrr}}, \bibinfo {author}
  {\bibfnamefont {A.~R.}\ \bibnamefont {Davila}}, \bibinfo {author}
  {\bibfnamefont {D.}~\bibnamefont {Ding}}, \bibinfo {author} {\bibfnamefont
  {J.}~\bibnamefont {Doi}}, \bibinfo {author} {\bibfnamefont {E.}~\bibnamefont
  {Drechsler}}, \bibinfo {author} {\bibnamefont {Drew}}, \bibinfo {author}
  {\bibfnamefont {E.}~\bibnamefont {Dumitrescu}}, \bibinfo {author}
  {\bibfnamefont {K.}~\bibnamefont {Dumon}}, \bibinfo {author} {\bibfnamefont
  {I.}~\bibnamefont {Duran}}, \bibinfo {author} {\bibfnamefont
  {K.}~\bibnamefont {EL-Safty}}, \bibinfo {author} {\bibfnamefont
  {E.}~\bibnamefont {Eastman}}, \bibinfo {author} {\bibfnamefont
  {P.}~\bibnamefont {Eendebak}}, \bibinfo {author} {\bibfnamefont
  {D.}~\bibnamefont {Egger}}, \bibinfo {author} {\bibfnamefont
  {M.}~\bibnamefont {Everitt}}, \bibinfo {author} {\bibfnamefont {P.~M.}\
  \bibnamefont {Fern{\'a}ndez}}, \bibinfo {author} {\bibfnamefont {A.~H.}\
  \bibnamefont {Ferrera}}, \bibinfo {author} {\bibfnamefont {A.}~\bibnamefont
  {Frisch}}, \bibinfo {author} {\bibfnamefont {A.}~\bibnamefont {Fuhrer}},
  \bibinfo {author} {\bibfnamefont {M.}~\bibnamefont {GEORGE}}, \bibinfo
  {author} {\bibfnamefont {J.}~\bibnamefont {Gacon}}, \bibinfo {author}
  {\bibnamefont {Gadi}}, \bibinfo {author} {\bibfnamefont {B.~G.}\ \bibnamefont
  {Gago}}, \bibinfo {author} {\bibfnamefont {J.~M.}\ \bibnamefont {Gambetta}},
  \bibinfo {author} {\bibfnamefont {A.}~\bibnamefont {Gammanpila}}, \bibinfo
  {author} {\bibfnamefont {L.}~\bibnamefont {Garcia}}, \bibinfo {author}
  {\bibfnamefont {S.}~\bibnamefont {Garion}}, \bibinfo {author} {\bibfnamefont
  {J.}~\bibnamefont {Gomez-Mosquera}}, \bibinfo {author} {\bibfnamefont
  {S.}~\bibnamefont {de~la Puente~Gonz{\'a}lez}}, \bibinfo {author}
  {\bibfnamefont {I.}~\bibnamefont {Gould}}, \bibinfo {author} {\bibfnamefont
  {D.}~\bibnamefont {Greenberg}}, \bibinfo {author} {\bibfnamefont
  {D.}~\bibnamefont {Grinko}}, \bibinfo {author} {\bibfnamefont
  {W.}~\bibnamefont {Guan}}, \bibinfo {author} {\bibfnamefont {J.~A.}\
  \bibnamefont {Gunnels}}, \bibinfo {author} {\bibfnamefont {I.}~\bibnamefont
  {Haide}}, \bibinfo {author} {\bibfnamefont {I.}~\bibnamefont {Hamamura}},
  \bibinfo {author} {\bibfnamefont {V.}~\bibnamefont {Havlicek}}, \bibinfo
  {author} {\bibfnamefont {J.}~\bibnamefont {Hellmers}}, \bibinfo {author}
  {\bibfnamefont {{\L}.}~\bibnamefont {Herok}}, \bibinfo {author}
  {\bibfnamefont {S.}~\bibnamefont {Hillmich}}, \bibinfo {author}
  {\bibfnamefont {H.}~\bibnamefont {Horii}}, \bibinfo {author} {\bibfnamefont
  {C.}~\bibnamefont {Howington}}, \bibinfo {author} {\bibfnamefont
  {S.}~\bibnamefont {Hu}}, \bibinfo {author} {\bibfnamefont {W.}~\bibnamefont
  {Hu}}, \bibinfo {author} {\bibfnamefont {H.}~\bibnamefont {Imai}}, \bibinfo
  {author} {\bibfnamefont {T.}~\bibnamefont {Imamichi}}, \bibinfo {author}
  {\bibfnamefont {K.}~\bibnamefont {Ishizaki}}, \bibinfo {author}
  {\bibfnamefont {R.}~\bibnamefont {Iten}}, \bibinfo {author} {\bibfnamefont
  {T.}~\bibnamefont {Itoko}}, \bibinfo {author} {\bibfnamefont
  {A.}~\bibnamefont {Javadi-Abhari}}, \bibinfo {author} {\bibnamefont
  {Jessica}}, \bibinfo {author} {\bibfnamefont {K.}~\bibnamefont {Johns}},
  \bibinfo {author} {\bibfnamefont {T.}~\bibnamefont {Kachmann}}, \bibinfo
  {author} {\bibfnamefont {N.}~\bibnamefont {Kanazawa}}, \bibinfo {author}
  {\bibnamefont {Kang-Bae}}, \bibinfo {author} {\bibfnamefont {A.}~\bibnamefont
  {Karazeev}}, \bibinfo {author} {\bibfnamefont {P.}~\bibnamefont {Kassebaum}},
  \bibinfo {author} {\bibfnamefont {S.}~\bibnamefont {King}}, \bibinfo {author}
  {\bibnamefont {Knabberjoe}}, \bibinfo {author} {\bibfnamefont
  {A.}~\bibnamefont {Kovyrshin}}, \bibinfo {author} {\bibfnamefont
  {V.}~\bibnamefont {Krishnan}}, \bibinfo {author} {\bibfnamefont
  {K.}~\bibnamefont {Krsulich}}, \bibinfo {author} {\bibfnamefont
  {G.}~\bibnamefont {Kus}}, \bibinfo {author} {\bibfnamefont {R.}~\bibnamefont
  {LaRose}}, \bibinfo {author} {\bibfnamefont {R.}~\bibnamefont {Lambert}},
  \bibinfo {author} {\bibfnamefont {J.}~\bibnamefont {Latone}}, \bibinfo
  {author} {\bibfnamefont {S.}~\bibnamefont {Lawrence}}, \bibinfo {author}
  {\bibfnamefont {D.}~\bibnamefont {Liu}}, \bibinfo {author} {\bibfnamefont
  {P.}~\bibnamefont {Liu}}, \bibinfo {author} {\bibfnamefont {Y.}~\bibnamefont
  {Maeng}}, \bibinfo {author} {\bibfnamefont {A.}~\bibnamefont {Malyshev}},
  \bibinfo {author} {\bibfnamefont {J.}~\bibnamefont {Marecek}}, \bibinfo
  {author} {\bibfnamefont {M.}~\bibnamefont {Marques}}, \bibinfo {author}
  {\bibfnamefont {D.}~\bibnamefont {Mathews}}, \bibinfo {author} {\bibfnamefont
  {A.}~\bibnamefont {Matsuo}}, \bibinfo {author} {\bibfnamefont {D.~T.}\
  \bibnamefont {McClure}}, \bibinfo {author} {\bibfnamefont {C.}~\bibnamefont
  {McGarry}}, \bibinfo {author} {\bibfnamefont {D.}~\bibnamefont {McKay}},
  \bibinfo {author} {\bibfnamefont {S.}~\bibnamefont {Meesala}}, \bibinfo
  {author} {\bibfnamefont {M.}~\bibnamefont {Mevissen}}, \bibinfo {author}
  {\bibfnamefont {A.}~\bibnamefont {Mezzacapo}}, \bibinfo {author}
  {\bibfnamefont {R.}~\bibnamefont {Midha}}, \bibinfo {author} {\bibfnamefont
  {Z.}~\bibnamefont {Minev}}, \bibinfo {author} {\bibfnamefont
  {N.}~\bibnamefont {Moll}}, \bibinfo {author} {\bibfnamefont {M.~D.}\
  \bibnamefont {Mooring}}, \bibinfo {author} {\bibfnamefont {R.}~\bibnamefont
  {Morales}}, \bibinfo {author} {\bibfnamefont {N.}~\bibnamefont {Moran}},
  \bibinfo {author} {\bibfnamefont {P.}~\bibnamefont {Murali}}, \bibinfo
  {author} {\bibfnamefont {J.}~\bibnamefont {M{\"u}ggenburg}}, \bibinfo
  {author} {\bibfnamefont {D.}~\bibnamefont {Nadlinger}}, \bibinfo {author}
  {\bibfnamefont {G.}~\bibnamefont {Nannicini}}, \bibinfo {author}
  {\bibfnamefont {P.}~\bibnamefont {Nation}}, \bibinfo {author} {\bibfnamefont
  {Y.}~\bibnamefont {Naveh}}, \bibinfo {author} {\bibfnamefont
  {P.}~\bibnamefont {Neuweiler}}, \bibinfo {author} {\bibfnamefont
  {P.}~\bibnamefont {Niroula}}, \bibinfo {author} {\bibfnamefont
  {H.}~\bibnamefont {Norlen}}, \bibinfo {author} {\bibfnamefont {L.~J.}\
  \bibnamefont {O'Riordan}}, \bibinfo {author} {\bibfnamefont {O.}~\bibnamefont
  {Ogunbayo}}, \bibinfo {author} {\bibfnamefont {P.}~\bibnamefont
  {Ollitrault}}, \bibinfo {author} {\bibfnamefont {S.}~\bibnamefont {Oud}},
  \bibinfo {author} {\bibfnamefont {D.}~\bibnamefont {Padilha}}, \bibinfo
  {author} {\bibfnamefont {H.}~\bibnamefont {Paik}}, \bibinfo {author}
  {\bibfnamefont {S.}~\bibnamefont {Perriello}}, \bibinfo {author}
  {\bibfnamefont {A.}~\bibnamefont {Phan}}, \bibinfo {author} {\bibfnamefont
  {M.}~\bibnamefont {Pistoia}}, \bibinfo {author} {\bibfnamefont
  {A.}~\bibnamefont {Pozas-iKerstjens}}, \bibinfo {author} {\bibfnamefont
  {V.}~\bibnamefont {Prutyanov}}, \bibinfo {author} {\bibfnamefont
  {D.}~\bibnamefont {Puzzuoli}}, \bibinfo {author} {\bibfnamefont
  {J.}~\bibnamefont {P{\'e}rez}}, \bibinfo {author} {\bibnamefont {Quintiii}},
  \bibinfo {author} {\bibfnamefont {R.}~\bibnamefont {Raymond}}, \bibinfo
  {author} {\bibfnamefont {R.~M.-C.}\ \bibnamefont {Redondo}}, \bibinfo
  {author} {\bibfnamefont {M.}~\bibnamefont {Reuter}}, \bibinfo {author}
  {\bibfnamefont {J.}~\bibnamefont {Rice}}, \bibinfo {author} {\bibfnamefont
  {D.~M.}\ \bibnamefont {Rodr{\'\i}guez}}, \bibinfo {author} {\bibfnamefont
  {M.}~\bibnamefont {Rossmannek}}, \bibinfo {author} {\bibfnamefont
  {M.}~\bibnamefont {Ryu}}, \bibinfo {author} {\bibfnamefont {T.}~\bibnamefont
  {SAPV}}, \bibinfo {author} {\bibnamefont {SamFerracin}}, \bibinfo {author}
  {\bibfnamefont {M.}~\bibnamefont {Sandberg}}, \bibinfo {author}
  {\bibfnamefont {N.}~\bibnamefont {Sathaye}}, \bibinfo {author} {\bibfnamefont
  {B.}~\bibnamefont {Schmitt}}, \bibinfo {author} {\bibfnamefont
  {C.}~\bibnamefont {Schnabel}}, \bibinfo {author} {\bibfnamefont
  {Z.}~\bibnamefont {Schoenfeld}}, \bibinfo {author} {\bibfnamefont {T.~L.}\
  \bibnamefont {Scholten}}, \bibinfo {author} {\bibfnamefont {E.}~\bibnamefont
  {Schoute}}, \bibinfo {author} {\bibfnamefont {I.~F.}\ \bibnamefont
  {Sertage}}, \bibinfo {author} {\bibfnamefont {K.}~\bibnamefont {Setia}},
  \bibinfo {author} {\bibfnamefont {N.}~\bibnamefont {Shammah}}, \bibinfo
  {author} {\bibfnamefont {Y.}~\bibnamefont {Shi}}, \bibinfo {author}
  {\bibfnamefont {A.}~\bibnamefont {Silva}}, \bibinfo {author} {\bibfnamefont
  {A.}~\bibnamefont {Simonetto}}, \bibinfo {author} {\bibfnamefont
  {N.}~\bibnamefont {Singstock}}, \bibinfo {author} {\bibfnamefont
  {Y.}~\bibnamefont {Siraichi}}, \bibinfo {author} {\bibfnamefont
  {I.}~\bibnamefont {Sitdikov}}, \bibinfo {author} {\bibfnamefont
  {S.}~\bibnamefont {Sivarajah}}, \bibinfo {author} {\bibfnamefont {M.~B.}\
  \bibnamefont {Sletfjerding}}, \bibinfo {author} {\bibfnamefont {J.~A.}\
  \bibnamefont {Smolin}}, \bibinfo {author} {\bibfnamefont {M.}~\bibnamefont
  {Soeken}}, \bibinfo {author} {\bibfnamefont {I.~O.}\ \bibnamefont {Sokolov}},
  \bibinfo {author} {\bibfnamefont {D.}~\bibnamefont {Steenken}}, \bibinfo
  {author} {\bibfnamefont {M.}~\bibnamefont {Stypulkoski}}, \bibinfo {author}
  {\bibfnamefont {H.}~\bibnamefont {Takahashi}}, \bibinfo {author}
  {\bibfnamefont {I.}~\bibnamefont {Tavernelli}}, \bibinfo {author}
  {\bibfnamefont {C.}~\bibnamefont {Taylor}}, \bibinfo {author} {\bibfnamefont
  {P.}~\bibnamefont {Taylour}}, \bibinfo {author} {\bibfnamefont
  {S.}~\bibnamefont {Thomas}}, \bibinfo {author} {\bibfnamefont
  {M.}~\bibnamefont {Tillet}}, \bibinfo {author} {\bibfnamefont
  {M.}~\bibnamefont {Tod}}, \bibinfo {author} {\bibfnamefont {E.}~\bibnamefont
  {de~la Torre}}, \bibinfo {author} {\bibfnamefont {K.}~\bibnamefont
  {Trabing}}, \bibinfo {author} {\bibfnamefont {M.}~\bibnamefont {Treinish}},
  \bibinfo {author} {\bibnamefont {TrishaPe}}, \bibinfo {author} {\bibfnamefont
  {W.}~\bibnamefont {Turner}}, \bibinfo {author} {\bibfnamefont
  {Y.}~\bibnamefont {Vaknin}}, \bibinfo {author} {\bibfnamefont {C.~R.}\
  \bibnamefont {Valcarce}}, \bibinfo {author} {\bibfnamefont {F.}~\bibnamefont
  {Varchon}}, \bibinfo {author} {\bibfnamefont {A.~C.}\ \bibnamefont
  {Vazquez}}, \bibinfo {author} {\bibfnamefont {D.}~\bibnamefont {Vogt-Lee}},
  \bibinfo {author} {\bibfnamefont {C.}~\bibnamefont {Vuillot}}, \bibinfo
  {author} {\bibfnamefont {J.}~\bibnamefont {Weaver}}, \bibinfo {author}
  {\bibfnamefont {R.}~\bibnamefont {Wieczorek}}, \bibinfo {author}
  {\bibfnamefont {J.~A.}\ \bibnamefont {Wildstrom}}, \bibinfo {author}
  {\bibfnamefont {R.}~\bibnamefont {Wille}}, \bibinfo {author} {\bibfnamefont
  {E.}~\bibnamefont {Winston}}, \bibinfo {author} {\bibfnamefont {J.~J.}\
  \bibnamefont {Woehr}}, \bibinfo {author} {\bibfnamefont {S.}~\bibnamefont
  {Woerner}}, \bibinfo {author} {\bibfnamefont {R.}~\bibnamefont {Woo}},
  \bibinfo {author} {\bibfnamefont {C.~J.}\ \bibnamefont {Wood}}, \bibinfo
  {author} {\bibfnamefont {R.}~\bibnamefont {Wood}}, \bibinfo {author}
  {\bibfnamefont {S.}~\bibnamefont {Wood}}, \bibinfo {author} {\bibfnamefont
  {J.}~\bibnamefont {Wootton}}, \bibinfo {author} {\bibfnamefont
  {D.}~\bibnamefont {Yeralin}}, \bibinfo {author} {\bibfnamefont
  {R.}~\bibnamefont {Young}}, \bibinfo {author} {\bibfnamefont
  {J.}~\bibnamefont {Yu}}, \bibinfo {author} {\bibfnamefont {C.}~\bibnamefont
  {Zachow}}, \bibinfo {author} {\bibfnamefont {L.}~\bibnamefont {Zdanski}},
  \bibinfo {author} {\bibfnamefont {C.}~\bibnamefont {Zoufal}}, \bibinfo
  {author} {\bibnamefont {Zoufalc}}, \bibinfo {author} {\bibnamefont
  {azulehner}}, \bibinfo {author} {\bibnamefont {bcamorrison}}, \bibinfo
  {author} {\bibnamefont {brandhsn}}, \bibinfo {author} {\bibnamefont
  {chlorophyll zz}}, \bibinfo {author} {\bibnamefont {dime10}}, \bibinfo
  {author} {\bibnamefont {drholmie}}, \bibinfo {author} {\bibnamefont
  {elfrocampeador}}, \bibinfo {author} {\bibnamefont {faisaldebouni}}, \bibinfo
  {author} {\bibnamefont {fanizzamarco}}, \bibinfo {author} {\bibnamefont
  {gruu}}, \bibinfo {author} {\bibnamefont {kanejess}}, \bibinfo {author}
  {\bibnamefont {klinvill}}, \bibinfo {author} {\bibnamefont {kurarrr}},
  \bibinfo {author} {\bibnamefont {lerongil}}, \bibinfo {author} {\bibnamefont
  {ma5x}}, \bibinfo {author} {\bibnamefont {merav aharoni}}, \bibinfo {author}
  {\bibnamefont {ordmoj}}, \bibinfo {author} {\bibnamefont {sethmerkel}},
  \bibinfo {author} {\bibnamefont {strickroman}}, \bibinfo {author}
  {\bibnamefont {sumitpuri}}, \bibinfo {author} {\bibnamefont {tigerjack}},
  \bibinfo {author} {\bibnamefont {toural}}, \bibinfo {author} {\bibnamefont
  {vvilpas}}, \bibinfo {author} {\bibnamefont {willhbang}}, \bibinfo {author}
  {\bibnamefont {yang.luh}}, \ and\ \bibinfo {author} {\bibnamefont
  {yotamvakninibm}},\ }\href {\doibase 10.5281/zenodo.2562110} {\enquote
  {\bibinfo {title} {Qiskit: An open-source framework for quantum computing},}\
  } (\bibinfo {year} {2019})\BibitemShut {NoStop}%
\bibitem [{\citenamefont {Guerreschi}\ \emph {et~al.}(2020)\citenamefont
  {Guerreschi}, \citenamefont {Hogaboam}, \citenamefont {Baruffa},\ and\
  \citenamefont {Sawaya}}]{Guerreschi2020}%
  \BibitemOpen
  \bibfield  {author} {\bibinfo {author} {\bibfnamefont {G.~G.}\ \bibnamefont
  {Guerreschi}}, \bibinfo {author} {\bibfnamefont {J.}~\bibnamefont
  {Hogaboam}}, \bibinfo {author} {\bibfnamefont {F.}~\bibnamefont {Baruffa}}, \
  and\ \bibinfo {author} {\bibfnamefont {N.~P.~D.}\ \bibnamefont {Sawaya}},\
  }\href {\doibase 10.1088/2058-9565/ab8505} {\bibfield  {journal} {\bibinfo
  {journal} {Quantum Sci. Technol.}\ }\textbf {\bibinfo {volume} {5}},\
  \bibinfo {pages} {034007} (\bibinfo {year} {2020})}\BibitemShut {NoStop}%
\end{thebibliography}%

\vspace{1cm}

\onecolumngrid

\appendix
\section*{Appendix}

\section{LDCA Details}\label{app:sec:ldca_details}
\begin{figure}[ht]
\centering
\includegraphics[width=0.8\textwidth]{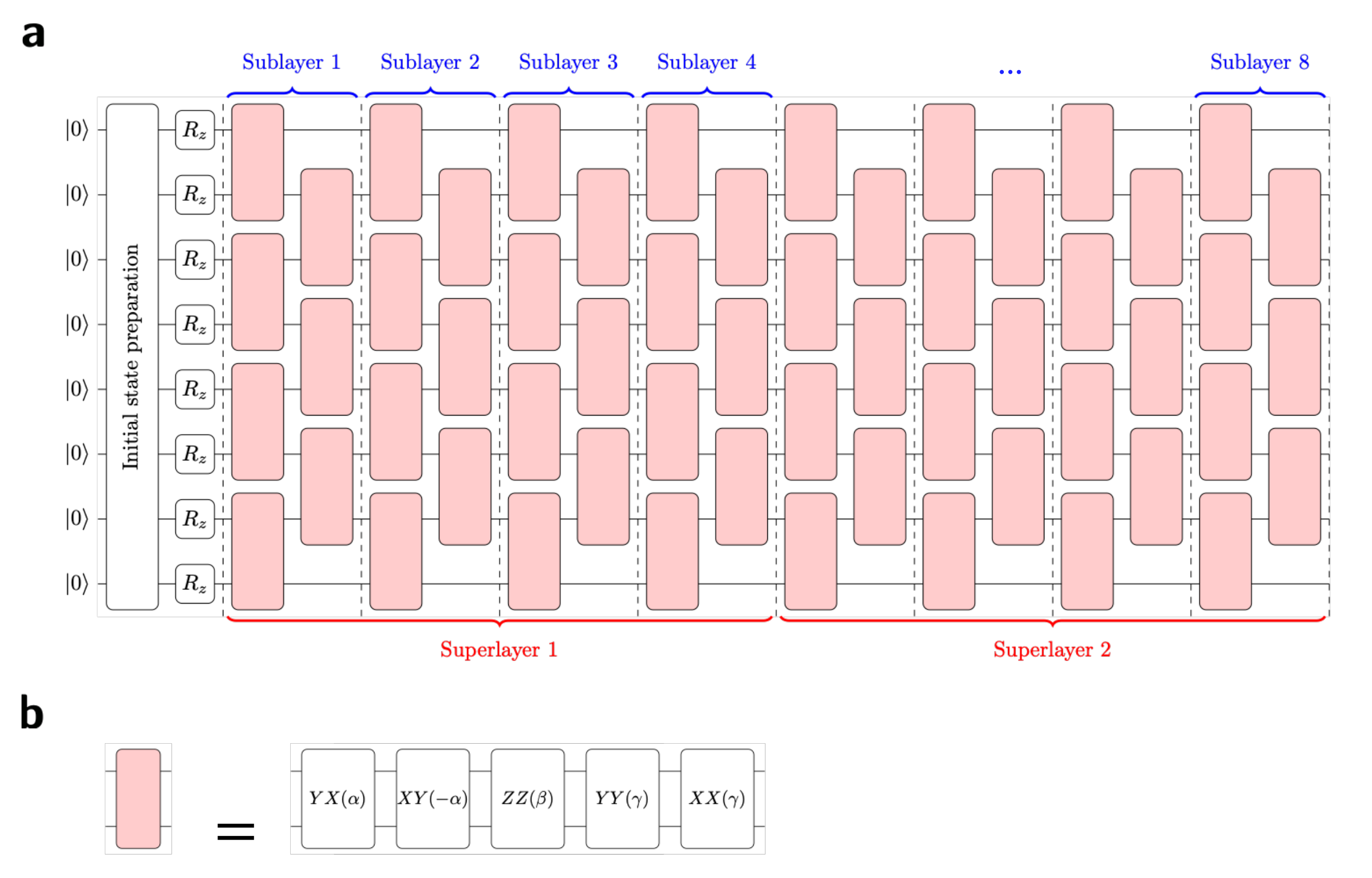}
\caption{
Gate composition of a simplified version of the low-depth circuit ansatz (LDCA) from Ref. \cite{Dallaire-Demers2018}. 
Circuit in (a) is an instance of LDCA for 8 qubits.
Each entangling operation in LDCA is composed of five two-qubit gate operations with three free parameters, as shown in (b).
The $YX$ and $XY$ operations sharing a parameter (with opposite signs) and $XX$ and $YY$ operations sharing another parameter leads to an ansatz that preserves particle number, assuming the initial state is e.g. the Hartree-Fock state.}
\label{fig:ldca_circuit}
\end{figure}

In this work, we consider three types of ansatze: (a) the unitary coupled-cluster ansatz including singles and doubles excitations (UCCSD), (b) a variant of UCC ansatz using chained generalized singles and paired doubles excitations called $k$-UpCCGSD \cite{Lee2019}, and (c) a simplified version of the low-depth circuit ansatz (LDCA) \cite{Dallaire-Demers2018}.
While gate compositions for (a) and (b) can be found in Refs. \cite{Romero2017strategies} and \cite{Lee2019} respectively, we provide the gate composition for the particle number conserving construction of LDCA in Figure \ref{fig:ldca_circuit}.
LDCA has a compact nested structure with several levels of layers.
Its outermost layer, or what we call ``superlayer,'' comprises $\ceil{\frac{n}{2}}$ ``sublayers,'' where $n$ is the number of qubits, and each ``sublayer'' comprises two layers of nearest neighbor two-qubit blocks. The first layer of two-qubit blocks starts with the first qubit, and the second layer starts with the second qubit. 
Each two-qubit block is composed of five two-qubit gate operations with three free parameters, as shown in Figure \ref{fig:ldca_circuit}(b).
While the original formulation of LDCA allowed five free parameters, one for each two-qubit gate, and included particle number constraints, our preliminary calculations (unreported) showed that using the particle number conserving version of LDCA led to significantly better convergence behaviors.
For each two-qubit gate operation labeled $P(\theta)$, the gate is defined as $\text{exp}(-i \frac{\theta}{2} P)$.

\section{Ansatz cost reductions for $\text{H}_2\text{O}$}\label{app:sec:h2o_reductions}
In Section \ref{sec:ansatz_reduction}, we provide average percent reductions in circuit resources, namely circuit depth and two-qubit gate counts, from using PECT as an optimization strategy.
For simulations estimating ground state energies of water, these quantities may not be as meaningful because optimizations of several bond lengths did not converge to chemically accurate energies.
To provide a more meaningful report of reductions in circuit resources, we provide tables with reduction numbers at bond lengths at which both non-PECT and PECT calculations produced final energies with errors below the chemical accuracy threshold.
We show resulting percent reductions in depths and two-qubit gate counts arriving from using PECT for UCCSD in Table \ref{table:h2o_uccsd_reductions}, 4-UpCCGSD in Table \ref{table:h2o_4_upccgsd_reductions}, and 5-UpCCGSD in Table \ref{table:h2o_5_upccgsd_reductions}.

\begin{table}[]
\centering
\def\arraystretch{1.2}\tabcolsep=8pt
\begin{tabular}{@{}c c c@{}}
\toprule
\textbf{\begin{tabular}[c]{@{}c@{}}Bond \\ length {[}\angstrom{]}\end{tabular}} & \textbf{\begin{tabular}[c]{@{}c@{}}\% reduced \\ depth\end{tabular}} & \textbf{\begin{tabular}[c]{@{}c@{}}\% reduced two-qubit\\ gate count\end{tabular}} \\ \midrule
0.8 & 10\% & 10\% \\
1.0 & 10\% & 10\% \\
1.2 & 15\% & 15\% \\
1.4 & 9\% & 9\% \\
1.6 & 21\% & 20\% \\
1.8 & 9\% & 8\% \\
2.6 & 24\% & 24\% \\ \bottomrule
\end{tabular}
\caption{Percent reductions in the circuit depths and two-qubit gate counts from utilizing PECT to optimize VQE calculations with the UCCSD ansatz at various O-H bond lengths of a water molecule. At these bond lengths, both non-PECT and PECT calculations achieved energy errors below the chemical accuracy threshold.}
\label{table:h2o_uccsd_reductions}
\end{table}

\begin{table}[]
\centering
\def\arraystretch{1.2}\tabcolsep=8pt
\begin{tabular}{@{}ccc@{}}
\toprule
\textbf{\begin{tabular}[c]{@{}c@{}}Bond \\ length {[}\angstrom{]}\end{tabular}} & \textbf{\begin{tabular}[c]{@{}c@{}}\% reduced \\ depth\end{tabular}} & \textbf{\begin{tabular}[c]{@{}c@{}}\% reduced two-qubit\\ gate count\end{tabular}} \\ \midrule
0.8 & 15\% & 32\% \\
1.2 & 17\% & 29\% \\
1.4 & 21\% & 33\% \\
1.6 & 18\% & 37\% \\
2.0 & 13\% & 25\% \\ \bottomrule
\end{tabular}
\caption{Percent reductions in the circuit depths and two-qubit gate counts from utilizing PECT to optimize VQE calculations with the 4-UpCCGSD ansatz at various O-H bond lengths of a water molecule. At these bond lengths, both non-PECT and PECT calculations achieved energy errors below the chemical accuracy threshold.}
\label{table:h2o_4_upccgsd_reductions}
\end{table}

\begin{table}[]
\centering
\def\arraystretch{1.2}\tabcolsep=8pt
\begin{tabular}{@{}ccc@{}}
\toprule
\textbf{\begin{tabular}[c]{@{}c@{}}Bond \\ length {[}\angstrom{]}\end{tabular}} & \textbf{\begin{tabular}[c]{@{}c@{}}\% reduced \\ depth\end{tabular}} & \textbf{\begin{tabular}[c]{@{}c@{}}\% reduced two-qubit\\ gate count\end{tabular}} \\ \midrule
0.8 & 17\% & 30\% \\
1.2 & 16\% & 36\% \\
1.4 & 18\% & 30\% \\
1.6 & 15\% & 27\% \\
2.0 & 19\% & 35\% \\ \bottomrule
\end{tabular}
\caption{Percent reductions in the circuit depths and two-qubit gate counts from utilizing PECT to optimize VQE calculations with the 5-UpCCGSD ansatz at various O-H bond lengths of a water molecule. At these bond lengths, both non-PECT and PECT calculations achieved energy errors below the chemical accuracy threshold.}
\label{table:h2o_5_upccgsd_reductions}
\end{table}

\section{Parameter dynamics of $k$-UpCCGSD and LDCA optimizations}\label{app:sec:param_dynamics_app}

\begin{figure}[ht]
\centering
\includegraphics[width=0.75\textwidth]{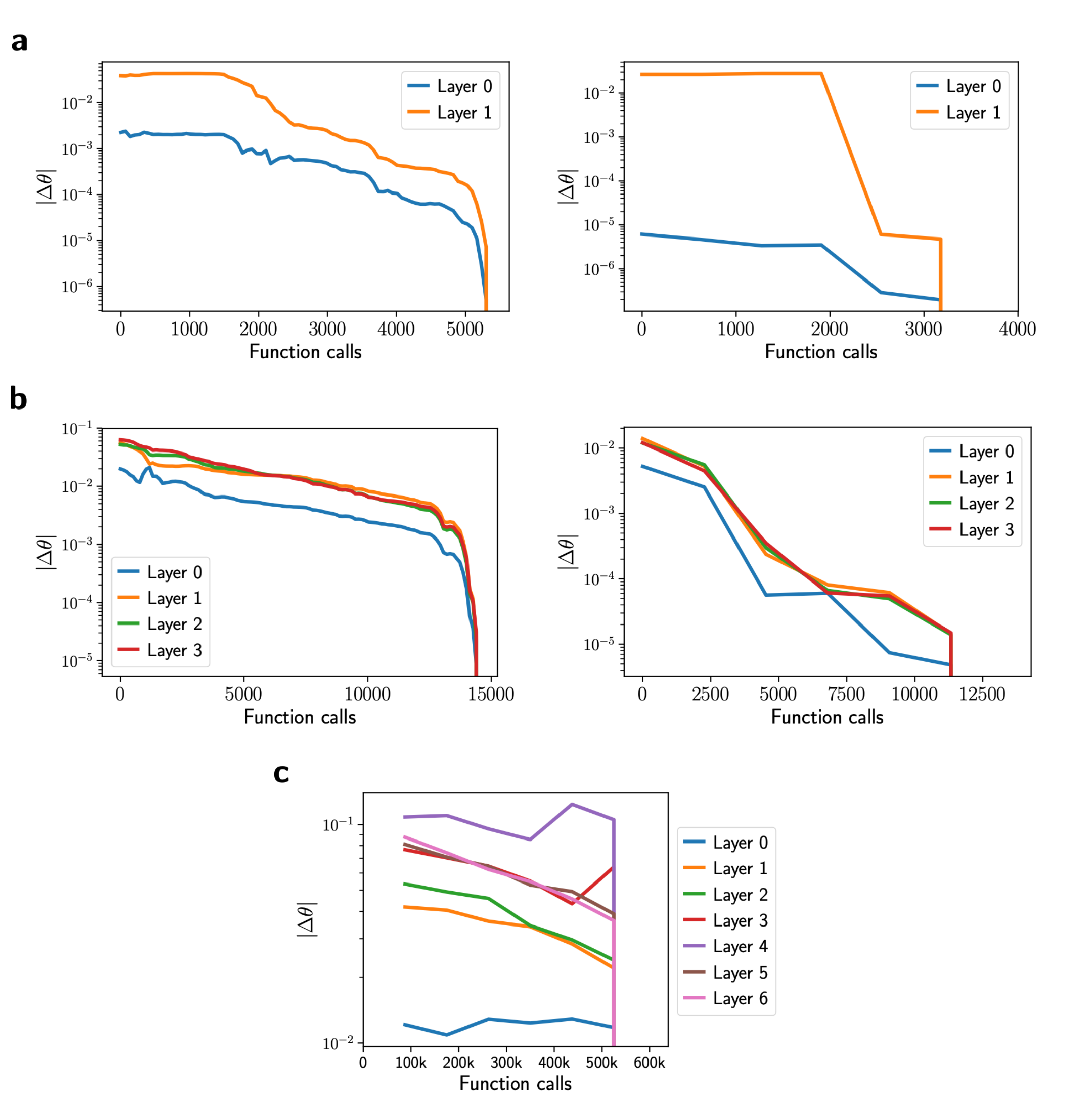}
\caption{
Absolute difference of parameter values with respect to their final values averaged over each circuit layer at each optimization iteration for $k$-UpCCGSD and LDCA circuits.
We compute the averaged parameter distance for:
(a) $2$-UpCCGSD circuits describing ground state of LiH without PECT (left) and with PECT (right), 
(b) $4$-UpCCGSD circuits describing ground state of $\text{H}_2\text{O}$ without PECT (left) and with PECT (right), and 
(c) LDCA circuit for describing ground state of $\text{H}_2\text{O}$ with PECT.
}
\label{fig:param_dynamics_appendix}
\end{figure}

In Section \ref{sec:param_dynamics}, we discussed how parameters of a multi-layered PQC evolve throughout an optimization.
For example, in the case of LDCA describing ground states of LiH, parameters of earlier layers started and generally remained closer to their final values compared to parameters of later layers.
In Figure \ref{fig:param_dynamics_appendix}, we provide further numerical evidence for the $k$-UpCCGSD and LDCA ansatze describing the ground state of LiH at $R=2.5 \angstrom$ and ground state of $\text{H}_2\text{O}$ at $R_{\text{O-H}}=2.4 \angstrom$.
In each plot, similar behavior in the average parameter movement is observed.
For certain optimization instances such as optimizing LDCA for $\text{H}_2\text{O}$ using PECT in Figure \ref{fig:param_dynamics_appendix}(c), we observed parameters of layer 6 changing less than parameters of layers 3, 4, and 5.
This appears to suggest that after some number of circuit layers, the order in the degree of movement becomes less structured. 
Nevertheless, plots in Figure \ref{fig:param_dynamics_appendix} support the idea that parameters of especially the first few layers should be carefully initialized to facilitate the tuning of later circuit layers.

\section{Estimating the potential energy surface for $\text{H}_2\text{O}$}\label{app:sec:h2o_pes}
\begin{figure}[ht]
\centering
\includegraphics[width=0.8\textwidth]{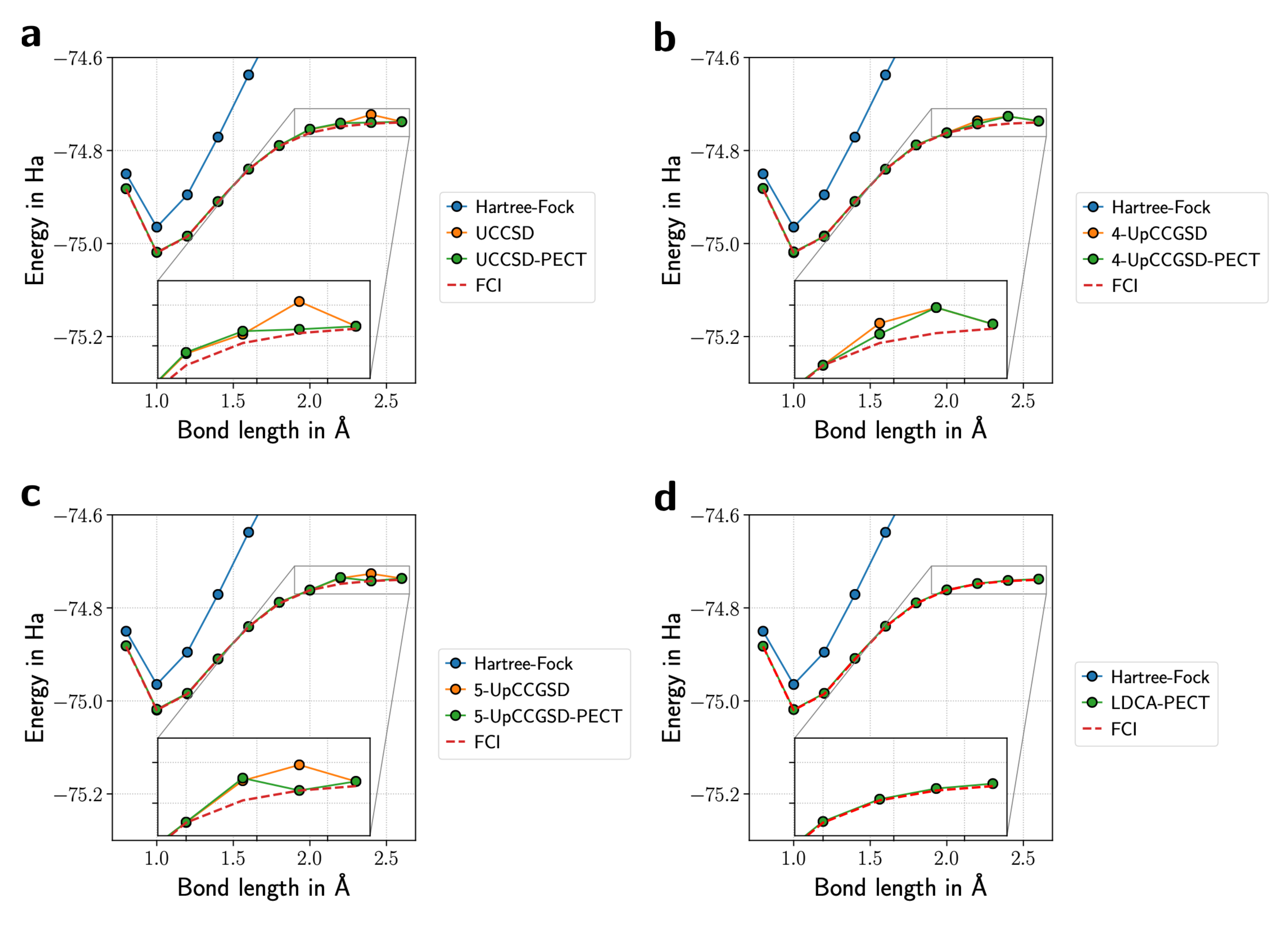}
\caption{
Potential energy surfaces of H$_2$O constructed using optimized (a) UCCSD, (b) 4-UpCCGSD, (c) 5-UpCCGSD, and (d) LDCA circuits.
We show the energies from optimizations without PECT in orange markers and ones from optimizations with PECT in green markers.
Inset in each plot zooms in on the same region from $R_{\text{O-H}} = 1.9$ to $2.65 \angstrom$ and energies from -74.77 to -74.71 Hartrees.
}
\label{fig:h2o_pes}
\end{figure}

We plot the VQE-optimized energies for water in Figure \ref{fig:h2o_pes}.
Since potential energy surfaces are often used as an input for small scale quantum dynamics simulations, the uniformity of the energy errors across different bond lengths (i.e. smoothness) is an important comparison metric for electronic structure methods.
We observe that using UCCSD with PECT, the resulting potential energy surface appears smoother than the one generated using UCCSD without PECT, which has a bump in the surface due to the high energy error at $R_{\text{O-H}} = 2.4 \angstrom$.
With $k$-UpCCGSD ansatze, the potential energy surfaces are less smooth due to inconsistent improvements in energies at particular bond lengths.
For smoother surfaces, we may need alternative methods for initializing $k$-UpCCGSD parameters.
With LDCA, we note that its PECT calculation yielded a smooth and accurate potential energy surface compared to resulting surfaces of other methods.

\end{document}